\documentclass[pra,longbibliography,showpacs,nofootinbib,superscriptaddress,notitlepage,accepted=2018-08-21]{quantumarticle}

\usepackage[bookmarks=false,linkcolor=blue,urlcolor=blue,colorlinks,citecolor=blue]{hyperref}
\usepackage{algorithm}
\usepackage{algpseudocode}
\usepackage{graphicx}  
\usepackage{epstopdf}
\usepackage{dcolumn}   
\usepackage{bm}        
\usepackage{amsmath}
\usepackage{amssymb}   
\usepackage{wrapfig}
\usepackage{array}
\usepackage[caption=false]{subfig}
\usepackage{exscale,relsize}
\usepackage[usenames, dvipsnames]{color}
\usepackage{times, verbatim}
\usepackage{pstricks}
\usepackage{bbold}

\usepackage{pst-all}

\newcommand{\bra}[1]{\langle #1|}
\newcommand{\ket}[1]{|#1\rangle}

\begin{document}

\title{Modeling noise and error correction for Majorana-based quantum computing}

\author{Christina Knapp}
\affiliation{Department of Physics, University of California, Santa Barbara,
	California 93106 USA}

\author{Michael Beverland}
\affiliation{Station Q Quantum Architectures and Computation Group, Microsoft Research, Redmond, Washington  98052 USA}

\author{Dmitry I. Pikulin}
\affiliation{Station Q, Microsoft Research, Santa Barbara, California 93106-6105 USA}
	
\author{Torsten Karzig}
\affiliation{Station Q, Microsoft Research, Santa Barbara, California 93106-6105 USA}

\begin{abstract}
Majorana-based quantum computing seeks to use the non-local nature of Majorana zero modes to store and manipulate quantum information in a topologically protected way.  While noise is anticipated to be significantly suppressed in such systems, finite temperature and system size result in residual errors.  In this work, we connect the underlying physical error processes in Majorana-based systems to the noise models used in a fault tolerance analysis.  Standard qubit-based noise models built from Pauli operators do not capture leading order noise processes arising from quasiparticle poisoning events, thus it is not obvious {\it a priori} that such noise models can be usefully applied to a Majorana-based system.  We develop stochastic Majorana noise models that are generalizations of the standard qubit-based models and connect the error probabilities defining these models to parameters of the physical system.  Using these models, we compute pseudo-thresholds for the $d=5$ Bacon-Shor subsystem code. Our results emphasize the importance of correlated errors induced in multi-qubit measurements. Moreover, we find that for sufficiently fast quasiparticle relaxation the errors are well described by Pauli operators. This work bridges the divide between physical errors in Majorana-based quantum computing architectures and the significance of these errors in a quantum error correcting code.
\end{abstract}

\maketitle

\section{Introduction}\label{sec:motivation}

Topological phases of matter provide an attractive platform for quantum computation due to the possibility of manipulating information stored in the non-local degenerate state space of non-Abelian anyons. This gives rise to the idea of topological quantum computation~\cite{Kitaev03,Nayak08}. Any local probe of the system does not provide information about the qubit state; this {\it topological protection} of the encoded information is anticipated to suppress decoherence from the environment, thereby leading to exceptionally long qubit lifetimes.  At present, the most promising approach for topological quantum computing uses the non-Abelian properties of Majorana zero modes (MZMs) that can be localized at the ends of topological superconducting wires~\cite{Kitaev01,Lutchyn10,Oreg10,Mourik12,Rokhinson12,Deng12,Das12,Finck12,Churchill13,Albrecht16,Deng16,Lutchyn17}\footnote{While strictly speaking MZMs in a superconductor are not dynamical and therefore are defects rather than quasiparticles, they can be utilized for topological quantum computing in much the same way as non-Abelian anyons.}.   

Each MZM is described by a Hermitian operator, ${\gamma_a=\gamma_a^\dagger}$, which obeys fermionic anticommutation relations ${\{\gamma_a,\gamma_b\}=2\delta_{a,b}}$.  A pair of MZMs corresponds to a single fermionic state with occupation number ${c^\dagger c=\left(1+i\gamma_a\gamma_b\right)/2}$, where $c=\left(\gamma_a+i\gamma_b\right)/2$ is a Dirac fermion operator. Quantum information can be encoded through the occupation of the fermionic state that can be unoccupied (even parity) or occupied (odd parity).  Application of one of the Majorana operators to the state flips the fermion parity.  When using a modular approach where each qubit is encoded in a separate superconducting island, at least four MZMs are required to store a qubit~\cite{Landau2016}.  If the total fermion parity of the superconducting island is fixed to be even, then the qubit states correspond to both pairs of MZMs having even parity or both pairs having odd parity.  
It is a choice of basis how to pair the MZMs ({\it e.g.}, pairing $\gamma_1$ with $\gamma_2$ and $\gamma_3$ with $\gamma_4$ as opposed to $\gamma_1$ with $\gamma_3$ and $\gamma_2$ with $\gamma_4$); which means that unlike other qubit schemes, phase errors and bit-flip errors result from the same physical processes.  That is, for one encoding choice, the operator $\gamma_1\gamma_3$ flips the qubit state, while for the other, it results in a phase error.  
Physically, errors described by Majorana operators are due to, for instance, thermally excited quasiparticles, finite hybridization of MZMs, or external quasiparticle poisoning of the island. In the absence of non-equilibrium noise sources, these errors will generally be exponentially suppressed in ratios involving the physical parameters of the system that are expected to be large, but cannot be made arbitrarily small for practical reasons~\cite{Goldstein11,Budich12,Rainis12,Mazza13,Hu15,Pedrocchi15,Ippoliti16,Knapp2017}.

The small error rates discussed above set an upper bound on the qubit lifetime.  In order to store information for a longer time, it is necessary to perform quantum error correction on the system.  
For a given quantum error correcting code, if the error rate is below a particular value, known as the code's pseudo-threshold, the qubit's lifetime is increased.  Studying the effectiveness of quantum error correcting codes for Majorana-based systems is a relatively recent and important development in the field of topological quantum computation~\cite{Bravyi10,Terhal12,Brell14,Vijay15,Li16,Vijay16,Landau2016,Plugge2016,Li17,Litinski2017,Litinski17b,Hastings17,Vijay2017,Litinski18}. Thus far, the majority of studies (with the exceptions of Refs.~\onlinecite{Bravyi10,Brell14,Hastings17,Vijay2017}) have assumed qubit-based noise models, which approximate noise in the system by Pauli errors and measurement bit-flips.   For a MZM system, errors are naturally modeled by products of Majorana operators.  Some of the most important types of errors ({\it e.g.}, quasiparticle poisoning~\cite{Goldstein11,Budich12,Rainis12}) involve an odd number of Majorana operators, which take the system out of the computational subspace and are therefore not described by Pauli operators.  Thus, unless the effect of these errors can be captured with measurement bit-flips, the noise affecting a MZM system is outside the scope of a qubit-based noise model.  

The purpose of this paper is to connect the underlying physical error processes in a Majorana-based quantum computing architecture to a noise model that can be used to analyze fault tolerance of the system. To this end, we develop stochastic Majorana noise models from physical considerations of proposed MZM systems and discuss the parameters that control the probabilities of applying different products of Majorana operators and the probability of measurement bit-flips. These noise models reduce to qubit-based noise models when the probabilities of an odd number of Majorana operators being applied is set to zero.  By analyzing these noise models for a small Bacon-Shor subsystem code~\cite{Bacon06,Kribs05,Shor96}, we find that correlated errors induced through multi-qubit measurements are most problematic for fault tolerance, and therefore, would be most important to minimize in a Majorana-based quantum computing architecture. We find that for charging-energy-protected MZM qubits~\cite{Plugge2017,Karzig17}, the pseudo-threshold values calculated with a Majorana noise model are well approximated by a qubit-based noise model for finite, but sufficiently small, odd-Majorana error probabilities. More generally, our work provides and exemplifies a framework for analyzing Majorana-based error correction that can be extended to other physical MZM architectures.  

\subsection{Guide to the Reader}

The remainder of this paper is organized as follows.  In Section~\ref{sec:noise-models}, we present four stochastic Majorana noise models in order of simplest to most realistic.  We define exactly what is meant by a stochastic Majorana noise model and justify why stochastic noise is an appropriate approximation of the errors occurring in a Majorana-based system.  We further discuss in what limit our models reduce to analogous qubit-based noise models.  In Section~\ref{sec:probabilities}, we use physical considerations to motivate the errors included in the noise models.  We focus on a particular Majorana-based qubit proposal \cite{Karzig17} and estimate the dependence of error rates on physical parameters in this system.  A summary of how these physical considerations relate to the parameters of the noise models is given in Table~\ref{fig:prob-summary}.  In Section~\ref{sec:threshold}, we apply the Bacon-Shor subsystem code to our noise models to estimate pseudo-thresholds and the relative importance of the different error processes from an error correction viewpoint. One motivation for using the Bacon-Shor subsystem code is that it can be implemented using only two-qubit measurements, which we anticipate to be easier to perform experimentally on a MZM system than the higher-weight measurements required for more standard error-correcting codes ({\it e.g.}, the surface code)\footnote{Higher-weight measurements could alternatively be built from sequences of two-qubit measurements, but this is likely to introduce additional errors.}.  We discuss experimental implications of our analysis at the end of this section.  In Section~\ref{sec:extensions}, we discuss extensions of our analysis to other MZM qubit proposals and error correcting codes. Finally, in Section~\ref{sec:conclusions}, we conclude and identify future directions.  Details of the various discussions are relegated to the appendices.  

This paper is intended for both a condensed matter and a fault tolerance audience; as such it contains review material that readers from either community may find superfluous.  Readers from a fault tolerance background may wish to skip Sections~\ref{sec:stochastic-qubit}, \ref{sec:subsystem}, and \ref{sec:BS}.
For those already familiar with topological superconductivity, Sections~\ref{sec:physical} and \ref{sec:error-probabilities} may be skimmed.  Additionally, those not interested in the technical aspects of fault tolerance might wish to skim Sections~\ref{sec:subsystem} and \ref{sec:fault-tolerant}, while those who are not invested in the physical noise processes affecting a MZM system should skim Sections~\ref{sec:physical} and \ref{sec:error-probabilities}.   The main results of this paper are contained in Sections~\ref{sec:stochastic-MZM} and \ref{sec:results}.

\section{Stochastic Majorana Noise Models}\label{sec:noise-models}

In this section, we develop stochastic Majorana noise models analogous to the standard qubit-based noise models.  There are several motivations for tailoring a noise model to a system of MZMs. (1) In general, the physical sources of errors are best understood in terms of interactions of the environment with the MZMs; a Majorana noise model is therefore more transparently connected to the physical system, affords a more precise description of the noise that can lead to more realistic quantum error correction simulations, and can be applied independently of the encoding of quantum information.  (2) Majorana-based quantum computing architectures do not necessarily group MZMs into qubits; as such, noise models that describe environmental effects as qubit errors are not applicable to all MZM systems. For instance, a Majorana fermion code~\cite{Bravyi10,Hastings17,Vijay2017} could not be fully analyzed with a qubit noise model. (3) Even when MZMs are arranged into qubits, some of the most common types of errors take the system out of the computational subspace ({\it e.g.}, quasiparticle poisoning~\cite{Goldstein11,Budich12,Rainis12}), and are therefore not captured by the probabilistic application of Pauli errors.

Throughout this paper, we consider a set of $2n$ Majorana zero modes (MZMs), with corresponding operators $\gamma_1,\gamma_2,\dots \gamma_{2n}$.  Noise models discretize time into {\it time steps}.  In a stochastic Majorana noise model, after a time step $\tau$, a probabilistically generated string of Majorana operators, $\gamma_1^{a_1}\dots \gamma_{2n}^{a_{2n}}$, for $\vec{a}\in \left\{ 0,1\right\}^{2n}$, is applied to the state $\rho_\text{M}$ of the MZM subsystem.  
Additionally, the noise models allow for measurement errors that modify the binary vector $\vec{m}=(m_1,\dots,m_N)$ of the time step's $N$ measurement outcomes by bitwise addition of the probabilistically generated vector $\vec{b}=(b_1,\dots,b_N)$, for $\vec{m},\vec{b}\in \{0,1\}^{N}$.  The noise model only tracks operators applied to either the MZM subsystem or the measurement outcomes.  Considering operators acting on the full system ({\it i.e.}, MZM subsystem and its environment) enables us to identify which operators to include in the noise model.

More explicitly, when the full system (MZMs plus environment) begins in a product state $\rho_\text{M}\otimes \ket{e_0}\bra{e_0}$, the time-evolved projected density matrix can be written as~\cite{Nielsen11}
\begin{align} \label{eq:unitaryevolution}
 \sum_j \bra{e_j} U \left[ \rho_\text{M}\otimes \ket{e_0}\bra{e_0}\right] U^\dagger \ket{e_j} &= \sum_j \varepsilon_j \rho_\text{M} \varepsilon_j^\dagger, 
\end{align}
where $U$ denotes the unitary evolution and $\varepsilon_j\equiv \bra{e_j}U\ket{e_0}$ is the projection of the environmental noise processes onto the MZM subsystem.  The operator $\varepsilon_j$ is therefore some combination of Majorana operators:
\begin{align}
\varepsilon_j &= \sum_{\vec{a}}O^j_{\vec{a}}=\sum_{\vec{a}} o^j_{a_1\dots a_{2n}} (\gamma_1)^{a_1}\dots (\gamma_{2n})^{a_{2n}}.
\end{align}
Instead of considering the density matrix $\rho_\text{M}$, Eq.~\eqref{eq:unitaryevolution} can equivalently be seen as a quantum trajectory where during the time step the pure state $|\psi\rangle_\text{M}$ transforms as
\begin{align}
|\psi\rangle_\text{M}&\overset{\tau}{\rightarrow} \sum_{\vec{a}} O_{\vec{a}}^j\, |\psi\rangle_\text{M} \label{eq:generalstochastic1comp}
\end{align}
with some probability $P_j$. Noise described by Eq.~\eqref{eq:generalstochastic1comp} depends on the $2^n$ coefficients $o^{l}_{a_1\dots a_{2n}}$, which renders numerical simulations of large systems intractable.

Fortunately, Eq.~\eqref{eq:generalstochastic1comp} can be greatly simplified by noting that decoherence processes such as energy relaxation and phonons will destroy the coherence between different products of Majorana operators.  In other words, local noise processes do not result in superpositions of products of Majorana operators (as opposed to non-local operations on the computational state that allow coherent superpositions of Majorana operators to be maintained over long time periods). 
Moreover, error correction itself separates many of the linear combinations in Eq.~\eqref{eq:generalstochastic1comp}~\cite{Nielsen11}.  
Given these considerations, we can replace the intractable model of Eq.~\eqref{eq:generalstochastic1comp} with a simpler, stochastic Majorana noise model. Then each time step gives:
\begin{align}
\ket{\psi}_\text{M}  &\overset{\tau}{\rightarrow} (\gamma_1)^{a_1} \dots (\gamma_{2n})^{a_{2n}} \ket{\psi}_\text{M} \label{eq:generalstochastic1a} \\
\vec{m}&\overset{\tau}{\rightarrow} \vec{m}\oplus\vec{b}, \label{eq:generalstochastic1b}
\end{align}  
with some probability $\text{Pr}(\vec{a},\vec{b})$. The order of Majorana operators in Eq.~\eqref{eq:generalstochastic1a} is unimportant as it only contributes to the overall phase of the error operator.

The noise described by Eq.~\eqref{eq:generalstochastic1a} is unitary, and thus does not include an amplitude damping channel or an erasure channel. This is not necessarily a crucial limitation since a pessimistic estimate can be obtained by sufficiently strong noise that randomly flips the qubits. This is the standard approach for the qubit-based noise models reviewed in the next section. As a consequence, however, the latter fails to take into account the relaxation time $T_1$ characterizing the timescale during which the system relaxes to the lower-energy qubit state. In a MZM system, there is no such time scale, since in practice the temperature is larger than the degeneracy splitting of the qubit states. Thus, the assumption that noise has unit amplitude provides a more accurate description of the physical noise processes for Majorana-based qubits.

The content of different stochastic models is contained entirely in the probability distribution $\{\text{Pr}(\vec{a},\vec{b})\}$.  Given this distribution, the errors in the system propagate classically and can be efficiently simulated using standard Monte Carlo techniques by tracking the net Majorana operators applied at any given time.  For this reason, when a model of the type given in Eqs.~(\ref{eq:generalstochastic1a}) and (\ref{eq:generalstochastic1b}) mimics the actual noise in a physical system, it is extremely useful for studying quantum error correction. 

In the following, a {\it noise event} refers to the application of one of the operations of the right hand side of Eqs.~\eqref{eq:generalstochastic1a} or \eqref{eq:generalstochastic1b}.  For simplicity of relating the probabilities defining our noise models to physical processes, we include noise events that apply the same Majorana operator twice (therefore not causing an error).   The following presentation of the noise models is tailored for conceptual ease; in Section~\ref{sec:threshold} and Appendix~\ref{app:code} we give a more explicit description of how one can simulate these noise models.

\subsection{Qubit-based stochastic noise models}\label{sec:stochastic-qubit}

We first review three well-known qubit-based stochastic noise models, all of which are built from Pauli operators and measurement bit-flips. In each case, we consider a scenario consisting of a sequence of time steps, where a set of single- and multi-qubit measurements and/or gates are applied in each step.

Throughout the paper, the error probabilities of the Pauli noise models are slightly reweighted compared to their standard presentation by also including the identity operator as a possible error. This allows for an easier comparison with Majorana noise models where noise events can lead to the application of $\gamma_a^2=\mathbb{1}$.
\\

\noindent {\bf Pauli noise} (or {\it code capacity noise}).  For a given time step and for each qubit:
\begin{enumerate}
	\item Apply one single-qubit operator (either $\mathbb{1}$, $X$, $Y$, or $Z$, chosen uniformly) with probability $p$; otherwise do nothing.
	\item Apply all measurement projectors perfectly.
\end{enumerate}
More explicitly, Pauli operators $X,$ $Y$, or $Z$ are applied with probability $p/4$ and identity is applied with probability $1-3p/4$.  As noted above, the non-standard normalization is chosen for ease of comparison with the Majorana noise models introduced in the following section.  For the remaining noise models, we will not explicitly write ``otherwise do nothing.'' 

Pauli noise is defined by the single parameter $p$.  While the model is too simple to provide realistic estimates of an error-correcting code's performance, it serves as a quick first test of any code. \\

\noindent {\bf Pauli noise with bit-flip measurement} (or {\it phenomenological noise}).  For a given time step, for each qubit apply step 1 of Pauli noise, then:
\begin{enumerate}
\setcounter{enumi}{1}
	\item Apply all measurement projectors perfectly, then flip each measurement outcome with probability $p_{\text{mst}}$.
\end{enumerate}
Note that flipping the measurement outcome does not change the state of the system, only our information about the system.

Pauli noise with bit-flip measurement is defined by two parameters, $\{p,p_{\text{mst}}\}$, which may be taken to be equal, $p=p_\text{mst}$,
for a simple estimate of a code's pseudo-threshold.  We emphasize that the measurement projections are still applied exactly, but the classical bit which stores the measurement outcome can be flipped.  This model is motivated on the grounds that qualitatively different error correction approaches are required to handle faulty measurements in addition to errors on the encoded information alone, making this minimal addition to Pauli noise useful for discriminating between codes.  
\\

\noindent {\bf Pauli circuit noise} (or {\it circuit-level noise}) extends the previous two models to account for the different noise processes affecting a qubit during an operation (unitary gate or measurement). For a given time step, each qubit is involved in a $k$-qubit operation, where $k=0$ for an idle qubit. For all sets of qubits involved in the same $k$-qubit operation:
\begin{enumerate}
\item Do the following:
\begin{enumerate}
\item For each qubit in the set, apply one single-qubit operator (either $\mathbb{1}$, $X$, $Y$, or $Z$, chosen uniformly) with probability $p^{(k)}$. \label{step:1a}
\item Apply a $k$-qubit Pauli operator with probability $p_\text{cor}^{(k)}$.  For $k=2$, this is any element of the set of 16 operators $\{Z\otimes X, \mathbb{1}\otimes Y,\dots\}$.  For $j\leq1,$ $p_\text{cor}^{(j)}=0$. \label{step:1b}
\end{enumerate}
\item Apply the measurement projector perfectly, then flip the $k$-qubit measurement outcome with probability $p_{\text{mst}}^{(k)}$.  For an idle qubit, do nothing.
\end{enumerate}
For step~\ref{step:1a}, $X$, $Y$, or $Z$ are applied with probability $p^{(k)}/4$ and identity is applied with probability $1-3p^{(k)}/4$.  For step~\ref{step:1b}, any given non-trivial Pauli operator is applied with probability $p^{(k)}_\text{cor}/16$ and identity is applied with probability $1-15p^{(k)}_\text{cor}/16$.  Again, the non-standard normalization is chosen to simplify comparison with the Majorana noise models in the following section. 

Pauli circuit noise is defined by the set of probabilities $\{p^{(0)},p^{(k)}, p_\text{cor}^{(2)},p_{\text{mst}}^{(k)}\}$ for $k\in\{1,2\}$.  It is for this noise model (slightly renormalized\footnote{Note that because of the renormalization of probabilities to include the identity operation, this result is found for $p=p^{(k)}_\text{mst}=3/4 p^{(0)}=3/4 p^{(1)}=15/16 p_\text{cor}^{(2)}$.}) with the probability of a single-qubit, two-qubit, and measurement bit-flip error equally likely, that the well-known result is found that the qubit surface code has an error threshold value of $p_{\text{th}}\approx 1\%$~\cite{Raussendorf07,Wang11}. This means that a quantum state can be reliably stored in an (arbitrarily large) surface code for an indefinite period of time for qubits subjected to circuit-level noise with $p<p_\text{th}$.
\\ 

\subsection{Stochastic Majorana noise models}\label{sec:stochastic-MZM}

We now present four stochastic Majorana noise models in order of increasing complexity.  The first three are analogous to the qubit-based models reviewed above.  The fourth is motivated by a particular physical implementation and measurement protocol of a Majorana-based quantum computing architecture~\cite{Plugge2017,Karzig17}.

\begin{figure}
	\includegraphics[width=0.95\columnwidth]{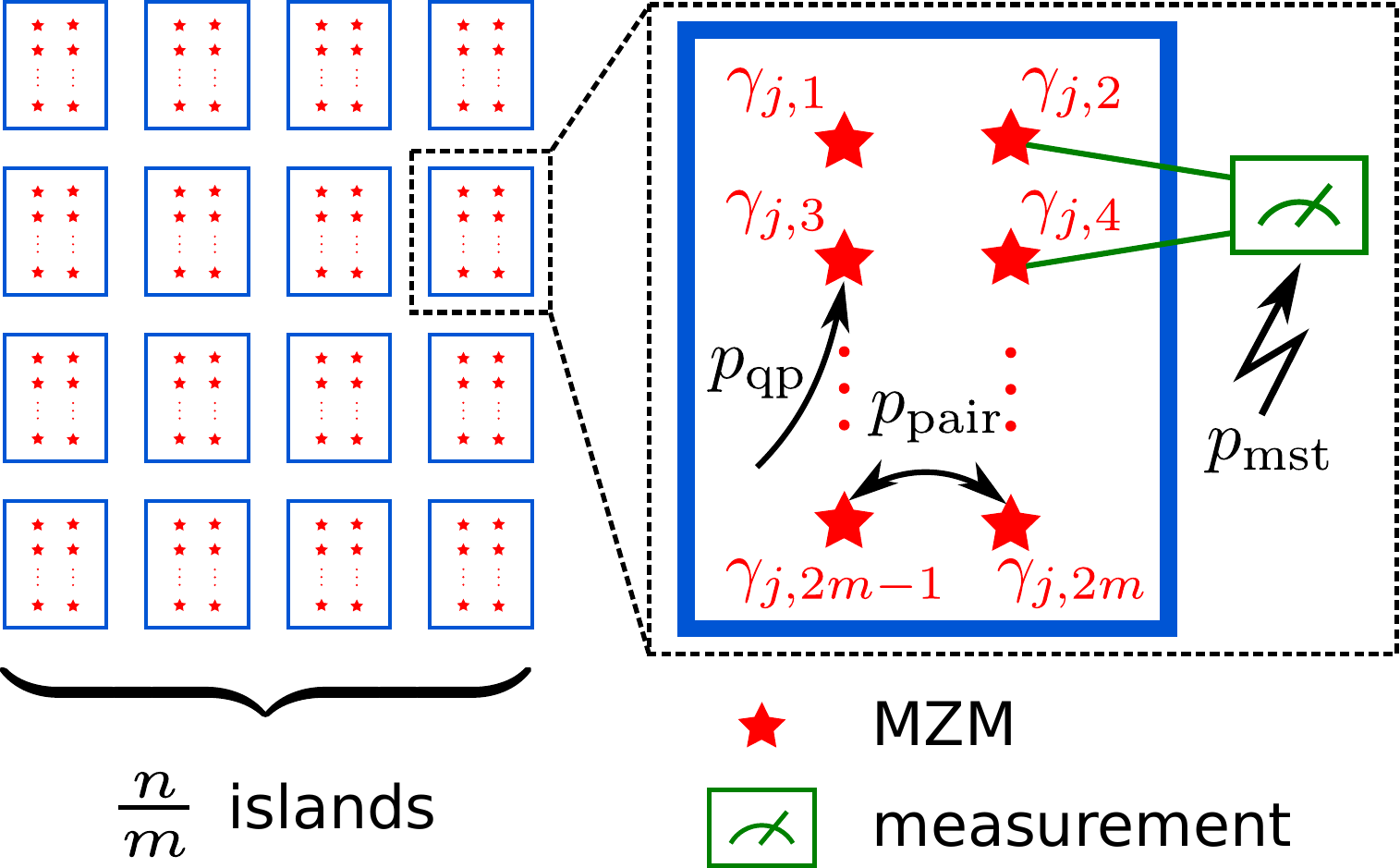}
	\caption{Schematic of a quantum computing architecture with $2n$ MZMs (red stars) equally divided among $n/m$ islands (blue boxes). The inset zooms in on the $j$th island.  As an example, we indicate a two-MZM measurement and the possible noise events for the Majorana noise model QpBf when the island begins a time step in the even parity state.}
	\label{fig:fig1}
\end{figure}

Naively, the simplest stochastic Majorana noise model to consider would simply apply the Majorana operator $\gamma_i$ with probability $p$ for each $i\in\{1,2,\dots,2n\}$ in each time step, followed by perfect measurements.  However, such a model would on average spend an equal amount of time in an even total MZM parity state as in an odd total MZM parity state.  In the case of superconducting islands with charging energy, {\it e.g.}, for the system described in Section~\ref{sec:physical}, the energy separation of these states is large (on the order of the superconducting gap or the charging energy of the island) and thus physically we would expect the system to spend much more time in the lower-energy state corresponding to a specific MZM parity.  To more accurately describe this situation, we go beyond the naive model and introduce the concept of MZM islands.  

We assume that the MZMs are naturally split into $n/m$ subsets, each of which contains $2m$ MZMs belonging to the same superconducting island (see Fig.~\ref{fig:fig1}). We assume that the initial $2m$-MZM parity on each island is even so that, for a given time step, an island has \textit{even parity} if an even number of Majorana operators have been applied in its history, and has \textit{odd parity} otherwise. By keeping track of the parity of the islands, the probability of applying an odd number of Majorana operators can be adjusted depending on whether this would relax the system back to the ground state or lead to an excited state. This is captured by an additional step 0 in Majorana noise models that is not required for the qubit ones.  

Each noise model contains up to four types of noise events:
\begin{itemize}
\item {\it Quasiparticle} event: application of a single Majorana operator: $\ket{\psi}_\text{M}\to \gamma_{j,a} \ket{\psi}_\text{M}$.
\item {\it Pair-wise dephasing} event: application of a pair of Majorana operators belonging to the same island: ${\ket{\psi}_\text{M} \to \gamma_{j,a}\gamma_{j,b}\ket{\psi}_\text{M}}$.
\item {\it Correlated} event: application of Majorana operators from multiple islands involved in the same measurement (see later discussion or Table~\ref{fig:table} for examples).
\item {\it Measurement bit-flip}: flipping of the classical bit storing the outcome of a $2k$-MZM parity measurement: {\it e.g.}, $\vec{m}\to \vec{m}\oplus(0,1,0,\dots,0)$.
\end{itemize}
The naming of the noise events will become clear in Section~\ref{sec:probabilities} when we describe the physical processes contributing to each error-type. To simplify combinatorial prefactors, we allow the same Majorana operator to be used multiple times in a given noise event ({\it e.g.}, pair-wise dephasing includes the identity operator $\gamma_a^2$).  We also keep track of ordering, so that applying the pairs $\gamma_a\gamma_b$ and $\gamma_b\gamma_a$ are considered different noise events (multiple noise events contribute to the same type of error).  Unless otherwise noted, we assume that the Majorana operators corresponding to a given noise event are chosen uniformly over all MZMs on the island.

We note that even if subsets of MZMs are combined into physical qubits, errors involving an odd number of Majorana operators on any given island cannot be described by Pauli operators, motivating the consideration of stochastic Majorana noise models.
\\

\begin{table*}
\scriptsize
\begin{center}
\begin{minipage}{.42\columnwidth}
\vspace{0pt}
\centering
	\includegraphics[width=\columnwidth]{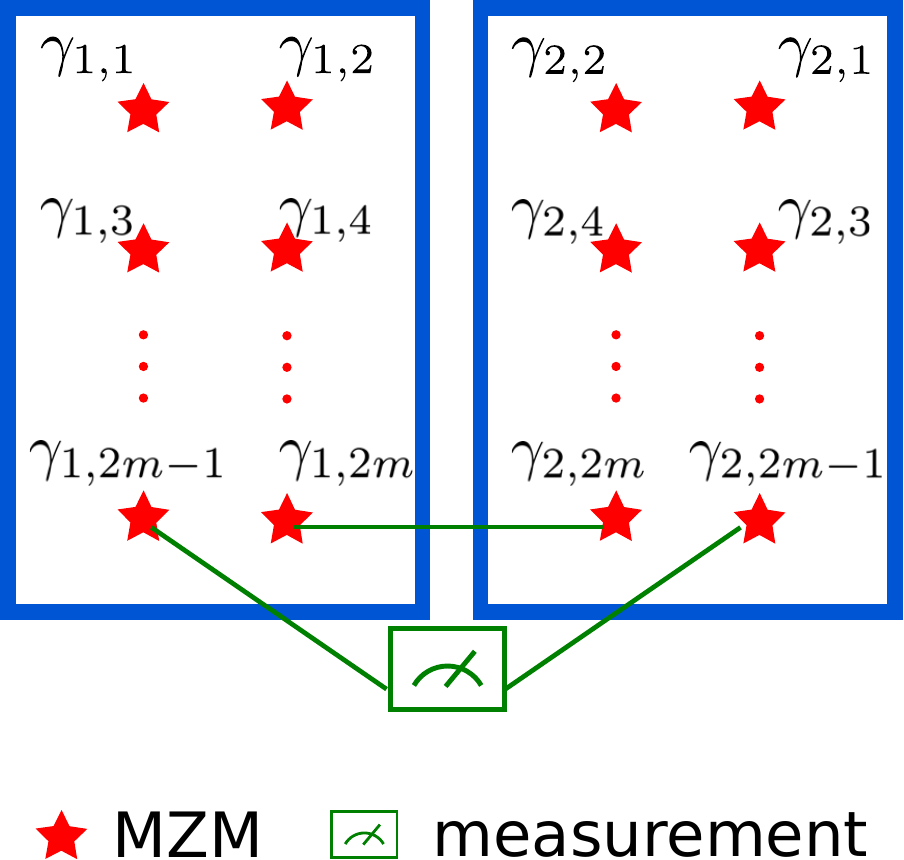}
\end{minipage}
\begin{minipage}{1.62\columnwidth}
\vspace{0pt}
\centering
	\begin{tabular}{|c|c|c|c|c|c|} \hline 
		Event &  Operator & Qp&QpBf & MC~ &  ~PMC~ \\\hline
		~qp~& $\gamma_{1,1}$ &  $p_{\text{qp}}/(2m)$& $p_{\text{qp}}/(2m)$&  $p_{\text{qp}}^{(2)}/(2m)$&  $p_{\text{qp}}^{(0)}/(2m)$\\  \cline{2-6}
		& $\gamma_{1,2m}$ & $p_{\text{qp}}/(2m)$ & $p_{\text{qp}}/(2m)$ & $p_{\text{qp}}^{(2)}/(2m)$ & $p_{\text{qp}}^{(2)}/(2m)$ \\  \cline{2-6}
		\hline
		pair & $\gamma_{1,1}\gamma_{1,2}$ &    $p_{\text{pair}}/(2m)^2$ &  $p_{\text{pair}}/(2m)^2$ &  $p_{\text{pair}}^{(2)}/(2m)^2$ & $p_{\text{pair}}^{(0)}/(2m)^2$ \\  \cline{2-6}
		& $\gamma_{1,1}\gamma_{1,2m}$&    $p_{\text{pair}}/(2m)^2$ &    $p_{\text{pair}}/(2m)^2$ &   $p_{\text{pair}}^{(2)}/(2m)^2$& $p_{\text{pair}}^{(2)}/(2m)^2$ \\ \hline
		cor& $\gamma_{1,1}\gamma_{1,2m} \gamma_{2,{2m}}$ & $p_{\text{pair}}p_{\text{qp}}/(2m)^3$ & $p_{\text{pair}}p_{\text{qp}}/(2m)^3$ & $p_\text{cor,odd}/N_\text{odd} $ & $p_{\text{cor,odd}}^{(2)}/(8m)$ \\ \cline{2-6}
		&$\gamma_{1,1}\gamma_{1,2m}\gamma_{2,{2m}}\gamma_{2,2}$ &  $p_{\text{pair}}^{2}/(2m)^4$ & $p_{\text{pair}}^{2}/(2m)^4$ & $p_\text{cor,even}/N_\text{even}$ &  $p_{\text{cor,even}}^{(2)}/(16m^2)$\\  \hline
		mst &$ b^{(2)}$ & N/A &  $p_{\text{mst}}$ & $p_{\text{mst}}^{(2)}$ & $p_{\text{mst}}^{(2)}$ \\ \hline 
	\end{tabular}
\end{minipage}
	\caption{{\it Left panel}: 
MZMs (red stars) are grouped into sets of $2m$ on an island (blue box).  The operator of $a$th MZM on the $j$th island is $\gamma_{j,a}$.  For the time step considered here, the two islands are involved in a four-MZM measurement. 
{\it Right panel}: Example noise event probabilities when both islands begin the time step with even parity. The table could equivalently be understood as noise event probabilities after the initial step 0 of each noise model accounting for the asymmetry between even and odd parity islands.  The left-most column labels the error type, with the abbreviations meaning quasiparticle, pair-wise dephasing, correlated, and measurement bit-flip, respectively.  The operators for correlated noise events for models MC and PMC could be applied from two independent single-island noise events, analogously to models Qp and QpBf, or from a single correlated event; for simplicity we only write the probability of the latter.
The parameters $N_\text{even}=\left((2m)^2 (2m+1)^{m-2} \right)^2$ and $N_\text{odd}=N_\text{even}/2m$ in the fifth column are defined to be the number of different odd correlated events and even correlated events, respectively, in the model MC.  Note that in model PMC, correlated events must involve a pair of MZMs connected by the measurement, which reduces the combinatorial factors in the denominators.  For instance, an odd correlated event has $4m$ possibilities for the initial excitation ($\gamma_{1,1}$ in the table) and only two choices for the pair of MZMs that transfers the excitation to the second island ($\gamma_{1,2m}\gamma_{2,2m}$ in the table). 
}\label{fig:table}
\end{center}
\end{table*}


\noindent{\bf Quasiparticle noise} (Qp).  In a given time step, implement the following sequence for each island:
\begin{enumerate}
\setcounter{enumi}{-1}
\item If the island begins the time step with odd parity, apply a quasiparticle event with probability $p_\text{odd}$.  \label{step:Qp0}
\item Apply one single-island noise event: either a quasiparticle event with probability $p_\text{qp}$ or a pair-wise dephasing event with probability $p_\text{pair}$. \label{step:Qp1}
\item Apply all measurement projectors perfectly.
\end{enumerate}
We count $2m\times 2m$ different pairs of Majorana operators per island.  In step~\ref{step:Qp1}, the operator $\gamma_{j,a}$ is applied with probability $p_\text{qp}/2m$, the operator $\gamma_{j,a}\gamma_{j,b}$ ($a\neq b$) is applied with probability $p_\text{pair}/(2m)^2$, and identity is applied with probability ${1-p_\text{qp}-3 p_\text{pair}/4}$ (because of pair-wise dephasing events with $\gamma_{j,a}^2=\mathbb{1}$). 

The set of probabilities $\{ p_\text{odd},p_\text{qp},p_\text{pair}\}$ defines this model.  In an encoding where four MZMs define a physical qubit, Qp reduces to the qubit-based model {\it Pauli noise} when ${p_\text{qp}=0}$ (with $p\to p_\text{pair}$).  

As mentioned above, step~\ref{step:Qp0} accounts for the energy difference between an island with odd parity and even parity: when $p_\text{qp}\ll p_\text{odd}$, each island in the system spends on average very little time in the odd MZM parity state.  We will return to this discussion in Section~\ref{sec:physical}.   
\\

\noindent {\bf Quasiparticle noise and bit-flip measurement} (QpBf). In a given time step, implement steps 0 and 1 from model (Qp) for each island, then:
\begin{enumerate}
\setcounter{enumi}{1}
\item Apply all measurement projectors perfectly, then independently flip each classical bit storing a measurement outcome with probability $p_\text{mst}$.
\end{enumerate}
This model is defined by the set of probabilities $\{p_\text{odd},p_\text{qp},p_\text{pair},p_\text{mst}\}$.  In an encoding where four MZMs define a physical qubit, when $p_\text{qp}=0$, QpBf reduces to the qubit-based model Pauli noise and bit-flip measurement (with $p\to p_\text{pair}$ and the same $p_\text{mst}$).   Example noise events and their corresponding probabilities are schematically depicted in Fig.~\ref{fig:fig1} and listed in Table~\ref{fig:table} for an island beginning a time step in the even parity state.  
\\ 

\noindent {\bf Correlated events}.
Models Qp and QpBf do not distinguish the probabilities of noise events involving MZMs on idle islands from those on measured islands.  We would now like to account for these differences.
Define $\gamma_{j,a}$ to be the operator corresponding to the $a$th MZM on the $j$th island.  We consider two types of correlated events possible in a multiple-island measurement:
\begin{itemize}
\item {\it Odd correlated event}: application of a string of three or more Majorana operators involving at least two islands, such that an odd number (up to $m-1$) of Majorana operators are applied to one of the islands involved in a $k$-island measurement. There is an even number (up to $m$) of Majorana operators applied to the remaining $k-1$ islands involved in the measurement. For example, an odd correlated event in a two-island measurement of islands $i$ and $j$ results in ${\ket{\psi}_\text{M}\to \gamma_{i,a}\gamma_{j,b}\gamma_{j,c}\ket{\psi}_\text{M}}$.
\item {\it Even correlated event}: application of a string of four or more Majorana operators involving at least two islands, such that an even number (up to $m$) of Majorana operators are applied to all the islands involved in the $k$-island measurement.  For example, an even correlated event in a two-island measurement of islands $i$ and $j$ results in  ${\ket{\psi}_\text{M}\to \gamma_{i,a}\gamma_{i,b}\gamma_{j,c}\gamma_{j,d}\ket{\psi}_\text{M}}$.
\end{itemize}
In the following, we assume that during a time step each island is either idle or involved in a single measurement.  The spread of correlated events can be mitigated by restricting the number of islands involved in a measurement.  For instance, for the Bacon-Shor code studied in Section~\ref{sec:threshold}, correlated events only involve nearest neighbor islands.
\\

\noindent{\bf Majorana circuit noise} (MC).  In a given time step, for a set of islands involved in the same $k$-island measurement ($k=0$ for an idle island), implement the following sequence:
\begin{enumerate}
\setcounter{enumi}{-1}
\item \label{hop}For each island that begins the time step with odd parity, apply a quasiparticle event with probability $p_\text{odd}^{(k)}$.
\item \label{err} For the set of islands involved in the same $k$-island measurement, do the following:
\begin{enumerate}
\item For each island in the set, apply one single-island noise event: either a quasiparticle event with probability $p_\text{qp}^{(k)}$ or a pair-wise dephasing event with probability $p_\text{pair}^{(k)}$. \label{step:MC1a}
\item Apply a correlated event to the set: either an odd correlated event with probability $p_\text{cor,odd}^{(k)}$ or an even correlated event with probability $p_\text{cor,even}^{(k)}$.  For ${j\leq 1},$ $p_\text{cor,odd}^{(j)}=p_\text{cor,even}^{(j)}=0$.   \label{step:MC1b}
\end{enumerate}
\item Apply the measurement projector perfectly, then flip the classical bit storing the measurement outcome with probability $p_\text{mst}^{(k)}$.  For an idle island, do nothing.
\end{enumerate}
MC is defined by the probability set ${\{p_\text{odd}^{(k)},p_\text{qp}^{(k)},p_\text{pair}^{(k)},p_\text{mst}^{(k)},p_\text{cor,odd}^{(k)},p_\text{cor,even}^{(k)}\}}$ for $k\leq k_\text{max}$, where $k_\text{max}$ is the maximum number of islands involved in a measurement.  
MC has the same action on an idle island ($k=0$) as Qp, with the probability set ${\{p_\text{odd},p_\text{qp},p_\text{pair}\}}\to{ \{ p_\text{odd}^{(0)},p_\text{qp}^{(0)},p_\text{pair}^{(0)}\}}$.
MC has the same action on an island involved in a single-island measurement ($k=1$) as QpBf, 
 with the probability set ${\{p_\text{odd},p_\text{qp},p_\text{pair},p_\text{mst}\} }\to {\{p_\text{odd}^{(1)},p_\text{qp}^{(1)},p_\text{pair}^{(1)},p_\text{mst}^{(1)}\}}$. In an encoding where four MZMs define a physical qubit and if the probability of any error involving an odd number of Majorana operators on a given island is set to zero ({\it i.e.}, ${p_\text{qp}^{(k)}=p_\text{cor,odd}^{(k)}=0}$), MC reduces to the qubit-based model Pauli circuit noise, with the probabilities related by ${\{p^{(k)},p_\text{cor}^{(2)},p_\text{mst}^{(k)}\}}\to {\{ p_\text{pair}^{(k)},p_\text{cor,even}^{(2)},p_\text{mst}^{(k)}\}}$.

 In Table~\ref{fig:table}, we compare the probabilities of noise events in step \ref{err} of models Qp, QpBf, and MC.  We see that MC allows for the possibility that noise events affecting multiple islands connected by a measurement happen with greater probability than independent noise events on the islands ({\it e.g.}, when $p_\text{cor,even}^{(2)}>p_\text{pair}^2$).
\\

\noindent {\bf Physical Majorana circuit noise} (PMC). This model refines MC by considering a specific physical implementation and measurement protocol of the MZM system~\cite{Plugge2017,Karzig17}. Focusing on a specific measurement protocol allows us to drop many of the correlated events included in MC, as well as to separate noise events involving measured MZMs from those only involving unmeasured MZMs.  These two modifications enable a more accurate description of this physical system, see Section~\ref{sec:physical} for a description of the underlying causes of errors in such a system.  

We assume that our measurement protocol allows parity measurements of two MZMs belonging to the same island and joint parity measurements of a set of four MZMs on two islands.  During a given time step, each island is now either idle  $(k=0)$, involved in a two-MZM measurement $(k=1)$, or  involved in a four-MZM measurement $(k=2)$.

The model follows the same steps as MC, with slight modifications of steps 0 and 1(a) for islands that are involved in a measurement, to account for whether particular MZMs are connected to the measurement apparatus or not:
\begin{enumerate}
\setcounter{enumi}{-1}
\item For each island that begins the time step with odd parity, apply a quasiparticle event corresponding to:   
\begin{itemize}
\item A MZM not involved in the measurement with probability $\frac{2m-2}{2m} p_\text{odd}^{(0)}$.
\item A MZM involved in the measurement with probability $\frac{2}{2m} p_\text{odd}^{(k)}$.
\end{itemize}
\item (a) For islands in the set not involved in a measurement ($k=0$) apply either a quasiparticle of pair-wise dephasing event with respective probabilities $p_\text{qp}^{(0)}$ and $p_\text{pair}^{(0)}$.

For each island in the set with $k\geq 1$, apply either a quasiparticle event corresponding to:  
\begin{itemize}
\item A MZM not involved in the measurement with probability $\frac{2m-2}{2m} p_\text{qp}^{(0)}$.
\item A MZM involved in the measurement with probability $\frac{2}{2m} p_\text{qp}^{(k)}$.
\end{itemize}
or a pair-wise dephasing event corresponding to:  
\begin{itemize}
\item A pair of MZMs not involved in the measurement with probability $\frac{(2m-2)^2}{(2m)^2} p_\text{pair}^{(0)}$.
\item A pair of MZMs, with at least one of them involved in the measurement, with probability $\frac{4(2m-1)}{(2m)^2} p_\text{pair}^{(k)}$.
\end{itemize}

\end{enumerate}
Furthermore, we consider a restricted set of correlated events in step 1(b) so that 
odd correlated events are strings of three Majorana operators $\gamma_{i,a}\gamma_{i,b}\gamma_{j,b}$ and even correlated events are strings of four Majorana operators $\gamma_{i,a}\gamma_{i,b}\gamma_{j,b}\gamma_{j,c}$, such that $i\neq j$ and the indices are chosen such that $\gamma_{i,b}$ and $\gamma_{j,b}$ are involved in a measurement.

PMC is defined by the same set of probabilities as MC, with $k_\text{max}=2$.  While MC treated all MZMs on a given island identically, PMC distinguishes between the measured and unmeasured MZMs within the island for single-qubit noise events.  Furthermore, correlated events in PMC always involve a pair of MZMs directly coupled by the measurement.  
In Table~\ref{fig:table}, we compare noise event probabilities for all four models.  Only a restricted set of correlated events are considered in PMC, which changes the combinatorial prefactors of the probabilities of these events between MC and PMC.


\section{Physical System}\label{sec:probabilities}

\begin{figure}[t]
\includegraphics[width=\columnwidth]{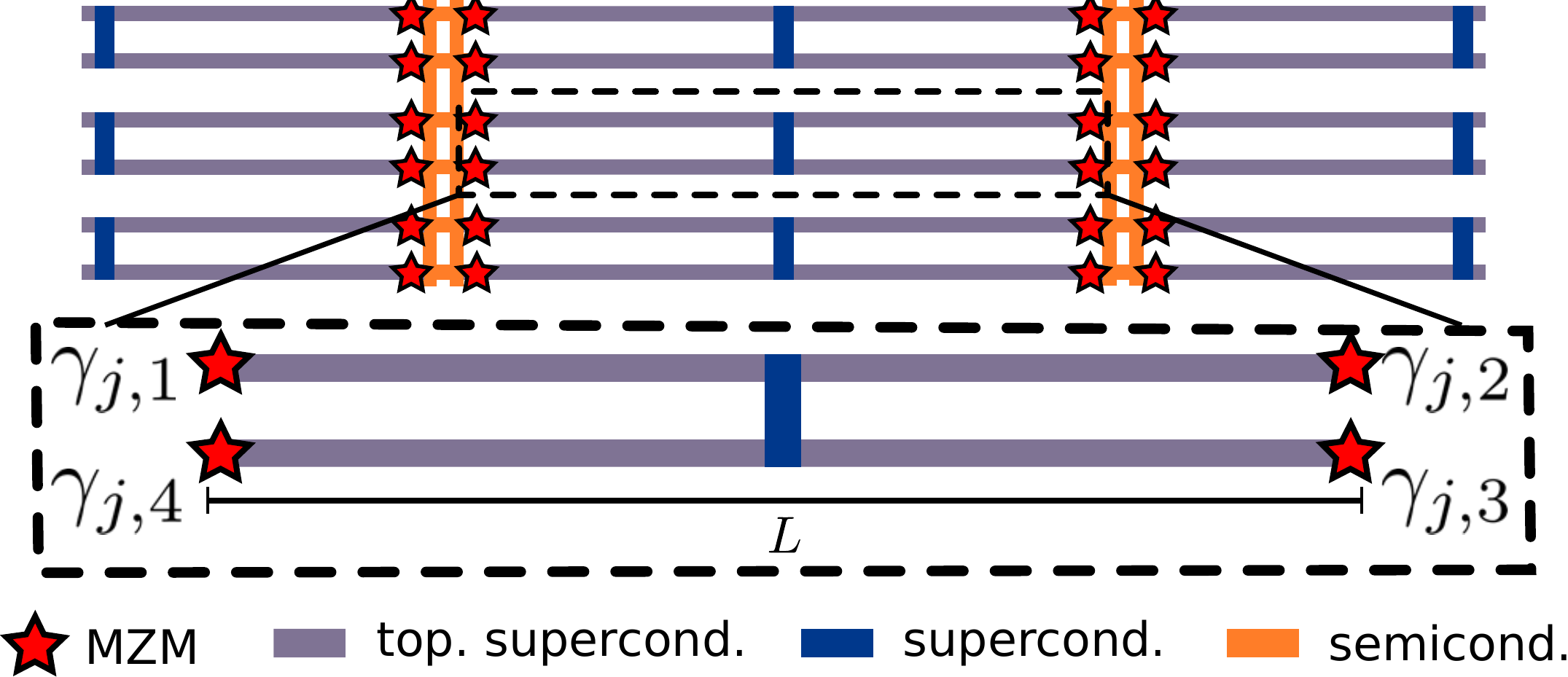}
\caption{
Example physical system: an array of charging energy-protected islands with four-MZMs (tetrons) connected by a network of semiconducting wires (orange lines).  The $j$th tetron, shown in the inset, is a superconducting island formed out of two topological superconducting wires (gray lines) connected by a regular superconducting bridge (blue line).  A MZM (red star) is localized at the ends of each topological superconductor.  In terms of the noise models of the previous section, each tetron corresponds to a single island with $2m=4$ MZMs.
 }
\label{fig:tetron}
\end{figure}

In the following, we use a recently proposed implementation of a measurement-based topological quantum computer~\cite{Karzig17} as an example to illustrate the different types of errors and to connect them to our noise models. The system consists of an array of four-MZM islands, \emph{tetrons}, depicted in Fig.~\ref{fig:tetron}.
In this proposal, the MZM islands have a finite charging energy $E_C$. Each tetron corresponds to a physical qubit, whose states are stored in the  two nearly-degenerate ground states within a total parity subspace. In the absence of quasiparticle poisoning, the latter is fixed to be either even or odd.
The inset of Fig.~\ref{fig:tetron} shows the details of a tetron implementation based on a single superconducting island composed of two topological superconducting wires of length $L$  (gray lines), connected by a conventional superconducting bridge (blue line).  There are four MZMs (red stars), localized at the end points of the topological superconductors. Multiple tetrons are connected to each other by semiconducting wires (orange lines), which may be gated (not shown) to allow for measurement by quantum dots \cite{Karzig17}.  The Hilbert space of a single tetron contains many of the features applicable to a large class of MZM systems, however the measurement protocol is specific to this qubit proposal.  

\subsubsection{Spectrum and single MZM excitations}

In this section, we identify the types of errors that appear in a Majorana-based quantum computing architecture and discuss how the probabilities of these errors depend on the physical parameters of the system.  For ease of explanation, we focus on a particular Majorana-based qubit proposal, reviewed in Section~\ref{sec:physical}. In Section~\ref{sec:error-probabilities}, we discuss the different error processes and corresponding transition rates.   In Section~\ref{sec:prob-summary}, we connect the parameters of the noise models with a small set of physical parameters that can be expressed in terms of the transition rates.
When possible, we distinguish between the features that are generic to systems of MZMs and those that are specific to the example system we consider.  The former will motivate the errors included in MC, while the latter will explain the refinement of MC to PMC.

\subsection{Example system: Tetron array}\label{sec:physical}

\begin{figure}[t!]
	\includegraphics[width=\columnwidth]{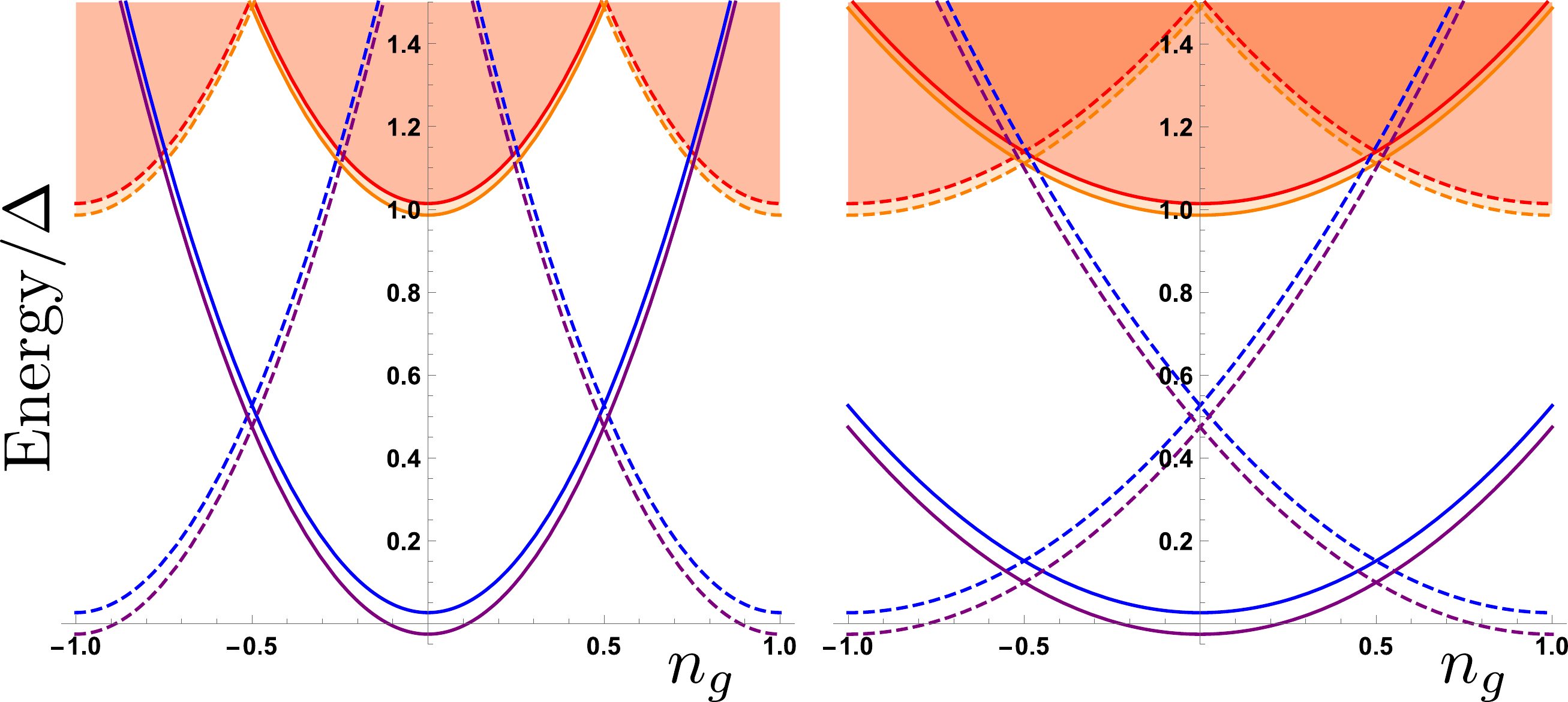}
	\caption{Cartoon of the energy levels of a single tetron against dimensionless gate voltage for $E_C=2\Delta$ ({\it left panel}) and $E_C=\Delta/2$ ({\it right panel}), where $E_C$ is the charging energy of the island and $\Delta$ is the superconducting gap.  
		For fixed gate volgate $-0.5 < n_g < 0.5$, the system has a nearly-degenerate ground state with an even four-MZM parity. The degeneracy is broken by the MZM hybridization $\delta E_{ab}$, leading to distinct states $\ket{g}\ket{\psi_i}_\text{M}$ denoted by solid blue and purple curves. The energy bands bordered by solid orange and red curves correspond to the states $\ket{e_{\Delta}}\ket{\bar{\psi}_i}_\text{M}$ that have total fermion parity even, and odd four-MZM parity.  The shaded regions indicate that the bands contain many discrete energy levels, with level spacing $\delta_{\text{is}}$. The dashed blue and purple curves correspond to the states $\ket{e_{C,\pm}}\ket{\bar{\psi}_i}_\text{M}$ that have total fermion parity odd, and odd four-MZM parity. Comparing the right and left panels near $n_g\approx 0$, we see that when $E_C$ is larger (smaller) than $\Delta$, the quasiparticle-poisoned states with one extra or one fewer electron on the superconducting island, $\ket{e_{C,\pm}}\ket{\bar{\psi}_i}_\text{M}$, are higher (lower) in energy than the lowest energy states with a thermally excited quasiparticle, $\ket{e_\Delta}\ket{\bar{\psi}_i}_\text{M}$.
	}
	\label{fig:energies}
\end{figure}

The $j$th tetron can be described by the Hamiltonian\footnote{In Eq.~(\ref{eq:tetron-Ham}) we assumed that ``mutual charging energy'' terms that could couple pairs of MZMs belonging to different tetrons are perfectly quenched. We will comment more these terms in Section~\ref{sec:other}.}
\begin{equation}\label{eq:tetron-Ham}
H= E_C\left(\hat{n}_s -n_g\right)^2 +\sum_k E_k\, \hat{n}_{\Delta,k}+\sum_{a\neq b} \delta E_{ab}\, i\gamma_{j,a}\gamma_{j,b},
\end{equation}
where $a,b =1,2,3,4$. The last two terms (BCS Hamiltonian and MZM hybridization) are present in all MZM systems, while the first term (charging energy Hamiltonian) is present in systems for which the superconducting island is Coulomb-blockaded ({\it i.e.}, not grounded).  

In the charging energy Hamiltonian, the operator $\hat{n}_s$, with integer eigenvalues $n_s$, is the number operator for electrons on the superconducting island. Generally the energy will be minimized when $n_s$ is the integer closest to $n_g$, the dimensionless gate voltage applied to the superconducting island. When $n_g$ is an integer, adding an electron to or subtracting an electron from the island costs a charging energy $E_C$, which is set by the capacitance of the superconducting island. 

The BCS Hamiltonian is written in terms of the quasiparticle number operator $\hat{n}_{\Delta,k}$, with integer eigenvalues $n_{\Delta,k}$.  More specifically, $\hat{n}_{\Delta,k}$ counts the number of above-gap quasiparticles on the superconducting island with crystal momentum $k$,  occupying the state with energy
\begin{equation}
E_k=\sqrt{\left( \frac{k^2}{2m} -\mu \right)^2 +\Delta^2 },
\end{equation} 
where $\Delta$ is the superconducting gap and $\mu$ is the chemical potential.  Due to the finite size of the island, only a discrete set of momenta, $\{k_i\}$, is allowed.  The level spacing of the island, $\delta_{\text{is}}$, is the separation between adjacent energies: ${\delta_{\text{is}} = E_{k}-E_{k-\delta k}}$.  

The last term of Eq.~(\ref{eq:tetron-Ham}) describes the MZM hybridization energy, $\delta E_{ab}$, between MZMs $\gamma_{j,a}$ and $\gamma_{j,b}$.  Generally, $\delta E_{ab}$ is set by the wavefunction overlap, resulting in a length dependence $\delta E_{ab}\propto e^{-L_{ab}/\xi}$, where $L_{ab}$ is the distance separating $\gamma_{j,a}$ and $\gamma_{j,b}$ and $\xi$ is the superconducting coherence length. The topological protection of the system is manifested as an exponentially small ground state degeneracy splitting and requires ${L_{ab}\gg \xi}$.  

We denote the eigenstates of Eq.~(\ref{eq:tetron-Ham}) using the basis $\ket{n_s,\{n_{\Delta,k}\}_\Delta}\ket{i\gamma_{j,1}\gamma_{j,2},i\gamma_{j,3}\gamma_{j,4}}$. The first set of quantum numbers describes the non-topological degrees of freedom: charge or quasiparticle excitations. The second set denotes the (almost) degenerate MZM subspace. The pairing of Majorana operators into $i\gamma_{j,1}\gamma_{j,2}$ and $i\gamma_{j,3}\gamma_{j,4}$ is an arbitrary choice. Note that the total fermion parity of the island, $2(n_s \mathrm{mod}\, 2)-1$, equals the product of the quasiparticle and four-MZM parity.  

To discuss the leading excitations, we compare the energy levels of the system for $E_C=2\Delta$ and for ${E_C=\Delta/2}$ in the left and right panels of Fig.~\ref{fig:energies}. Solid curves correspond to an even number of particles $n_s$ on the superconducting island, dashed curves to odd $n_s$. The tetron is operated at an even integer value for the dimensionless gate voltage to maximize protection from quasiparticle poisoning by maximizing the energy separation between solid and dashed curves. Without loss of generality we assume $n_g\approx 0$. In this regime, there are two nearly degenerate ground states:
\begin{equation}\label{eq:g}
\ket{g}\ket{\psi_i}_{\text{M}} =\ket{0,\{0\}_\Delta}\ket{\psi_i}_{\text{M}},
\end{equation}
with $i=0,1$ and corresponding energy given by the solid blue and purple curves centered at $n_g=0$. We write $\{n_{\Delta,k}\}_\Delta=\{0\}_\Delta$ to denote that there are no quasiparticles occupying energy levels above the superconducting gap.  
The qubit is stored in the two MZM states $\ket{\psi_i}_{\text{M}}$, which are orthogonal linear combinations of the even four-MZM parity basis states ${\ket{i\gamma_{j,1}\gamma_{j,2}=\pm 1,i\gamma_{j,3}\gamma_{j,4}=\pm 1}}$. When the only non-vanishing hybridizations are $\delta E_{12}$ and $\delta E_{34}$, ${\ket{\psi_i}_{\text{M}}}$ is simply $\ket{\pm 1, \pm 1}$. 

There are two bands of lowest excited states with even $n_s$, that correspond to a single thermally excited quasiparticle. Their energies are shown in Fig.~\ref{fig:energies} by the overlapping shaded regions bordered by solid orange and red curves. 
Two-quasiparticle excitations require energies of at least $2\Delta$ and are therefore much less likely; in the following, we restrict our attention to $n_{\Delta,k}\in \{0,1\}$.  We denote a single excitation with energy larger than the gap $\Delta$ as $e_\Delta$.  The two bands are denoted by the states
\begin{equation}
\label{eq:eth}
\ket{e_\Delta}\ket{\bar{\psi}_i}_\text{M}=\ket{0,\{n_{\Delta,k}=1\}_\Delta}\ket{\bar{\psi}_i}_\text{M},
\end{equation}
with $i=0,1$. In the above, $e_\Delta$ may denote different $k$ states; this does not matter for our discussion as long as we focus on states with energy $\approx\Delta$. 
Here, the MZM states ${\ket{\bar{\psi}_i}_\text{M}}$ are orthogonal linear superpositions of the {\it odd} four-MZM fermion parity states ${\ket{i\gamma_{j,1}\gamma_{j,2}=\pm 1,i\gamma_{j,3}\gamma_{j,4}=\mp1}}$. 

Depending on whether $n_g$ is positive or negative, the two lowest excited states with odd total fermion parity contain either an extra electron or one fewer electron, respectively.  Such states are quasiparticle-poisoned. The states with one extra (one fewer) electron are written as
\begin{equation}
\label{eq:eqp}
\ket{e_{C,\pm}}\ket{\bar{\psi}_i}_\text{M}=\ket{\pm 1,\{0\}_\Delta}\ket{\bar{\psi}_i}_\text{M},
\end{equation}
with energy levels shown in the dashed blue and purple parabolas centered about $n_g= \pm 1$.

The state of the MZMs is very similar for the two excited states discussed above. In both cases, an excitation process exchanges an electron between the MZMs and other fermionic modes represented by either the excited quasiparticles or by an external environment.  The excitation acts by applying a single Majorana operator to the state $|\psi_i\rangle_\text{M}$. The ground states $\ket{g}\ket{\psi_i}_\text{M}$ and the excited states ${\ket{e_\Delta}\ket{\bar{\psi}_i}_\text{M}, \ket{e_{C,\pm}}\ket{\bar{\psi}_i}_\text{M}}$ are therefore distinguished by their four-MZM parity, with odd parity states separated from the even parity ground states by an energy $\min\left(\Delta,E_C\right)$. In our stochastic Majorana noise models, we focus only on these four lowest energy states and can therefore use the four-MZM parity as a measure of whether an island is in a ground or an excited state. In order to include higher excited states, we would need to separately track the four-MZM parity and the quantum numbers $n_s$ and $\{n_{\Delta,k}\}$.

\subsubsection{Measurements}

\begin{figure}[t]
\includegraphics[width=\columnwidth]{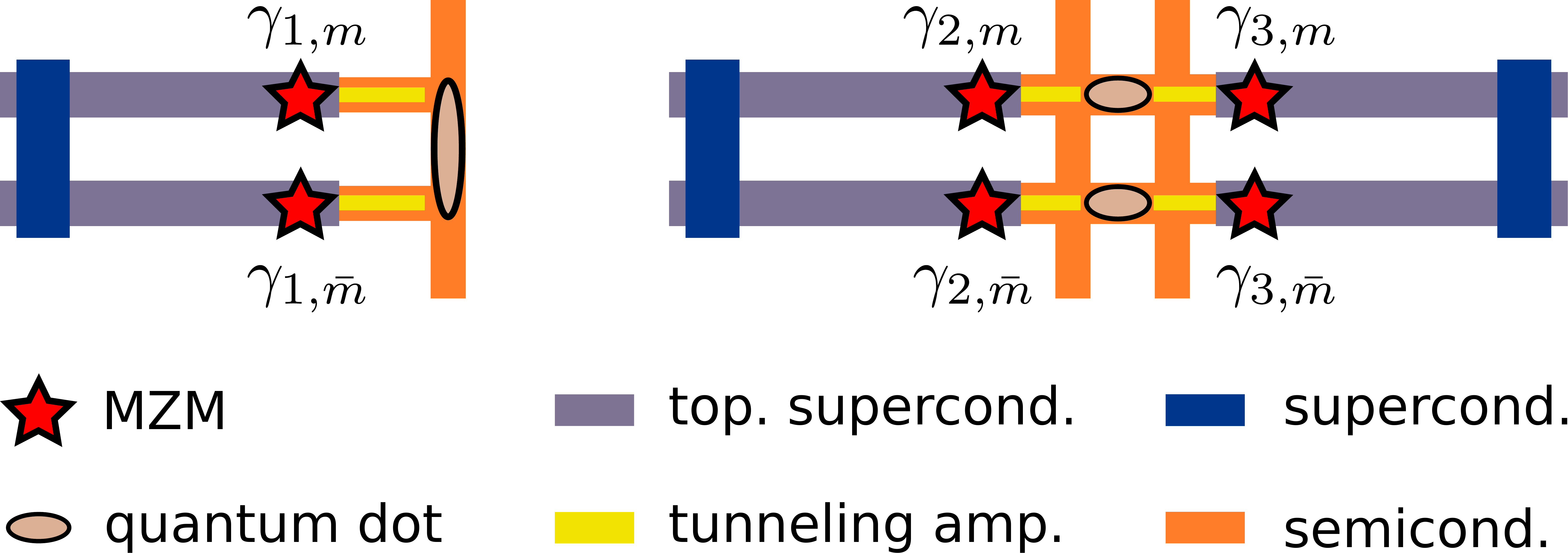}
\caption{MZM parity measurements using quantum dots.  {\it Left panel}: parity measurement of $i\gamma_{1,m}\gamma_{1,\bar{m}}$ by tunnel coupling (yellow line) the corresponding MZMs to a quantum dot.  {\it Right panel}: parity measurement of ${-\gamma_{2,m}\gamma_{2,\bar{m}}\gamma_{3,m}\gamma_{3,\bar{m}}}$ by tunnel coupling the corresponding MZMs to two quantum dots.  
For both panels, we show only half of the tetron(s) involved in the measurement.  The tunnel couplings enable the transfer of electrons between the quantum dots and the MZMs connected to it. Because the MZM islands have charging energy, an electron originating from a quantum dot returns to the quantum dot with high probability. Importantly, an electron can return either by (1) traveling through a subset of the measured MZMs then backtracking its way back to the quantum dot or (2) traversing the full loop formed by the tunnel couplings, the quantum dot(s), and the superconducting island(s). In (1), an even number of each Majorana operator is applied, while in (2) a product of all Majorana operators involved in the measurement loop are applied.  The processes of type (2) result in a measurement through a shift of the ground state properties of the quantum dot depending on the parity of the involved MZMs. 
}
\label{fig:measurement}
\end{figure}

For an array of tetrons, parity measurements can be implemented by coupling the MZMs to quantum dots~\cite{Karzig17} or by a conductance-based readout scheme \cite{Landau2016,Plugge2017}. Most of the measurement concepts can be used interchangeably between the two approaches. In the following, we focus on quantum-dot based measurements. The most common examples are measuring the parity of a pair of MZMs on a single island, or measuring a four-MZM parity composed out of two MZM-pairs from separate islands. The latter can induce correlated excitations of different islands. Other Majorana-based quantum computing architectures employ different measurement schemes, which in general will change the correlated events between islands being measured. In Section~\ref{sec:extensions}, we comment on how the different measurement schemes of other Majorana-based quantum computing architectures affect which types of correlated events the noise model should include.

To measure the parity of a pair of MZMs $\gamma_{1,m}$ and $\gamma_{1,\bar{m}}$ belonging to the same tetron, electrostatic gates in the semiconducting wire adjacent to the MZMs are tuned to form a quantum dot tunnel-coupled to $\gamma_{1,m}$ and $\gamma_{1,\bar{m}}$, as shown in the left panel of Fig.~\ref{fig:measurement}.  To measure the parity of four MZMs (two pairs on two different superconducting islands), the electrostatic gates are tuned to form two quantum dots, each of which is tunnel-coupled to a MZM on each island, as shown in the right panel of Fig.~\ref{fig:measurement}.   

An isolated quantum dot is described by a charging-energy Hamiltonian 
\begin{equation}\label{eq:QD-Ham}
H_D=E^\text{QD}_C \left(d^\dagger d -N_g\right)^2,
\end{equation}
where $d$ is the fermionic annihilation operator for the quantum dot and $N_g$ is the dimensionless gate voltage on the dot. Writing the eigenvalues of the number operator for quantum dot levels as $N_d$, the quantum dot Hamiltonian is spanned by the occupation basis $\ket{N_d}_{d}$.  We assume that the quantum dot is in the spin-polarized regime such that the available states are $\ket{0}_{d},\ket{1}_{d}$.  When performing a measurement, a tunneling Hamiltonian couples Eqs.~(\ref{eq:tetron-Ham}) and (\ref{eq:QD-Ham}).
For instance, for the two-MZM measurement, the tunneling term takes the form 
\begin{equation}\label{eq:Ht}
H_t = -a^- \left( t_{m} d^\dagger \gamma_{1,m} +t_{\bar{m}} d^\dagger \gamma_{1,\bar{m}}\right)+\text{H.c.},
\end{equation}
where $t_m$ is the tunneling amplitude between MZM $\gamma_{1,m}$ and the quantum dot, and $a^-$ is a bosonic operator removing a single electron charge from the island.   Equation~(\ref{eq:Ht}) hybridizes the two quantum dot states $\ket{0}_d$ and $\ket{1}_d$ in a two-MZM parity-dependent manner.  By measuring the energy levels, charge occupation, or quantum capacitance of the quantum dot, one can extract the two-MZM parity $i\gamma_{1,m}\gamma_{1,\bar{m}}$ of the system.  Such a two-MZM measurement can be used to infer whether the tetron is in computational state $\ket{\psi_0}_M$ or $\ket{\psi_1}_M$.  For a more-detailed discussion, see Ref.~\onlinecite{Karzig17}.

In the remainder of this section, we make the following gauge choice: all fermion operators are neutral and the charge on the system is accounted for by the bosonic operator $\hat{n}_s$.  We define the neutral creation operator of an above-gap quasiparticle with crystal momentum $k$ as $c_k^\dagger$ ({\it i.e.}, ${c_k^\dagger \ket{n_s,\{n_{\Delta,k}=0\}_\Delta}=\ket{n_s,\{n_{\Delta,k}=1\}_\Delta}}$).  When an electron enters or leaves the superconducting island, the eigenvalue $n_s$ changes by 1.  We account for this change with the ladder operators $a^\pm$, which raise or lower $n_s$.

\subsection{Error processes}\label{sec:error-probabilities}

\begin{figure}[t]
\begin{center}
	\includegraphics[width=.7\columnwidth]{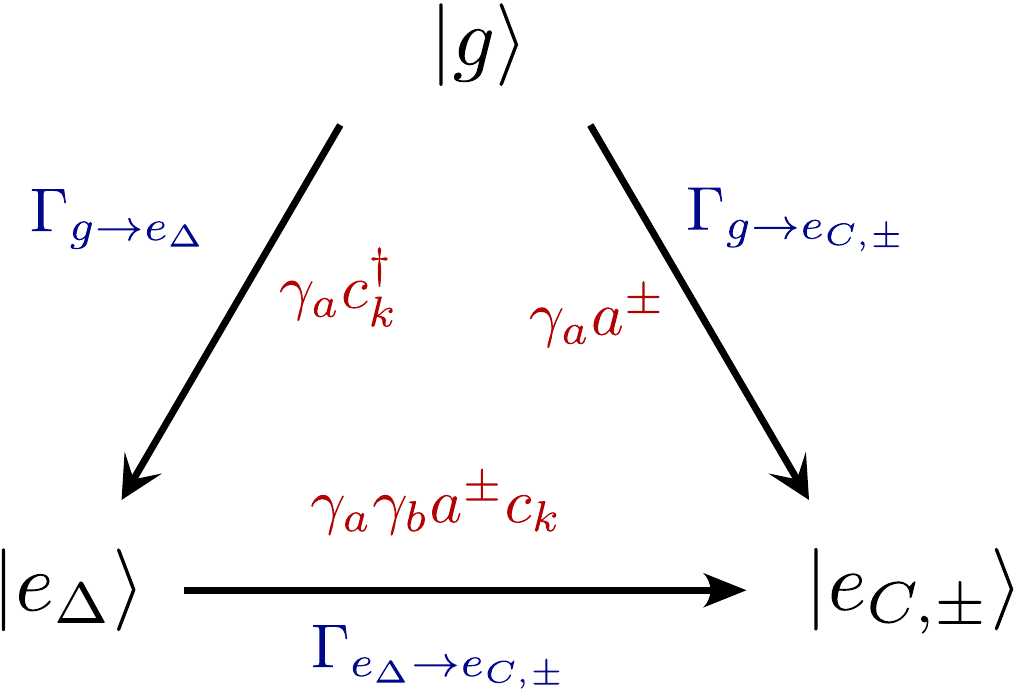}
	\caption{ Quasiparticle events for a single MZM island.  The corners of the triangle correspond to the three energy states we consider: the ground state with even MZM parity (corresponding to the computational states) $\ket{g}=\ket{0,\{0\}_\Delta}$, the thermal excited state with an above-gap quasiparticle $\ket{e_\Delta}=\ket{0,\{1\}_\Delta}$, and the quasiparticle-poisoned states $\ket{e_{C,\pm}}=\ket{\pm 1,\{0\}_\Delta}$.   The edges of the triangle denote the quasiparticle event that transitions the system between the given states  (thermal excitation of an above-gap quasiparticle or extrinsic quasiparticle poisoning), and are labeled by  the corresponding transition rate for that process (in blue) and the operators that act on the system for that given process (in red).  Arrows can be reversed by conjugating the operators, but the rates for the opposite processes can be drastically different. }
	\label{fig:diagram}
\end{center}
\end{figure}

In this section, we discuss the main physical processes contributing to errors in Majorana-based quantum computing architectures.  The corresponding error rates will be independent of which computational (MZM) state the system is in, as such we simplify notation by dropping the MZM labels $\ket{\psi}_\text{M}, \ket{\bar{\psi}}_\text{M}$ and only writing the energy state $\ket{g}, \ket{e_\Delta},$ or $\ket{e_{C,\pm}}$. When an effect is independent of which excited state the system is in, we will simply write $\ket{e}$.  

There are higher energy states, for instance, the state with both an extra electron and an above-gap quasiparticle on the island, that we have not discussed. Throughout this work, our aim will be to identify the {\it lowest order errors}: the most prominent errors in the system.   If the probability of a given error is less than or equal to the product of the probabilities of other errors, and the effect on the Hilbert space of the MZMs is the same for that particular error as for the combination of the other errors, then we call such an error {\it higher order}.  Processes involving excited states other than $\ket{e_\Delta}$ and $\ket{e_{C,\pm}}$ can be described as higher-order errors, see Appendix~\ref{app:higher-order}.
Note that an error is classified as higher order solely from its probability; it is possible for the physical process causing a particular higher order error to be distinct from the physical processes causing lower order errors.

Figure~\ref{fig:diagram} schematically illustrates single island error processes involving quasiparticles.  A transition between the ground and an excited state corresponds to the application of a single Majorana operator and thus sets the probability of a quasiparticle event.  The combination of an excitation and a relaxation, or a transition between excited states, corresponds to the application of an even number of Majorana operators.  These processes therefore contribute to the probability of a pair-wise dephasing event.  We first quantify the transition rates involving quasiparticles in a single island, before discussing correlated events between islands connected during a measurement and error processes that do not involve quasiparticles. In the rates given below, we only quote the parametric dependence and ignore prefactors of $\mathcal{O}(1)$.

\subsubsection{Thermally excited quasiparticles}\label{sec:thermal-error}

Thermal excitation of an above-gap quasiparticle, the top left process in Fig.~\ref{fig:diagram} $\ket{g}\to\ket{e_\Delta}$, is present in all MZM systems.  
This process occurs, for instance, when a Cooper pair breaks into two electrons, one of which occupies one of the non-local fermionic states formed by the MZMs while the other occupies a state in the continuum above the superconducting gap.  
Thermal excitation of a quasiparticle preserves the total fermion parity of the island and can therefore occur in an isolated system.\footnote{There are also thermal excitations that conserve the parity of the MZMs by exciting a pair of quasiparticles to the continuum. These processes have an energy cost greater than twice the superconducting gap and are therefore less likely to occur.  In practice, there will be a competition between a power-law enhancement for creating a particle-hole pair since it can happen anywhere in the system, and the additional exponential suppression due to the higher energy cost. Thermally excited particle-hole pairs do not cause errors by themselves, but increase the chance of subsequently transitioning to states that flip the MZM parity.  Furthermore, pair-excitation can cause correlated events when nearby superconducting islands are coupled together for a measurement.  For now we assume that these effects can be qualitatively captured by the transition rate of a single quasiparticle excitation and the rates describing correlated events. }

Crucially, in equilibrium, thermal excitation of a single quasiparticle to the state $\ket{e_\Delta}$ is exponentially suppressed in the ratio $\Delta/T$, where $T$ is the temperature of the system. The excitation and relaxation rates take the form
\begin{align}
\Gamma_{g\to e_{\Delta}} &= \tau_0^{-1} \exp(-\Delta/T) \label{eq:Gamma0-1}
\\ \Gamma_{e_{\Delta}\to g} &= \tau_0^{-1}\label{eq:Gamma1-0},
\end{align}
where $\tau_0$ is a characteristic time scale describing the electron-phonon coupling of the system. For InAs wires with an Al half-shell, $\tau_0\sim 50$ns \cite{Knapp2017}. In the presence of non-equilibrium quasiparticles, the factor $\exp(-\Delta/T)$ is essentially replaced by the number of quasiparticles in the vicinity of the MZMs.

A single thermal excitation event takes the computational (MZM) state of the island from an even parity state to an odd parity state, while a relaxation event does the reverse.  Both thermal excitation and relaxation apply a single Majorana operator $\gamma_a$ to the computational state, although as can be seen from Eqs.~\eqref{eq:Gamma0-1} and \eqref{eq:Gamma1-0} the corresponding rates are significantly different.  This is the underlying reason why all of the noise models of Section~\ref{sec:noise-models} account for quasiparticle events with two different probabilities, $p_\text{odd}^{(j)}$ and $p_\text{qp}^{(j)}$: if the island begins a time step in an odd MZM parity state, a relaxation event is much more likely than an excitation event if the island has even MZM parity, indicating a separation of scales between $p_\text{odd}^{(j)}$ and $p_\text{qp}^{(j)}$.  Furthermore, the application of a single quasiparticle event significantly changes the probability of future quasiparticle events, which is why our noise models only allow for a single noise event to occur in step 0, and a single noise event to occur in step 1(a) (this approximation is justified if the errors are sufficiently rare).  

A pair-wise dephasing event can result from an excitation and subsequent relaxation of a thermal quasiparticle.  The relaxation event is equally likely for all Majorana operators on the island, regardless of which Majorana operator was applied for the excitation event: this is the underlying reason why we allow for {\it all} pairs of Majorana operators, including the same Majorana operator twice, when considering pair-wise dephasing in our noise models.  Finally, notice that the rates of thermal excitation and relaxation are unaffected by whether or not the superconducting island is involved in a measurement.

\subsubsection{Extrinsic quasiparticle poisoning}\label{sec:qpp}

Extrinsic quasiparticle poisoning, the top right process in Fig.~\ref{fig:diagram} $\ket{g}\to\ket{e_{C,\pm}}$, is when a quasiparticle tunnels onto or off of the superconducting island, thereby changing the total fermion parity and charge of the island. We focus on the lowest-energy poisoning events where quasiparticles tunnel onto (off of) the island and into (out of) one of the fermion states provided by the MZMs.
In the following discussion, we will assume $n_g=0$ so that adding an electron to, or taking an electron off of the superconducting island costs an energy $E_C$; the following expressions could be made more general by replacing $E_C$ with $E_C\left( 1\mp 2 n_g\right)$. 
For an external quasiparticle to enter the island, it has to overcome the energy barrier $E_C +\delta \mu$, where $\delta \mu=\mu_{\text{is}}-\mu_{\text{res}}$ denotes a possible difference in the chemical potentials (a voltage bias) of the reservoir and the island. 

In the case of tetrons, the most likely source of quasiparticles is the quantum dot explicitly coupled to the MZMs during a parity measurement. When the quantum dot is tuned to resonance\footnote{Tuning the quantum dot to resonance maximizes the measurement visibility, but is not a necessary condition.}, the energy difference $\delta \mu=0$. Transitions between $\ket{g}$ and $\ket{e_{C,\pm}}$ are then described by the rates 
\begin{align}
\Gamma^{(k)}_{g\to e_{C,\pm}}&= g_T(k) \delta \exp\left\{- E_C/T\right\} 
\label{eq:eqp-dot1}
\\ \Gamma^{(k)}_{e_{C,\pm}\to g} &= g_T(k) \delta  \label{eq:eqp-dot2}
\end{align}
where  $g_T(k)$ is the dimensionless conductance between the quantum dot and MZM island. The parameter $\delta$ is either the level spacing of the quantum dot $\delta_\text{dot}$ when an electron transitions from the quantum dot to the island, or the effective MZM level spacing $\Delta$ when an electron transitions from the island to the quantum dot.  Equations~\eqref{eq:eqp-dot1}-\eqref{eq:eqp-dot2} can also be generalized to conductance-based measurement schemes \cite{Landau2016,Plugge2017}. In that case, $\delta_\text{dot}$ is replaced by the temperature $T$ and $g_T(k)$ is the dimensionless conductance between the lead and the MZM island. In the case of a quantum dot connected to the superconducting island, only a single channel with transmission $\mathcal{T}^{(k)}$ contributes to the island-dot coupling, thus $g_T(k)=|\mathcal{T}^{(k)}|^2$.  The transmission $\mathcal{T}^{(k)}$ is essentially zero for $k=0$ ({\it i.e.}, no measurement), and becomes appreciable for $k=1,2$.  Notice also that during the measurement, the quantum dot is tuned close to the charge degeneracy point, thus we do not have to pay extra energy to remove a particle from it.  If the quantum dot is then tuned to the bottom of its charging parabola when a measurement is not being performed, then $E_C\to E_C+E_C^{\text{QD}}$ in Eq.~(\ref{eq:eqp-dot1}), further protecting the system from quasiparticle poisoning. It thus follows that quasiparticle poisoning from the quantum dot is significantly more likely during a measurement.  This motivates why the noise models MC and PMC allow for different probabilities of quasiparticle events and pair-wise dephasing when an island is involved in a measurement.

Alternatively, quasiparticles could come from nearby metallic gates, for example those used to tune the MZM island into the topological regime. The transition rates are now independent of whether a measurement is being performed, as this is not expected to change the coupling between the island and the gate. If the metallic gates are kept at a large voltage, the energy $|\delta \mu|$ may become much larger than $E_C$ so that it could  be favorable for an extra quasiparticle to be on the superconducting island. In this case, it is essential that the dimensionless conductance between the metallic gate and the island $g_{\text{env}}$ is sufficiently small so that the island is not constantly being poisoned by quasiparticles. Since an insulating barrier to the gates suppresses $g_\text{env}$ exponentially in the thickness of the barrier, sufficiently small values of $g_{\text{env}}$ are possible in practice. In the following, we assume that extrinsic quasiparticle poisoning rates are well approximated by  Eqs.~(\ref{eq:eqp-dot1}) and (\ref{eq:eqp-dot2}).

From Eq.~\eqref{eq:eqp-dot1}, we see the benefit of the island having a large charging energy is to suppress the rate of extrinsic quasiparticle poisoning events.  Majorana-based quantum computing proposals for which the superconducting island is grounded ($E_C\to 0$) are likely to be more susceptible to quasiparticle and pair-wise dephasing events. 

A comment is in order about the fast relaxation of Eq.~\eqref{eq:eqp-dot2}. A crucial requirement is that the environment can accept or provide an electron at no energy cost. This is always the case for conductance-based measurements in which the island is connected to leads with no charging energy. For quantum dot-based measurements, it is important to properly initialize and decouple the quantum dots. An example procedure that always allows for fast relaxation is a double-dot measurement (see Fig.~\ref{fig:measurement} right panel).  By initializing the quantum dots in the state $\ket{0}_d\ket{1}_d$ and tuning them so that each dot is at its charge degenerate point, the island can relax its charge state by transitioning to the $\ket{1}_d\ket{1}_d$ or $\ket{0}_d\ket{0}_d$ state. Alternatively, we can use the quantum dots to check if a poisoning event has occured during the measurement by performing an additional charge measurement of the quantum dot(s) after the island is decoupled.  This additional measurement could allow immediate correction of extrinsic quasiparticle poisoning events, in which case we would not need to include such events in the noise models.

In terms of the computational state of the island, extrinsic quasiparticle poisoning has the same effect as the corresponding thermal quasiparticle event: an excitation, $\ket{g}\to\ket{e}$, takes the MZMs from an even parity state to an odd parity state by applying a single Majorana operator $\gamma_a$; relaxation, $\ket{e}\to\ket{g}$, does the reverse; and the combination of an excitation and subsequent relaxation, $\ket{g}\to\ket{e}\to\ket{g}$, applies the pair of Majorana operators $\gamma_a\gamma_c$ where $a$ can equal $c$.  Transitions between the excited states $\ket{e_{C,\pm}}$ and $\ket{e_\Delta}$ have the same effect on the computational state as a combined excitation and relaxation.  Moreover, a transition between excited states requires two unlikely processes to occur: first, an island must be excited and second, the island must transition to the other excited state before relaxing to the ground state.  It follows that such transitions only contribute a small correction to the probability of pair-wise dephasing events and can therefore be neglected.

\subsubsection{Correlated events}\label{sec:correlated}

Quasiparticle poisoning and thermal excitations can lead to correlated events that transfer excitations between two (or more) islands connected during a measurement. An excited charge state can be transferred by a quasiparticle tunneling between  two islands through the low-energy subspace provided by the MZMs. A thermal excitation can be transferred without changing the net particle number by an above-gap quasiparticle tunneling between two islands and a corresponding reverse tunneling of an electron through the low energy subspace. Each time an island transitions between the ground and an excited state, a Majorana operator is applied to the computational state of that island.  Thus, a process in which two islands swap being in the ground and excited state, 
\begin{equation}
\ket{g}\otimes\ket{e}\to \ket{e}\otimes\ket{g},\label{eq:swap}
\end{equation} 
results in a Majorana operator being applied to each island.  The corresponding MZMs are tunnel-coupled to the same quantum dot, as shown in the right panel of Fig.~\ref{fig:measurement} (either $\gamma_{2,m}\gamma_{3,m}$ or $\gamma_{2,\bar{m}}\gamma_{3,\bar{m}}$).
When this process combines with excitations and relaxations on the two islands, it results in correlated events involving three or four Majorana operators between two islands, described, {\it e.g.}, by the probabilities $p_{\text{cor,odd}}^{(2)}$ and $p_{\text{cor,even}}^{(2)}$, respectively, in PMC.

The most prominent correlated events come from the processes
\begin{align}
\ket{g}\otimes\ket{g} &\to \ket{e_x}\otimes \ket{g} \to \ket{g}\otimes\ket{e_x} \to \ket{g}\otimes \ket{g} \label{eq:cor1}
\\ \ket{g}\otimes\ket{g}&\to \ket{e_x}\otimes \ket{g} \to \ket{g}\otimes\ket{e_x} \label{eq:cor2}
\\  \ket{e_x}\otimes \ket{g} &\to \ket{g}\otimes\ket{e_x} \to\ket{g}\otimes\ket{g} \label{eq:cor3}
\end{align}
where the excited states $\ket{e_x}\in\{\ket{e_{C,\pm}},\ket{e_\Delta}\}$. Equation~\eqref{eq:cor1} applies an even number of Majorana operators and is described by the probability $p_\text{cor,even}^{(2)}$ in PMC. If relaxation processes are fast compared to the time step $\tau$, this is the most probable type of correlated event. Equations~\eqref{eq:cor2} and \eqref{eq:cor3} describe odd correlated events captured by $p_\text{cor,odd}^{(2)}$ in PMC and can be seen as being part of an even correlated event split up over two time steps. Both odd correlated events have the same effect on the computational subspace and are equally likely. To simplify our noise models, we only take into account the process described by Eq.~\eqref{eq:cor2}.  When connecting the probabilities defining PMC to physical processes, we therefore have to overestimate the probability of the event described by Eq.~\eqref{eq:cor2} by at least a factor of two in order to avoid undercounting odd correlated events. Similarly to the previous discussion of single-island errors, we ignore higher order correlated events between the two islands, {\it e.g.}, $\ket{e_{C,\pm}}\otimes\ket{g} \to \ket{e_{\Delta}}\otimes\ket{e_\Delta}$, see Appendix~\ref{app:higher-order}.

Assuming the islands are identical, there is no energy difference between the states $\ket{e_x}\otimes\ket{g}$ and $\ket{g}\otimes\ket{e_x}$.  As such, the corresponding transition rates have no exponential suppression:
\begin{eqnarray}
\Gamma_{e_{e_C,\pm},g\to g,e_{e_C,\pm}} &=& g_{\text{is-is}} \Delta \label{eq:ex,g-g,ex1}\\
\Gamma_{e_{e_\Delta},g\to g,e_{e_\Delta}} &=& g_{\text{is-is}}^2 \Delta\delta_{\text{is}}/E_C,\label{eq:ex,g-g,ex2}
\end{eqnarray}
where $g_{\text{is-is}}$ is the dimensionless conductance between the two islands. The second line originates from the tunneling of an above-gap quasiparticle (level spacing $\delta_\text{is}$) and a tunneling through the MZMs (effective level spacing $\Delta$) within the time $1/E_C$ of the virtual charge and thermal excited state. Note that for weak coupling to the dots, the interisland conductance $g_{\text{is-is}}\ll g_T^{(k)}$, and grows quadratically with the tunneling amplitude to the dots. Despite the smaller prefactor of  Eqs.~(\ref{eq:ex,g-g,ex1}) and \eqref{eq:ex,g-g,ex2}, the corresponding rates are expected to be much larger than those given in Eqs.~(\ref{eq:Gamma0-1}) and (\ref{eq:eqp-dot1}) due to the absence of exponential suppression.  It follows that an excitation on one island moves to the other with higher probability than an independent relaxation of one island and excitation of the other; this is what is meant by a {\it correlated event.}

Processes in which $\ket{e_x}\otimes \ket{g}\to \ket{g}\otimes\ket{e_y}$ with ${x\neq y}$ are a combination of transitions between excited states ${\ket{e_{C,\pm}}\leftrightarrow \ket{e_\Delta}}$ and the same-state correlated events of  Eqs.~\eqref{eq:ex,g-g,ex1} and \eqref{eq:ex,g-g,ex2}. We thus expect the corresponding transition rates to only contribute a small correction to the correlated event probabilities in PMC, for the same reason that transitions between excited states only contribute a small correction to the probability of a pair-wise dephasing event.  

\subsubsection{Other errors}\label{sec:other}

Processes that do not involve quasiparticles can also contribute to the probabilities of pair-wise dephasing events and measurement bit-flips.  Finite MZM hybridization, $\delta E_{ab}$ in Eq.~(\ref{eq:tetron-Ham}), leads to dephasing of the MZMs associated with operators $\gamma_a\gamma_b$.   The MZM hybridization is caused by finite overlap of the MZM wavefunctions when the correlation length of the topological superconductor is not sufficiently smaller than the distance separating the MZMs.  The degeneracy splitting can have both a constant and fluctuating piece.  The former would be problematic for topological gates that rely on degeneracy of the computational subspace, but will not be explicitly considered here.  The latter is problematic even for memory storage.  Charge noise in the electrostatic gates and substrate on which the MZM island is sitting can cause fluctuations of the electric field at the topological wire, which in turn results in a noisy time-dependence of the MZM hybridization $\delta E_{ab}$.  We expect the rate of such errors to scale as 
\begin{equation}\label{eq:GammaEE}
\Gamma_{EE}\sim \sqrt{\int d\omega S_{EE}(\omega)} e^{-L/\xi},
\end{equation}
where $S_{EE}(\omega)$ is the spectral function of electric field fluctuations in the system and $L$ is the typical separation of MZMs in the island.  Any computing approach that involves tuning a superconductor into the topological regime with electrostatic gates will likely experience noise described by $\Gamma_{EE}$, however, as discussed in Ref.~\onlinecite{Knapp2017}, the time scales associated with this dephasing process are predicted to be quite long for reasonably-sized systems (order of minutes for $L/\xi\sim 30$). At these system sizes, the thermal excitations of Eq.~\eqref{eq:Gamma0-1} become the most important source of pairwise dephasing events. For concreteness, we use this limit and neglect hybridization-based errors for the probability estimates in the next section. 

Another source of error occurs when the classical bit storing the outcome of a $2k$-MZM parity measurement does not agree with the actual parity of the measured MZMs at the end of the time step.  Physically, the measurement is implemented by adding a term to the Hamiltonian with the $2k$ Majorana operators to be measured.  This term splits the energy of the two eigenstates of $2k$-MZM parity and the environment then quickly dephases these two states, collapsing the system into one of the eigenstates.  The measurement outcome is obtained by gathering statistics throughout the time step as to which eigenstate the system has collapsed.
 
There are two ways for measurement outcomes to disagree with the state of the system at the end of the time step: (1) the classical bit storing the measurement is flipped, or (2) the state of the system changes between the measurement projection and the end of the time step.  Errors from case (1) are described by  the probability $p_\text{mst}^{(k)}$ in MC and PMC (probability $p_\text{mst}$ in QpBf).  This case either results from statistical error (the integration time of the measurement was too short), or from classical noise in the measurement flipping the bit's outcome. Classical noise is strongly dependent on the physical implementation of the measurement, as such, we do not further analyze these processes other than to note that a measurement error will be more probable for smaller measurement signals, likely to occur when multiple islands are involved in a single measurement.

Due to our convention that measurement projections are performed at the end of a time step, our models do not describe case (2).  In reality, a measurement is performed by collecting statistics throughout the time interval, thus the measurement result will likely reflect a noise event that occurs near the beginning of the time step, but not one that occurs at the end.  The latter are instead reflected in the next measurement that occurs.  In an error correction protocol in which almost every island is measured in each time step, it is only a small effect (for a large number of time steps) whether we apply projectors at the beginning or the end of a time step, as for one convention the noise events will be reflected in the measurement associated with that time step, and in the other convention the noise events will be reflected in the measurements associated with the subsequent time step.  
\\

\subsection{Error probabilities}\label{sec:prob-summary}

We now use the transition rates discussed in the previous section to estimate how the error probabilities in the stochastic Majorana noise model PMC depend on physical parameters of the system.  These estimates serve the double purpose of connecting the noise models of the previous section to the underlying physical system, and of simplifying the simulations of pseudo-thresholds in the next section by reducing the number of independent parameters in PMC.  At the end of this section, we discuss the other three Majorana noise models Qp, QpBf, and MC, and the limit in which they reduce to the analogous qubit-based noise models.  

\begin{figure}[t]
	\includegraphics[width=\columnwidth]{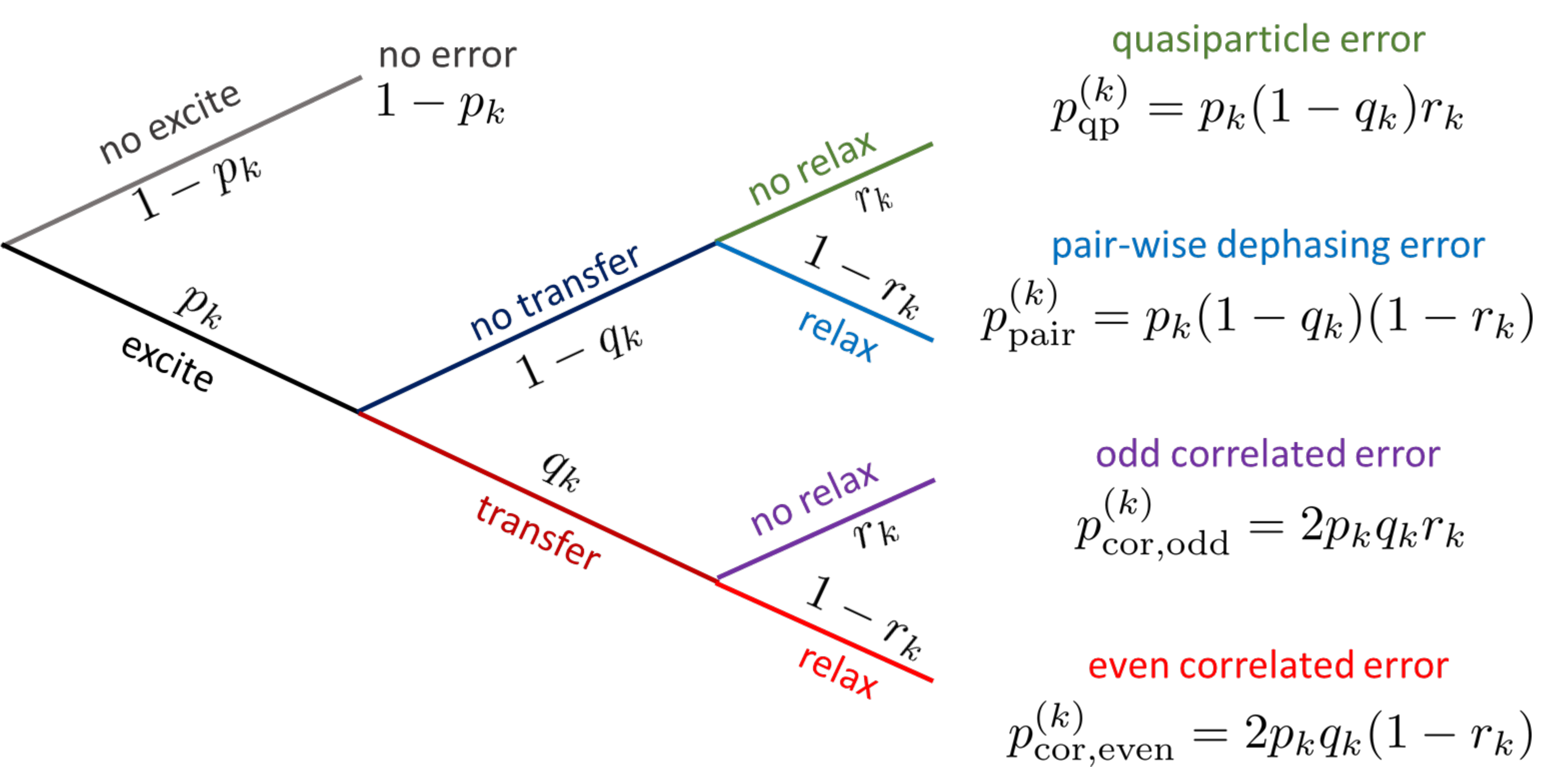}
	\caption{ Probability tree for the different error types involving quasiparticles for Majorana noise model PMC: $p_k$ is the probability that a MZM on an island in a $k$-island measurement is excited, $q_k$ is the probability that this excitation is transferred to a different island in the measurement, and $r_k$ is the probability that this excitation does not relax.  The probabilities $p_\text{qp}^{(k)}$ and $p_\text{pair}^{(k)}$ are for single islands, while $p_\text{cor,odd}^{(k)}$ and $p_\text{cor,even}^{(k)}$ are for both islands in a measurement.  In the text, we relate these small, dimensionless parameters to the transition rates discussed in the previous section.  When $m=2$, corresponding to a tetron architecture, and $r_k=0$, corresponding to no odd Majorana operators, the model PMC reduces to the qubit-based model Pauli circuit noise.
	}
	\label{fig:prob-tree}
\end{figure}

In Fig.~\ref{fig:prob-tree}, we schematically show that for a $k$-island measurement, all errors involving quasiparticles are related by three small dimensionless parameters: $p_k$, the probability that a quasiparticle is excited on one of the islands involved in the measurement; $q_k$, the probability that the energy from this excitation is transferred to a different island involved in the measurement; and $r_k$, the probability that an excitation created during the time step does not relax.  These dimensionless parameters are related to the three types of transition rates involving quasiparticles discussed in the previous sections: excitations $\Gamma^{(k)}_{g\to e_x}$, energy transfers between islands $\Gamma^{(2)}_{e_x,g\to g,e_x}$, and relaxations $\Gamma^{(k)}_{e_x\to g}$, respectively.   More explicitly, denoting the length of the time step by $\tau$, we find
\begin{align}
p_k &= 2m\sum_{x}\Gamma_{g\to e_x}^{(k)} \tau \label{eq:p}
\\ r_k &=\max_x \left(\tau \Gamma^{(k)}_{e_x\to g}\right)^{-1} = \tau_\text{r}^{(k)}/\tau \label{eq:r}
\\ q_2&= \max_{x}\left(\text{exp}\left\{\Gamma^{(2)}_{e_x,g \to g,e_x}\tau_x^{(2)}\right\}-1\right). \label{eq:q}
\end{align}

The factor of $2m$ in Eq.~(\ref{eq:p}) comes from any of the $2m$ MZMs on an island initially being excited. We assume that excitations are rare, so that $\Gamma^{(k)}_{g\to e_x}\tau \ll 1$ and $p_k$ is a small parameter.  For an idle MZM, we expect extrinsic quasiparticle poisoning to be negligible, thus only $\Gamma_{g\to e_\Delta}^{(0)}$ will contribute to $p_0$.  Generally, we expect $\Gamma_{g\to e_\Delta}^{(0)}\approx \Gamma_{g\to e_\Delta}^{(1)}\approx \Gamma_{g\to e_\Delta}^{(2)}$ and  $\Gamma_{g\to e_C}^{(1)}\lesssim \Gamma_{g\to e_C}^{(2)}$, so that $p_0<p_1\lesssim p_2$.

In Eq.~(\ref{eq:r}), $\tau_\text{r}^{(k)} \equiv \max_{x}\left[1/\Gamma^{(k)}_{e_x\to g}\right]$ is the typical time scale over which the longest living excited state relaxes to the ground state. We assume $\tau_\text{r}^{(k)}\ll\tau$ based on physical considerations.  The parameter $\tau$ is bounded from below by the longest measurement time, a four-MZM parity measurement in the current discussion.   We expect this measurement time to be $1-10~\mu$s for high-fidelity measurements, while typical time scales for the electron-phonon coupling facilitating quasiparticle relaxation are $\sim 0.1~\mu$s~\cite{Lutchyn2007}.  Physically, we expect $0<r_k<1/10$ and $r_0 \approx r_1 \approx r_2$. The probability of an excitation, created at the beginning of the time step, to not relax is exponentially small in $1/r_k$ and will be neglected through the remainder of the paper. Equation~\eqref{eq:r} then follows from the probability of a given excitation to happen withing the window $\tau_\text{r}^{(k)}$ just before the end of the time step. When $r_k=0$, an island never ends (or begins) a time step in an excited state; for $m=2$, the noise model then reduces to the qubit-based model Pauli circuit noise. If the assumption of $r_k\ll 1$ is violated, excitations become long-lived. In this  case, excitations can travel long distances before relaxing, thereby significantly degrading the effectiveness of error correction. We discuss noise models describing long-lived excitations in Appendix~\ref{app:long-lived}. 

For simplicity, and to obtain an upper bound on the correlated errors in Eq.~\eqref{eq:q}, we maximize the probability for transferring energy over the excited states.
Since $q_2$ is not exponentially suppressed and can therefore be $\mathcal{O}(1)$ when islands involved in a measurement are well-coupled, we did not use the approximation for small transition rates.  Physically, we expect the probability of an energy transfer between the two islands to be smaller than the probability that the excitation remains on the same island, so we anticipate $0<q_2<1/2$.  For an idle island or single-island measurement, $q_0=q_1=0$. 

In the following, we describe the noise model PMC using the parameters introduced above.
\\

\noindent {\it Quasiparticle events.}  The probability $p_\text{qp}^{(k)}$ that a single Majorana operator is applied to an island involved in a $2k$-MZM measurement can be obtained from considering an initial excitation ($p_k$) of a quasiparticle that does not relax ($r_k$) and does not transition to another island ($1-q_k$). We therefore find
\begin{align}
p_\text{qp}^{(k)} 
&\approx  \,p_k \left(1-q_k\right) r_k\,. \label{eq:p-qp}
\end{align}

The probability $p_\text{odd}^{(k)}$ of an initially excited state to relax during the time step is given by 
\begin{align}
p^{(k)}_{\text{odd}}& \approx 1-e^{-1/r_k} \\
p^{(k)}_{\text{odd}}& > 1-p_{\text{qp}}^{(k)}. \label{eq:p-odd}
\end{align}
The bound in Eq.~\eqref{eq:p-odd} reflects the assumption of fast relaxation. In that case, it is more likely for an excitation to relax and get re-excited, than for the excitation to remain over the entire time step. Within the fast relaxation limit, we can use the upper bound for $1-p_\text{odd}^{(k)}=p_\text{qp}^{(k)}$. We discuss how the noise models can be generalized beyond the fast relaxation limit in Appendix~\ref{app:long-lived}. 
\\

\noindent {\it Pair-wise dephasing events.}  Pair-wise dephasing events can occur from quasiparticle excitations and relaxations, or from non-quasiparticle processes such as MZM hybridization discussed in Section~\ref{sec:other}: 
\begin{equation}
\begin{split}\label{eq:p-pair}
p_\text{pair}^{(k)}
 & \approx p_k \, \left(1-q_k\right)\,\left(1-r_k\right) +m\,\Gamma_{EE}\,\tau.
\end{split}
\end{equation}
Here, $(1-q_k)(1-r_k)$ is the probability of an excited quasiparticle to relax within the same island and thus contribute to pair-wise dephasing. The second term arises from noise in the hybridization, see Eq.~\eqref{eq:GammaEE}. The combinatoric prefactor of $m$ counts the number pairs of MZMs whose wavefunction overlap contributes to pair-wise dephasing. In the following, we will assume the MZMs are sufficiently separated that we can drop the second term.
\\

\noindent {\it Correlated events.}  An odd correlated event occurs when one island involved in a measurement is excited and transfers its excess energy to a second island, which remains in the excited state for the rest of the time step.  Similar to quasiparticle events, only excitations within a fraction of $r_k$ of the time step contribute significantly to odd correlated events.  More explicitly, the probability of an odd correlated event is
\begin{equation}\label{eq:p-cor,odd}
\begin{split}
p_\text{cor,odd}^{(2)} 
 &\approx  2\, p_2\, q_2\, r_2.
\end{split}
\end{equation}
The factor of two is a result of the initial quasiparticle excitation occurring on either of the two islands involved in the measurement. 

An even correlated event occurs when a quasiparticle excitation in one island is transferred to the other island, which subsequently relaxes: 
\begin{equation}\label{eq:p-cor,even}
\begin{split}
p_\text{cor,even}^{(2)} 
 &\approx 2 p_2 \, q_2 \, (1-r_2).
\end{split}
\end{equation}
The factor of two again comes from either island initially being excited.
When hybridization errors are subleading ({\it i.e.}, the MZMs are sufficiently well-separated), an even correlated event is essentially a pair-wise dephasing event interrupted by an energy transfer.  
\\

\noindent {\it Measurement bit-flip}.   A measurement bit-flip depends only on classical and statistical noise, and is thus independent of the other probabilities in the noise model. The signal-to-noise ratio of a measurement typically increases as a square root of the integration time \cite{Clerk10}. We assume that $\tau$ is essentially given by the integration time of the measurement. The confidence of the measurement outcome therefore behaves as 
\begin{equation}\label{eq:p-mst}
p_{\text{mst}}^{(k)} \approx e^{-\tau/\tau_\text{mst}^{(k)}},
\end{equation}
where $\tau_\text{mst}$ is a characteristic time scale that gives a signal-to-noise ratio of 1. Four-MZM measurements are expected to have lower visibility and thus be more susceptible to statistical and classical error, {\em i.e.}, $p_{\text{mst}}^{(1)}<p_{\text{mst}}^{(2)}$.
\\

\begin{table}[t!]
\scriptsize
	\begin{tabular}{|c|l|} \hline 
	~Model~&~Probabilities~ \\ \hline 
	~ Qp~ & ~$p_\text{pair} = p\, (1-r)$,~$p_\text{qp} = p\,r$,~~$ p_\text{odd} = 1- p_\text{qp}$ \\ \hline
	~QpBf~ & ~$p_\text{pair} = p\, (1-r)$,~$p_\text{qp} = p\,r$,~~$ p_\text{odd} = 1- p_\text{qp}$,~~$p_\text{mst}$~ \\ \hline
	MC  & ~$p_\text{pair}^{(k)} = p_k \, (1-q_k) \, (1-r)$,~~$p_\text{qp}^{(k)} = p_k \, (1-q_k)\, r$\\
	\& &~ $p_\text{odd}^{(k)} = 1- p_\text{qp}^{(k)}$,~~$ p_\text{cor,even}^{(2)} = 2p_2 \, q_2 \, (1-r)$\\
	PMC&~ $p_\text{cor,odd}^{(2)} = 2p_2 \, q_2 \, r$,~~$p_\text{mst}$\\ \hline 
	\end{tabular}
	\caption{
	Probabilities for pair-wise dephasing, quasiparticle, and correlated events in the four stochastic Majorana noise models can be written in terms of an excitation parameter $p_k$ ($p$), a relaxation parameter $r$, and a correlation parameter $q_k$ (with $q_0=q_1=0$), see Eqs.~(\ref{eq:p}-\ref{eq:q}). The expressions above neglect the contribution of hybridization to pair-wise dephasing events, and assume that quasiparticle relaxation times are fast, {\em i.e}, $p_\text{qp}^{(k)}>\exp(-1/r)$ and similar regardless of whether an island is involved in a measurement.  Model PMC is physically motivated when $0\leq r\lesssim 1/10$ and $0 \leq q_2\leq 1/2$.  When $m=2$, the limit of $r=0$ corresponds to the qubit model Pauli circuit noise.  The expressions in this table are used in the pseudo-threshold calculations in Section~\ref{sec:threshold}.    
	}
\label{fig:prob-summary}
\end{table}

We have shown that the probabilities defining model PMC can be defined in terms of excitation parameters $p_k$, relaxation parameter $r_k$, and a correlation parameter $q_2$.  Varying these parameters over the appropriate ranges and studying the effect on the pseudo-threshold for the system should inform us about the relative importance of single-island vs. two-island errors, as well as errors with an odd vs. even number of Majorana operators.  A natural choice for the three other stochastic Majorana noise models is to assume a similar dependence of the probabilities on these parameters.  In the next section, for ease of comparing the effects of different errors, we will set $r_k=r$ and use the lower bound of the relaxation probability $p_\text{odd}^{(k)}=1-p_\text{qp}^{(k)}$. We summarize the scaling relations between error probabilities for the four models in Table.~\ref{fig:prob-summary}.  

Finally, note that if we set $p_0=p_2$ and restrict our attention to tetrons ($m=2$), the noise models MC and PMC differ only in that some of the odd correlated events in MC do not occur in PMC.
For both MC and PMC, there are only 16 computationally distinct\footnote{Because the total MZM parity of an island is fixed, some Majorana operators are computationally equivalent, {\it e.g.} $\gamma_{j,1}\gamma_{j,2}$, $\gamma_{j,2}\gamma_{j,1}$, $\gamma_{j,3}\gamma_{j,4}$, and $\gamma_{j,4}\gamma_{j,3}$.} even correlated events, each occurring with equal probability.  In contrast, there are 16 computationally distinct odd correlated events that can occur in PMC with equal probability, and 32 computationally distinct odd correlated events that can occur in MC with equal probability.  For larger values of $m$, we expect MC and PMC to differ more substantially.

\section{Application of noise models}\label{sec:threshold}

We now apply the noise models presented in Section~\ref{sec:noise-models} to analyze the error correction performance of the Bacon-Shor code~\cite{Bacon06,Kribs05,Shor96} on a small system of tetrons.   As explained in Section~\ref{sec:probabilities}, a tetron consists of four MZMs, and therefore stores a single qubit of information in the overall even parity state.  Although our noise models can be applied to any quantum error correction scheme, there are several motivations for considering this small subsystem code rather than (1) a Majorana fermion code specifically designed to correct for quasiparticle events~\cite{Bravyi10,Vijay2017,Hastings17}, or (2) a qubit-based stabilizer code ({\it e.g.}, the surface code).  (1)  A Majorana fermion code either requires the ability to measure the total parity of all MZMs on an island, or the ability to dynamically adjust the number of MZMs on a given island. We discuss in Section~\ref{sec:extensions} why it is experimentally challenging to satisfy these conditions.  Furthermore, the codes discussed in Ref.~\onlinecite{Vijay2017} are categorized with a notion of code distance in which the most probable noise events are quasiparticle events, which is not expected to be the case for a tetron architecture.  Rather, for low temperatures ($T\ll E_C, \Delta$) and fast relaxation ($r_k\ll1$), quasiparticle events in a system of tetrons are converted by the environment into Pauli errors, which in turn are correctable by qubit-based codes.  As discussed in Section~\ref{sec:physical}, the odd total-MZM parity state of an island is highly excited above the (even MZM parity) ground state; as such, the environment relaxes the system back to the ground state.  In the language of Refs.~\onlinecite{Hastings17,Vijay2017}, the environment measures the missing stabilizers of the Majorana fermion code needed to detect errors involving odd numbers of Majorana operators.  

Regarding point (2), in the Bacon-Shor code, error correction is built out of two-qubit measurements, which can be simply implemented for tetrons with four-MZM parity measurements.  In contrast, typical stabilizer codes involve measurements of at least four-qubit (eight Majorana) operators, which are expected to be more difficult to implement experimentally.  While in principle it is possible to implement higher-weight measurements from a sequence of smaller-weight measurements, the extra operations can significantly increase the noise.  For example, a six-MZM measurement for a system of tetrons could be performed using  two ancilla tetrons with six four-MZM and four two-MZM measurements~\cite{Karzig17}, or alternatively by preparing ancilla tetrons in a cat state, then doing a sequence of four-MZM measurements~\cite{Litinski2017}.  Adding ancilla tetrons and additional measurements provides new locations and opportunities for noise to occur before it can be corrected by the code.  Additionally, as is further discussed in Section~\ref{sec:conclusions}, higher-weight measurements can result in a higher probability of correlated errors, with negative effects on the pseudo-threshold.  Circumventing higher-weight measurements using subsystem codes is therefore a natural starting point for error correction in Majorana-based quantum computing architectures.  Determining optimal error correction procedures by weighing the advantages and disadvantages of higher-weight measurements is an interesting direction for future research.

We review error correction with subsystem codes, focusing on the Bacon-Shor code and how it applies to a system of tetrons.  We then analyze the conditions for fault tolerance ({\it i.e.}, compute pseudo-thresholds) for each of the noise models presented in Section~\ref{sec:noise-models}.  The analysis of this section is restricted to quantum memory error correction; analyzing the fault tolerance of logical gates is an important subject that we relegate to future studies.  The choice of error correcting code and error correction protocol have not been optimized to minimize the number of resources~\cite{Chao17,Chamberland18} or to maximize the fault tolerance threshold~\cite{Fowler09}.  
As such, the pseudo-threshold values reported here are more informative in their relative magnitude (within a given noise model for different parameter values) than in their absolute value.  At the end of this section, we separately discuss what experimental implications can be drawn from our analysis.

\subsection{Subsystem error correcting codes} \label{sec:subsystem}

As subsystem codes are less commonly studied compared to stabilizer codes \cite{Nielsen11}, we first review the standard application of a subsystem code to a qubit noise model ({\it i.e.}, errors involving an even number of Majorana operators in a tetron architecture).  For a more extensive discussion of subsystem codes, see Ref.~\onlinecite{Poulin05}.  We then discuss how a subsystem code can also correct for errors involving an odd number of Majorana operators in a tetron architecture.  We address measurement bit-flips later when discussing the application of the Bacon-Shor code to the noise models of Section~\ref{sec:noise-models}. An example illustrating the formal concepts introduced below is given in Section~\ref{sec:BS}. 

Subsystem codes are generalizations of stabilizer codes in which information is encoded in a {\it subsystem} of a {\it subspace} of the Hilbert space, rather than a {\it subspace}.  More explicitly, the Hilbert space of physical qubits $\mathcal{H}$ can be decomposed into a code space $\mathcal{H}_\mathcal{C}$ and the perpendicular subspace of errors, ${\mathcal{H}_\mathcal{E}=\mathcal{H}_\mathcal{C}^\perp}$:
\begin{equation}
\mathcal{H}=\mathcal{H}_\mathcal{C}\oplus\mathcal{H}_\mathcal{E}.
\end{equation}
The code space can be further decomposed into a logical subsystem $\mathcal{H}_\mathcal{L}$, the Hilbert space of the logical qubits, and a gauge subsystem $\mathcal{H}_\mathcal{G}$:
\begin{equation}
\mathcal{H}_\mathcal{C} = \mathcal{H}_\mathcal{L}\otimes \mathcal{H}_\mathcal{G}.  
\end{equation}
In a stabilizer code, $\mathcal{H}_\mathcal{G}$ is trivial.  The benefit of having the gauge subsystem is that error correction only needs to correct an error modulo the gauge subsystem structure; a non-trivial action on the gauge subsystem does not affect the encoded quantum information.  As we detail below, this extra degree of freedom can allow the error correction procedure to be implemented directly with smaller measurements, but comes at the cost of a reduction of the number of inequivalent code states.

\begin{figure}[t]
	\includegraphics[width=\columnwidth]{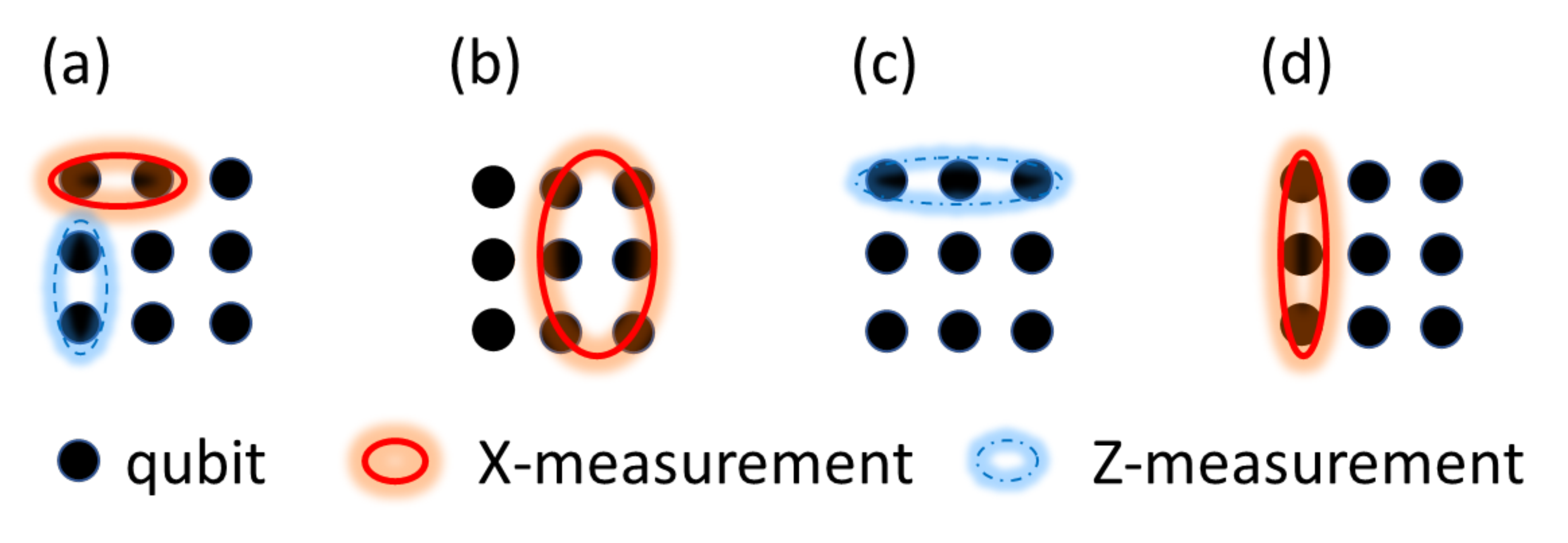}
	\caption{The $d=3$ Bacon-Shor code storing a single logical qubit is implemented by applying $X$-type (red solid ovals) and $Z$-type (blue dashed ovals) Pauli operators on a grid of $9$ qubits (black dots). $(a)$ Example of gauge generators. $X$-type ($Z$-type) gauge generators act on pairs of nearest neighbor qubits in the horizontal (vertical) direction. There are six different $X$ and $Z$ gauge generators. $(b)$ Example of an $X$-type stabilizer. $X$ stabilizers consist of pairs of $X$-type Pauli operators applied to an even number of columns. $Z$ stabilizers are pairs of $Z$-type Pauli operators applied to an even number of rows (not shown). $(c)$ The $Z$-type bare logical operator is applied to an odd number of rows.  $(d)$ The $X$-type bare logical operator is applied to an odd number of columns.  
	}
	\label{fig:BaconShor}
\end{figure}

More formally, we specify a subsystem code using subgroups of the Pauli group for $n$ physical qubits, $\mathcal{P}_n$:
\begin{itemize}
\item The {\it gauge group}, $\mathcal{G}$: a non-Abelian subgroup of $\mathcal{P}_n$ which defines the subsystem code.
\item The {\it stabilizer group}, $\mathcal{S}(\mathcal{G})$: the largest subgroup of $\mathcal{G}$, excluding $-\mathbb{1}$, consisting of elements which commute with every element of $\mathcal{G}$.  
\item The group of {\it bare logical operators}, $\mathcal{L}_B$: Pauli operators that commute with, but do not belong to, the gauge group $\mathcal{G}$.
\item The group of {\it dressed logical operators}, $\mathcal{L}_D=\langle \mathcal{L}_B,\mathcal{G}\rangle$.  
\end{itemize}
The \textit{code space}, $\mathcal{H}_\mathcal{C}(\mathcal{G})$, of a subsystem code with gauge group $\mathcal{G}$ is the $+1$ eigenspace of all stabilizer operators.  The {\it code distance}, $d$, is defined to be the minimum support of any logically non-trivial element of $\mathcal{L}_D$.  We use the standard qubit definition of code distance, not to be confused with the Majorana fermion code distance used in Refs.~\onlinecite{Bravyi10,Hastings17,Vijay2017}.

Error correction with a subsystem code involves using the measurement outcomes of the generators of the stabilizer group (stabilizers) to infer if any unwanted Pauli noise operators have been applied to the system\footnote{As stabilizer generators commute, they can be measured simultaneously.  This is not the case for gauge generators, which do not necessarily commute.}. One of the most appealing features of subsystem codes is that the eigenvalue of each stabilizer generator can be determined by multiplying together the eigenvalues of some of the gauge generators, which are often easier to measure. For an example of gauge generators and stabilizers, see Fig \ref{fig:BaconShor}.

To model memory error correction, the system is prepared in an eigenstate of one of the logical operators and is periodically measured to check that the state remains unchanged.  If at least one, but fewer than $d$, Pauli errors occur in an error correction round (measurement of all stabilizers), the system will no longer be in $\mathcal{H}_\mathcal{C}(\mathcal{G})$ and some of the stabilizer measurements will have outcome $-1$.  The {\it syndrome} of a given Pauli error is the corresponding set of stabilizer measurement outcomes. The error correction protocol uses the syndrome to infer which Pauli operators have been applied (under the assumption that the minimal number of errors corresponding to that syndrome has occurred). The errors can then be corrected by appropriate application of Pauli operators, returning the system to its intended state.  However, if more than $(d-1)/2$ Pauli errors occur in an error correction round, the correction procedure might incorrectly diagnose the error; in this case, the correction procedure might change the logical state of the encoded quantum information, resulting in a failure of the error correction protocol. 

\subsection{Bacon-Shor codes}\label{sec:BS}

One particular subsystem code, the distance-$d$ Bacon-Shor code, is implemented on a $d\times d$ grid of qubits, with $d$ odd~\cite{Bacon06,Kribs05}. Figure~\ref{fig:BaconShor} shows the relevant operators of the $d=3$ Bacon-Shor code.  The generators of the gauge group $\mathcal{G}$ are $XX$ acting on horizontal nearest neighbors and $ZZ$ acting on vertical nearest neighbors, as depicted in Fig.~\ref{fig:BaconShor}(a).  The stabilizer group $\mathcal{S}(\mathcal{G})$ has $d-1$ $X$-type and $d-1$ $Z$-type generators, where each $X$-type generator consists of $X$ applied to two columns ($2d$ qubits) and each $Z$-type generator consists of $Z$ applied to two rows ($2d$ qubits), see Fig.~\ref{fig:BaconShor}(b). A $Z$-type bare logical operator is a string of $Z$ operators applied to a row of qubits [Fig.~\ref{fig:BaconShor}(c)], and an $X$-type logical operator is a string of $X$ operators applied to a column of qubits [Fig.~\ref{fig:BaconShor}(d)].  Applying a stabilizer operator to a bare logical operator results in an equivalent bare logical operator.  Application of a gauge operator to a bare logical operator results in an equivalent dressed logical operator.

For the distance-$d$ Bacon-Shor code ($d=3$ in Fig.~\ref{fig:BaconShor}), a single $X$ error in a given row will anti-commute with any $Z$-type stabilizer with support in that row and will therefore have the same syndrome as all other $X$ errors in that row. 
This ambiguity does not cause any problem for error correction as $X$ errors belonging to the same row differ from each other by a gauge generator, and thus are corrected by the same procedure: applying a single $X$ operator to one qubit in the row.  
In order to find each stabilizer measurement outcome, it is enough to measure the $d$ two-qubit gauge generators and multiply the outcomes ({\it e.g.}, three two-qubit measurements for the $d=3$ code shown in Fig.~\ref{fig:BaconShor}).
Single-qubit $Z$ errors can similarly be identified by finding the eigenvalues of the stabilizer generators, and corrected by applying a single $Z$ operator to any qubit in the same column.  

In the following, we imagine a distance-$d$ Bacon-Shor code implemented in a tetron-like architecture (see Fig.~\ref{fig:tetron}) using $4 d^2$ MZMs. 
The $j$th qubit hosts four MZMs, $\gamma_{j,1},\,\gamma_{j,2},\, \gamma_{j,3},\,\gamma_{j,4}$, and we identify the corresponding Pauli operators as 
\begin{eqnarray}\label{eq:Pauli-M-mapping}
X_j \sim \gamma_{j,2}\gamma_{j,3}, ~~~ Y_j \sim \gamma_{j,1}\gamma_{j,3},  ~~~ Z_j \sim \gamma_{j,1}\gamma_{j,2}.
\end{eqnarray} 
(When analyzing noise model PMC, we will alter this definition to take into account the geometric arrangment of the MZMs on the island, see Fig.~\ref{fig:tetronBS}.  Specifically, we use the parity conservation of the ground state so that $X_j= \gamma_{j,1}\gamma_{j,4}$ or $\gamma_{j,2}\gamma_{j,3}$ depending on whether a tetron is in a joint measurement with its left or right neighbor; similarly $Z_j=\gamma_{j,1}\gamma_{j,2}$ or $\gamma_{j,3}\gamma_{j,4}$ depending on whether a tetron is in a joint measurement with its top or bottom neighbor.)
Using Eq.~\eqref{eq:Pauli-M-mapping}, we can map each Pauli operator to a MZM-parity measurement.  Furthermore, we assume that corrections are applied as {\it Pauli frame updates}: rather than  actually applying an operator to correct an error, we classically track measurement outcomes and appropriately reinterpret subsequent measurement outcomes.  As such, we can assume corrections are perfect, since faulty measurements are already taken into account. 

\begin{figure}[t]
\begin{center}
	\includegraphics[width=.8\columnwidth]{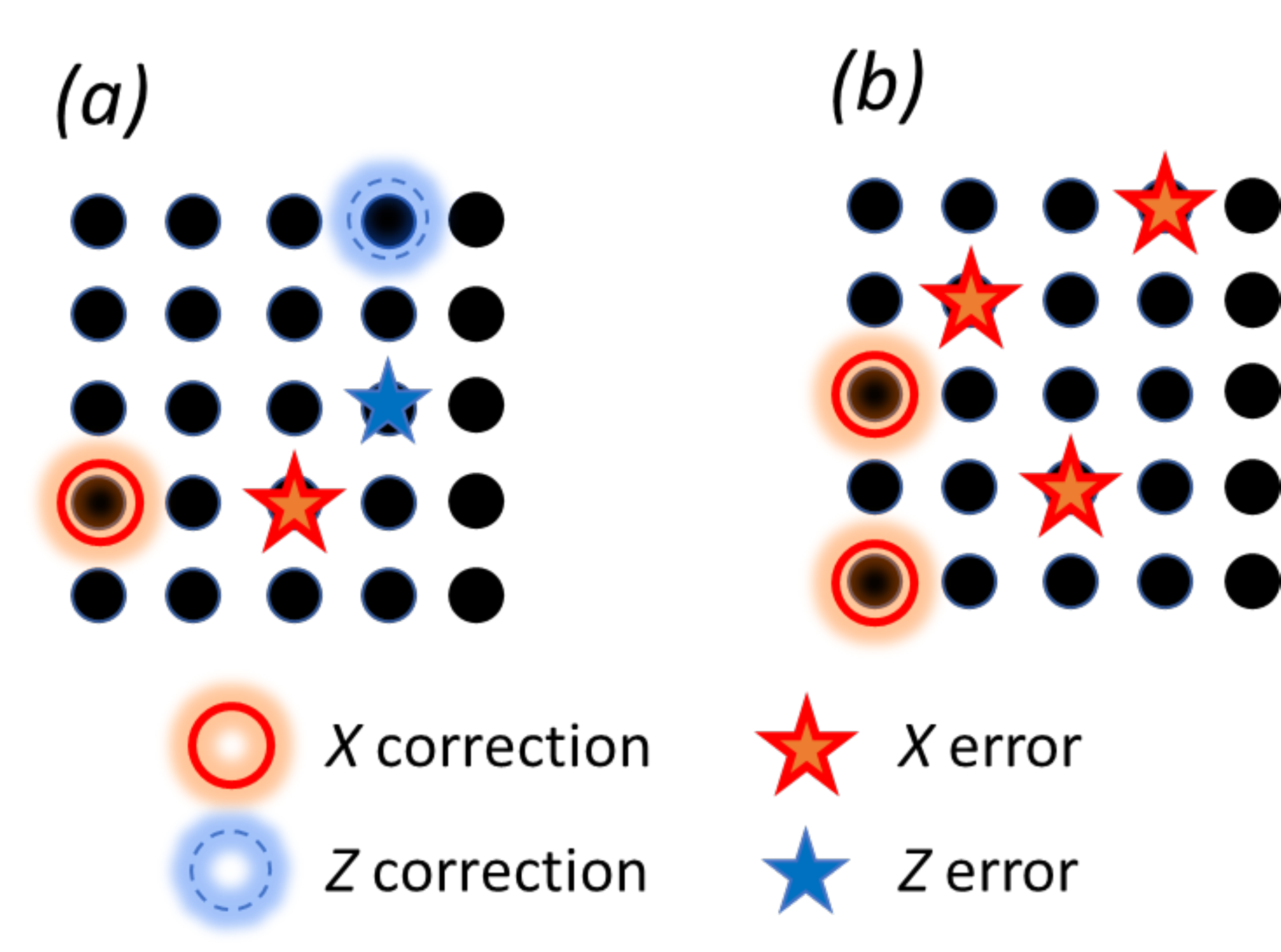}
	\caption{Two examples of Pauli errors (red and blue stars) and corresponding error correction (solid red and dashed blue circles).  All $X$ errors in the same row and all $Z$ errors in the same column have the same syndrome.  $(a)$ Example correction of a single $X$ and a single $Z$ error.  The net Pauli operator applied to the system (errors and correction) is in the gauge group and thus does not change the quantum information stored in the code.  $(b)$ Example failure of the code for three $X$ errors.  They syndrome of the three $X$ errors is the same as the syndrome for two $X$ errors in the third and fifth rows.  The correction thus applies $X$ operators to these rows, so that the net operator (error and correction) applied to the system is a dressed logical operator, which changes the quantum information.
	}
	\label{fig:decoding}
\end{center}
\end{figure}

We can choose to apply all $Z$ corrections (Pauli frame changes) to the qubit in the appropriate column in the top row and all $X$ corrections to the qubit in the appropriate row in the left column.  The stabilizer measurements are decoded by assuming that the minimal number of Pauli errors corresponding to a given syndrome have occurred (we will explain how to treat measurement bit-flips in the next section).  For the $d=5$ Bacon-Shor code, any two-qubit Pauli error can be corrected in this way~\cite{Bacon06}, see Fig.~\ref{fig:decoding}(a) for an example.  However, some three-qubit Pauli errors will be misdiagnosed by this decoding scheme, and can result in a logical operator being applied to the system after the correction step, see Fig.~\ref{fig:decoding}(b).  In this case, the quantum information stored in the code has changed, resulting in a failure of the protocol.

Note that the stabilizer measurements do not distinguish between the even and odd parity subspaces of the qubits: this can be seen with the mapping in Eq.~(\ref{eq:Pauli-M-mapping}) by noting that $\gamma_{j,4}$ does not change any of the stabilizer outcomes. Therefore, every error involving an odd number of Majorana operators has the same syndrome as some error involving an even number of Majorana operators.  Using the decoder described in the previous paragraph, all syndromes will be interpreted as corresponding to an error involving an even number of Majorana operators, thus the correction step (applying an $X$ operator to a qubit in the left column or a $Z$ operator to a qubit in the top row) will not return the system to the even parity subspace.  However, when the system relaxes back to the ground state in a later time step, it either does this through application of a Majorana operator that does not change the stabilizer measurement outcomes (in which case the environment has self-corrected), or the stabilizer measurements are altered and the error correction procedure now identifies the Pauli error resulting from the combined initial excitation-intermediate correction steps-relaxation processes. Note that when the excitation-relaxation process extends over more than one time step the outcome of subsequent stabilizer measurements might disagree. For example, an initial $\gamma_{j,2}$ excitation would be interpreted as a $Y_j$ error. If there is a subsequent relaxation to, say, $\gamma_{j,3}$ the combined process corresponds to an $X_j$ error. The above process can lead to misinterpretation of the error and can therefore be thought as contributing to measurement errors, which we discuss more in Section~\ref{sec:results}. We say that the protocol works if the correction step after the system has returned to the even parity subspace (possibly in a later time step) has applied a stabilizer operator.  Conversely, if the correction step after the qubit returns to the even parity subspace results in a logical error, the protocol fails. 

\subsection{Fault tolerant error correction}\label{sec:fault-tolerant}

A distance-$d$ error correcting code could be used to protect against up to $(d-1)/2$ errors with a single round of perfect measurements.  To account for the fact that some noise models can have imperfect measurements, a framework known as fault-tolerant error correction has been developed.  This framework distinguishes between {\it faults} and {\it errors}:
\begin{itemize}
\item A {\it fault} is any noise event that adversely disturbs the system or the measurements. 
\item An {\it error} is any non-trivial operator applied to a qubit.  
\end{itemize}
For a qubit-based model, a fault corresponds to any non-trivial noise event (including measurement bit-flips) and an error is some single-qubit Pauli operator (we will generalize the discussion to Majorana noise models at the end of this section and clarify what is meant by a noise event that ``adversely disturbs the system''). Faults are defined with respect to a given noise model; we illustrate this point by considering the qubit-based noise models Pauli noise with bit-flip measurement and Pauli circuit noise. For the former, a single fault is either a single-qubit Pauli operator or a measurement bit-flip.  For the latter, a single fault can be a single-qubit Pauli operator, a two-qubit Pauli operator, or a measurement bit-flip.  Notice that if a noise model includes correlated events, a single fault can result in multiple errors.

For a given qubit-based noise model, the requirement that the error correction procedure is {\it fault-tolerant} amounts to satisfying the following conditions~\cite{Gottesman09}:
\begin{enumerate}
	\item (EC A): For any initial state (irrespective of whether or not it is a code state), a single fault that can occur anywhere during the error correction procedure results in an output which differs from a code state by at most $(d-1)/2$ errors.
	\item (EC A'): For an initial error-free code state, a single fault that can occur anywhere during the error correction procedure outputs the error-free code state.
	\item (EC B): If the error correction proceeds without additional faults, any code state with $(d-1)/2$ errors will be corrected.
\end{enumerate}
If the noise is sufficiently weak, information stored in a fault-tolerant error correcting code is better-protected than information stored directly in a physical qubit.  The performance of a fault-tolerant error correction scheme can be quantified by the {\it pseudo-threshold}, defined for a noise model characterized by a single parameter $p$ by
\begin{equation}\label{eq:p-th}
p_\text{th}\equiv \max \{ p | p_\text{err}(p)\leq p\},
\end{equation}
where $p_\text{err}$ is the logical error rate, {\it i.e.}, the probability of an uncorrectable error remaining in the system after the application of strength-$p$ noise and error correction.  The pseudo-threshold will depend on the noise model, the error correcting code, and the error-correction protocol.  A pseudo-threshold is defined for a particular error correcting code on a fixed number of qubits, in contrast to a {\it threshold} which is defined for a family of error correcting codes, each for different system size ({\it i.e.}, a different number of physical qubits).  The threshold is the limit of the pseudo-threshold for infinitely large system size.  The Bacon-Shor code family does not have a finite threshold.

For a Majorana-based system, any non-trivial string of Majorana operators applied to a single island results in an error.  A fault is more subtle.  For instance, some noise events are a result of the system relaxing back to the even parity ground state.  This relaxation does not necessarily occur within the same time step as excitation, which is why our models include noise events involving an odd number of Majorana operators ({\it i.e.}, quasiparticle and odd correlated events).  When using a qubit-based code ({\it i.e.}, a code that only corrects for Pauli errors on the qubits) to error correct a Majorana-based system, it is essential for the environment to relax the system to the even parity ground state so that the net operator applied to any island involves an even number of Majorana operators (and therefore corresponds to some Pauli operator). This motivates a careful specification of what is considered ``adversely disturbing the system'' in the definition of a fault. If an island begins a time step in the odd parity state and a quasiparticle event relaxes the island back to the even parity state, this noise event is beneficial to the error correction protocol and is thus {\it not} considered a fault.  Conversely, if an island begins or ends a time step in the odd parity state, the lack of a relaxation {\it is} a fault.  For all the Majorana noise models presented in Section~\ref{sec:noise-models}, any non-trivial noise event applied during steps 1 or 2 correspond to a fault; a quasiparticle event  applied in step 0 is not a fault; and not applying a quasiparticle event in step 0 is a fault.  

With the above understanding of what constitutes a fault and what constitutes an error, fault tolerance conditions {(EC~A)}-{(EC~B)} apply to a Majorana-based system.  Note that ``a single fault'' in conditions {(EC~A)} and {(EC~A')} implies that if the fault corresponds to a quasiparticle or odd correlated event in a time step, the excitation is immediately relaxed in the subsequent time step.  Simiarly, ``without additional faults'' in condition {(EC~B)} implies that if the code state with $(d-1)/2$ errors is in an odd parity state, then the system is relaxed to the even parity state in the subsequent time step.

\subsection{Numerical results}\label{sec:results}

\begin{figure}[t]
	\includegraphics[width=1\linewidth]{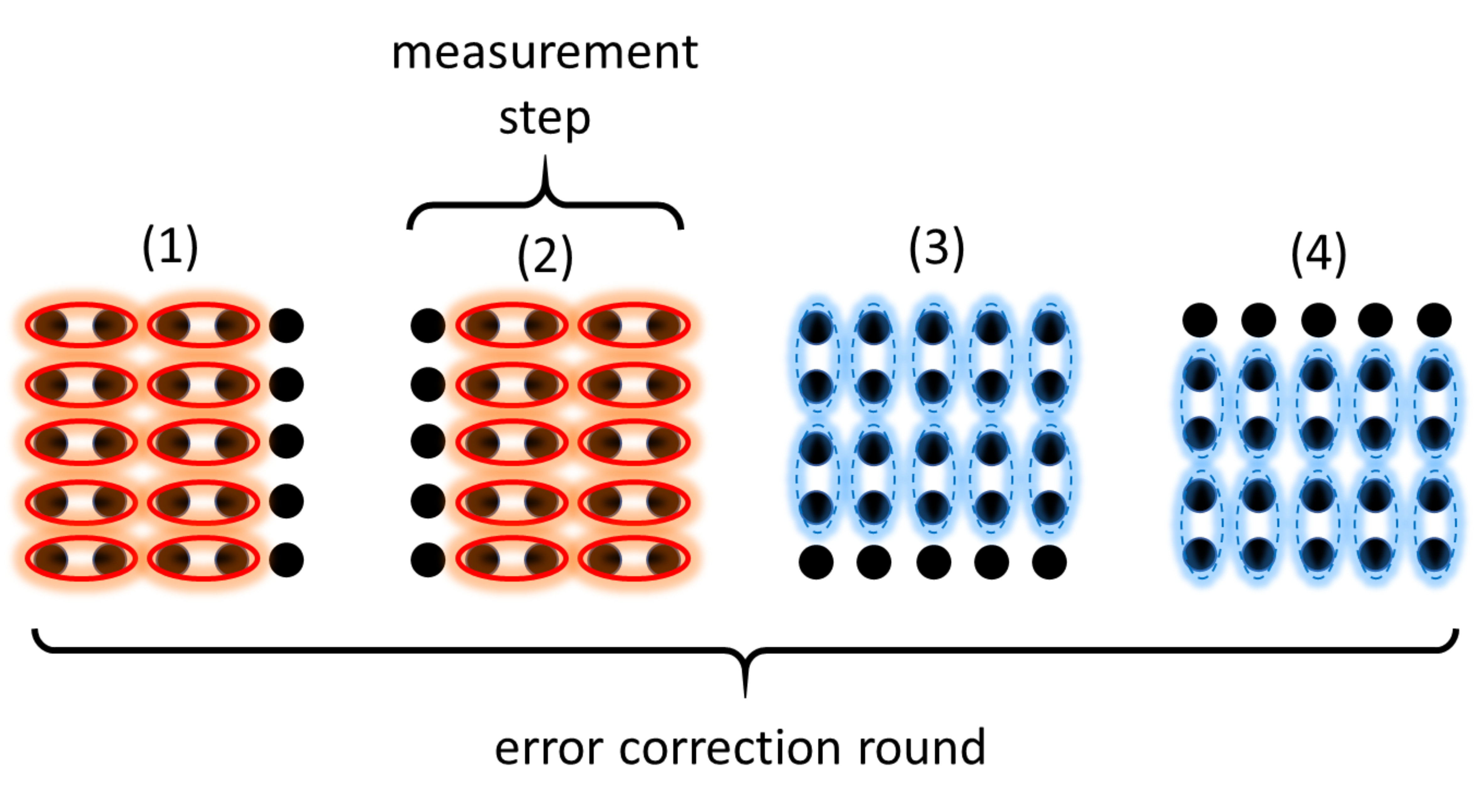}
	\caption{Measurement steps for a single error correction round of the $d=5$ Bacon-Shor code.  Same legend as Fig.~\ref{fig:BaconShor}, black dots correspond to qubits, red ovals denote $X$ measurements, blue dashed ovals denote $Z$ measurements.  Each step involves the measurement of two stabilizers built out of five gauge generator measurements.  Four measurement steps are necessary (for arbitrary $d$) for a single error correction round.  As discussed in the main text, faulty measurements can be corrected by repeating each error correction round at least four times.  One might, in principle, be able to do multiple measurements on a single qubit and thereby reduce the number of measurement steps required; however, this is inadvisable as it allows correlated errors to spread over longer distances. }\label{fig:measurement-steps}
\end{figure}

In this section, we analyze fault-tolerant error correction circuits for storing a qubit using the $d=5$ Bacon-Shor code under the Majorana noise models presented in Section~\ref{sec:noise-models}.  We define the parameters of each noise model according to Table~\ref{fig:prob-summary}, so that we can easily scale the probability of a logical error $p_\text{err}$ as a function of the probability $p_0$ of an excitation of an idle island.  
 The analogous qubit noise models are found by setting the probability of all errors involving an odd number of Majorana operators to zero (equivalently, by setting the relaxation parameter $r$ from Section~\ref{sec:prob-summary} to zero).  

An exhaustive simulation of all possible errors to find the exact pseudo-threshold $p_\text{th}^{\text{A}}$ of a noise model $A$ quickly becomes intractable. We therefore follow a Monte Carlo approach of simulating $n_\text{trials}$ runs for a single error correction round to estimate $p_\text{err}(p_0)$ from the fraction of trials which result in a logical error we extract
\begin{equation}
p_\text{err}(p_0)=\frac{n_\text{fail}}{n_\text{trials}} .
\end{equation}
The uncertainty of $p_\text{err}(p_0)$ for a given $p_0$ is of order $1/\sqrt{n_\text{trials}}$.  

The Majorana noise models Qp, QpBf, and MC are designed without respect to a specific physical measurement protocol.  Thus, when simulating pseudo-thresholds for these three models, we define measurements according to Eq.~\eqref{eq:Pauli-M-mapping}, {\it i.e.}, without considering the geometric arrangement of MZMs within a tetron.  In contrast, model PMC is designed with the measurement protocol of Section~\ref{sec:physical} in mind; as such when analyzing this model we redefine the measurements to account for a convenient physical implementation of the Bacon-Shor code.  

There are three main differences between our simulations of models MC and PMC and the most realistic approach to estimating optimal pseudo-threshold values.  (The simple models Qp and QpBf will clearly not estimate realistic pseudo-threshold values as they do not include all lowest order noise events in a Majorana-based system.)
Firstly, to help the simulations run faster, we do many trial runs of a single error correction round, rather than simulating many rounds and analyzing the frequency of logical errors being applied.\footnote{This approach allows us to speed up the run time by using importance sampling for the first error.}  The latter approach is a more accurate description of noise in an actual system because it feeds the errors from a previous error correction round into the next.  However, feeding errors into the next time step is only important when the system has the unlucky combination of an error going undetected ({\it e.g.}, from occuring in one of the later time steps of an error correction round and thus being identified as a measurement bit-flip) followed by enough additional errors occuring in the next error correction round to result in a logical error.  When the noise event rate is sufficiently small, the difference between simulating many error correction rounds and repeatedly simulating a single round will only result in a small correction to the pseudo-threshold estimates.  
Details of the code used for simulating the pseudo-thresholds are in Appendix~\ref{app:code}.  

Secondly, to correct for faulty measurements, we use the Shor error correction approach~\cite{Shor96} of repeating stabilizer measurements multiple times in a single error correction round.  Generally, Steane error correction~\cite{Steane97}, which uses entangling gates ({\it e.g.}, CNOTs) of the logical data qubit with logical ancilla qubits to locate errors, results in higher threshold estimates (in part because the effort of error correction is shifted from operations on the data qubits to preparation of ancillas~\cite{Gottesman09}).  
Steane error correction is only known to apply to stabilizer codes, which means that in order to do Steane error correction on the Bacon-Shor code we must fix the gauge subsystem and thereby lose the ability to build stabilizer measurements out of two-qubit measurements.  Furthermore, the logical ancillas (tripling the number of tetrons) and multiple CNOT gates (each using an additional ancilla tetron and multiple measurements~\cite{Karzig17}) needed for Steane error correction quickly complicate the pseudo-threshold simulations.

Finally, we include correlated events, which reduces the number of faults that can be corrected for models MC and PMC because two faults can result in a logical error ({\it e.g.}, a correlated event plus a pair-wise dephasing event can implement Pauli errors on three islands). We could have avoided correlated events reducing the code distance by using measurement gadgets~\cite{Gottesman09}, which use CNOT gates and ancilla tetrons to ensure that any correlated event only affects one data tetron.  However, measurement gadgets introduce more opportunities for a fault to occur within an error correction round due to both the introduction of ancilla tetrons and because CNOT gates require three measurements for a tetron architecture~\cite{Karzig17}. It is therefore not clear whether measurement gadgets would improve the error correction for tetron-based architectures. 

Given the above considerations, we believe the pseudo-threshold estimates for models MC and PMC in this section are below their optimal values and should not be taken as experimental targets. Our analysis should rather be taken as indication of the relative importance of the different noise events (quasiparticle, measurement, correlated, and pair-wise dephasing) on the pseudo-threshold.  In Section~\ref{sec:conclusions}, we identify future directions of study that could improve simulations of the noise models presented in this paper and result in more accurate pseudo-threshold estimates. We explicitly discuss experimental implications of our analysis at the end of this section.

\subsubsection{Quasiparticle noise with perfect measurement (Qp)}

\begin{figure}[t!]
	\includegraphics[width=\linewidth]{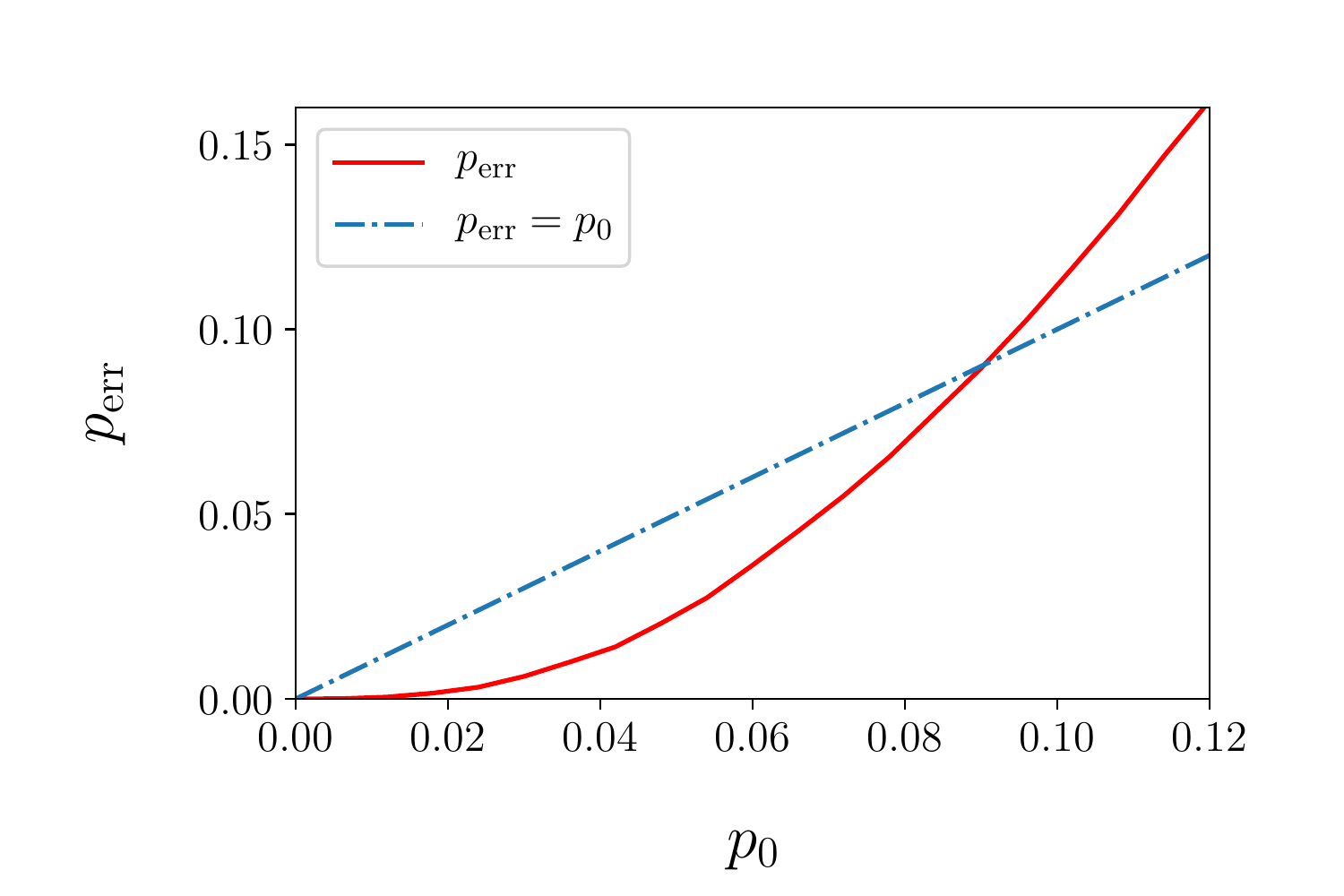}
	\caption{Logical error rate $p_\text{err}$ (solid red curve) as a function of the noise event rate $p_0$ for the Qp noise model (see Table~\ref{fig:prob-summary}). The dot-dashed blue line indicates no error correction. The pseudo-threshold is the intersection of the two curves:  ${p_\text{th}^{\text{Qp}}\approx 0.09}$. This plot was generated for $10^5$ Monte Carlo trials per data point, see Appendix~\ref{app:code} for simulation details. }\label{fig:Qp}
\end{figure}

Model Qp is the simplest stochastic Majorana noise model given in Section~\ref{sec:noise-models}, in which only pair-wise dephasing events or quasiparticle events can occur.  We define measurements according to Eq.~(\ref{eq:Pauli-M-mapping}).  The relative distribution of quasiparticle events and pair-wise dephasing events can be adjusted by varying the relaxation parameter $r$, see Table~\ref{fig:prob-summary}.   In the limit $r=0$, Qp reduces to the qubit model Pauli noise (or code capacity noise).

The standard procedure for calculating pseudo-thresholds for Pauli noise is to consider a noisy time step, followed by perfect application of all stabilizer measurements ({\it i.e.}, errors cannot occur between the four different measurement steps displayed in Fig.~\ref{fig:measurement-steps}).  We implement the same procedure for Qp, however, since the syndrome of a quasiparticle event after a single time step is identical to that of some pair-wise dephasing event, a single time step of Qp is not able to distinguish quasiparticle events from pair-wise dephasing events.  The $r$-dependence of the model will therefore only become apparent after simulating multiple time steps, for instance by considering several error correction rounds or by modeling application of a logical gate (not included in this paper).  
We do not simulate multiple time steps for Qp, as the simulations of QpBf and MC clearly indicate that the pseudo-threshold only has weak $r$ dependence.  The pseudo-threshold estimate under Qp (and Pauli noise) for a single time step is $p_\text{th}^{\text{Qp}}\approx 0.09 $ (see Fig~\ref{fig:Qp}).


\subsubsection{Quasiparticle noise with bit-flip measurement (QpBf)}

\begin{figure}[t!]
	\begin{minipage}{\columnwidth}
		\includegraphics[width=\linewidth]{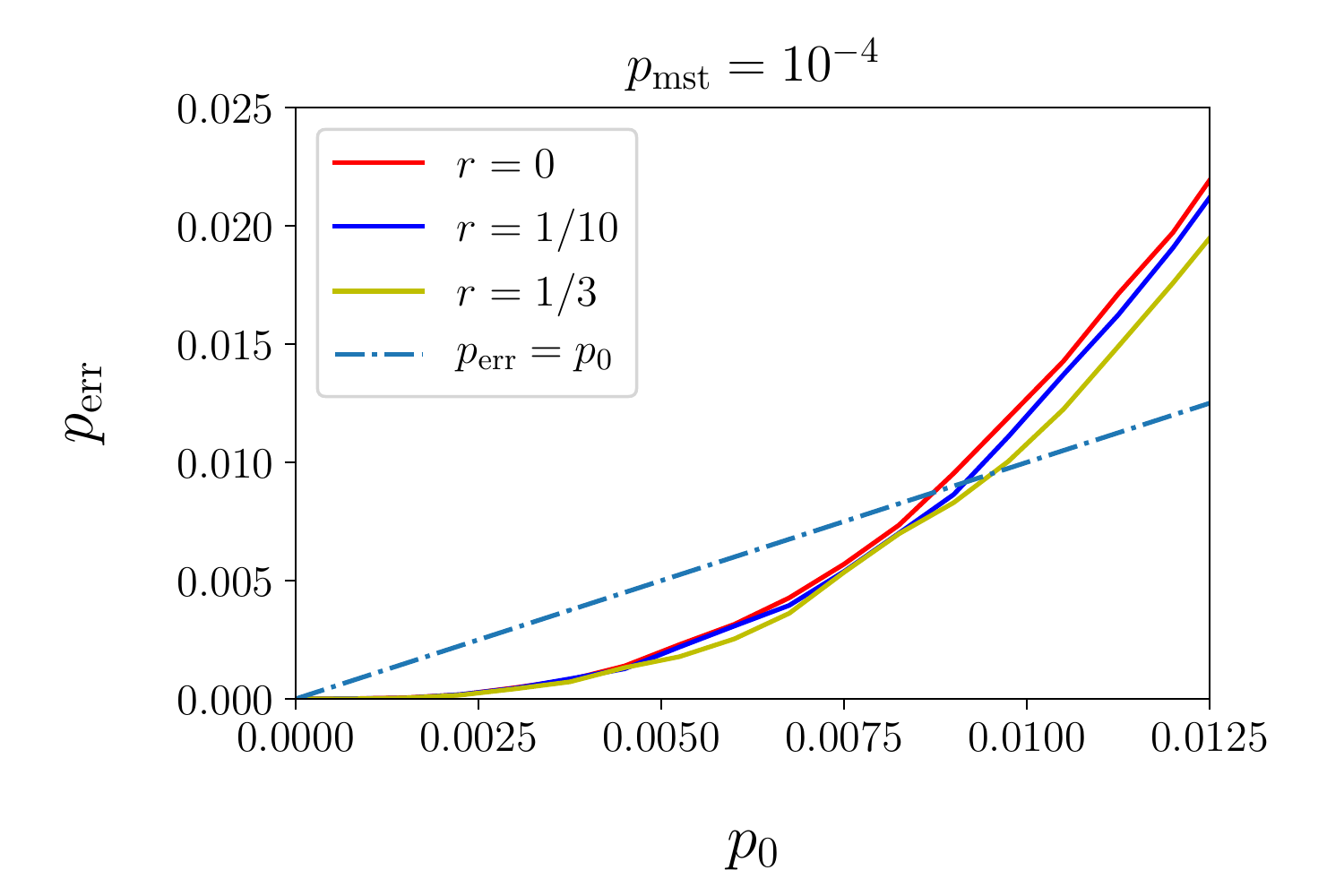}
	\end{minipage}
	\begin{minipage}{\columnwidth}
		\includegraphics[width=\linewidth]{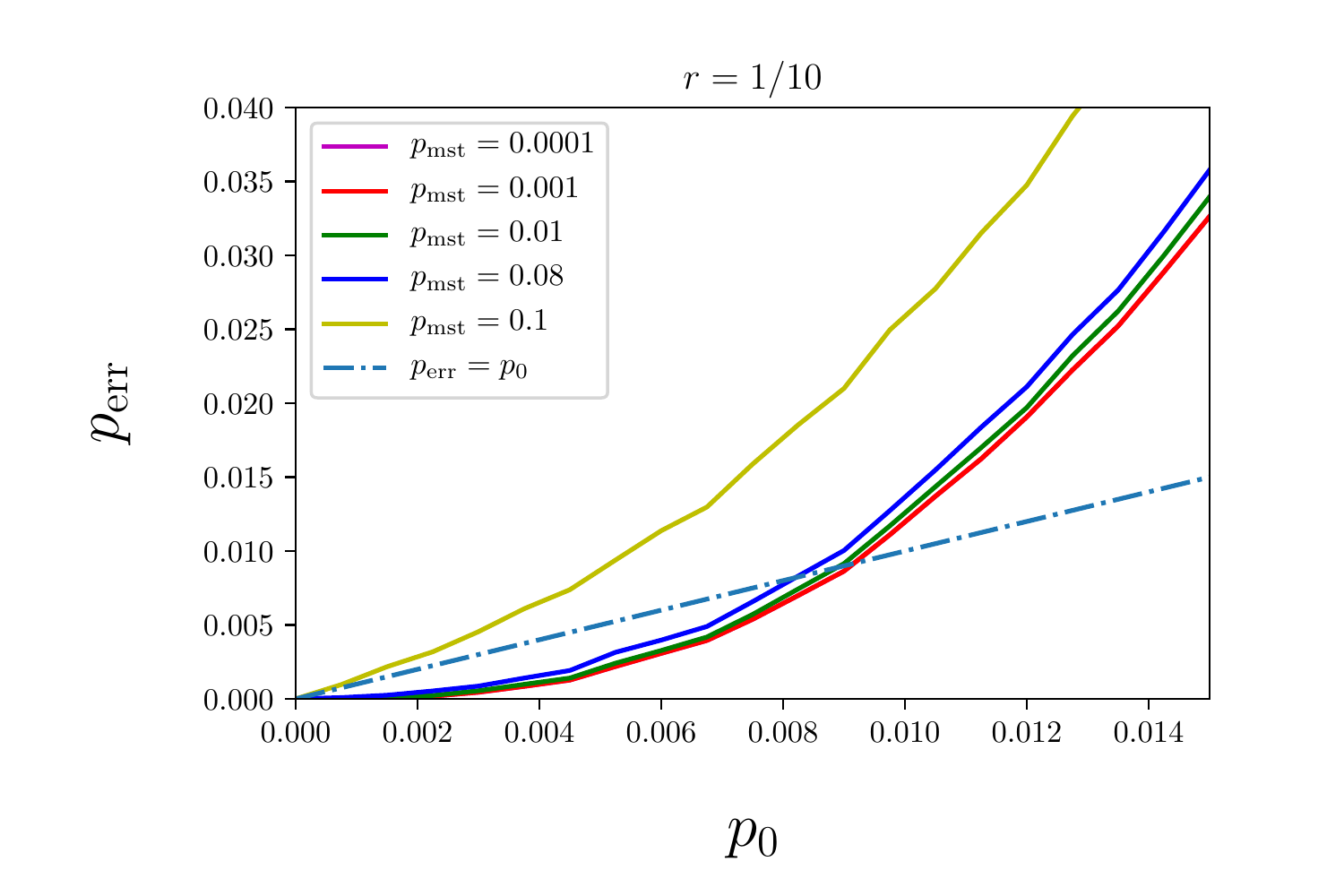}
	\end{minipage}
	\caption{Pseudo-threshold simulation results for the QpBf noise model varying $r$ ({\it top}) and $p_\text{mst}$ ({\it bottom}) (see Table~\ref{fig:prob-summary}).  Solid curves are the logical error rate as a function of the noise event rate $p_0$, dot-dashed blue lines indicate no error correction, and the pseudo-threshold for each choice of parameters $r$ and $p_\text{mst}$ is their intersection.  These plots were generated for $1.5\times 10^5$ Monte Carlo trials per data point, see Appendix~\ref{app:code} for simulation details.  See text for discussion of results. }\label{fig:QpBf}
\end{figure}

The model QpBf builds on the previous model Qp by allowing the classical bit storing the measurement outcome to be flipped.  We define the error probabilities in Table~\ref{fig:prob-summary} and the measurements according to Eq.~(\ref{eq:Pauli-M-mapping}).  For this noise model, we repeat the four measurement steps comprising a single round of error correction in order to avoid introducing errors into the code state from a single measurement bit-flip ({\it i.e.}, we use a Shor error correction scheme).  Our procedure is to consider a single (noisy) time step preceding each error correction round, repeated four times.  We again apply all stabilizer measurements simultaneously, as in the Qp simulation. If the repetition rounds result in different syndromes, we accept the last syndrome which repeats for two consecutive rounds.  When multiple errors occur, it is possible that no two consecutive rounds will have the same syndrome.  In this case, we assume the syndrome for the fourth round.

The pseudo-threshold dependence on $r$ and $p_\text{mst}$ is shown in Fig.~\ref{fig:QpBf}.  
In the top panel, we see weak dependence on $r$, with $p_\text{th}^\text{QpBf}\approx 8\times 10^{-3}$  for $0\leq r \leq 1/3$ and $p_\text{mst}=10^{-4}$.  It follows that under this noise model, the $d=5$ Bacon-Shor code is insensitive to the relative distribution of quasiparticle and pair-wise dephasing events.  Intuitively, we can understand this result as follows: in the parameter regime considered, the environment relaxes each island with high probability, thereby converting quasiparticle events in one time step to pair-wise dephasing events in the subsequent time step.  A pair-wise dephasing event in one time step does not have a significantly different effect than a pair-wise dephasing event broken into two time steps.  As noted in Section~\ref{sec:prob-summary}, larger $r$ is not captured by model QpBf and would be highly problematic (see Section~\ref{sec:extensions} and Appendix~\ref{app:long-lived}),  requiring a Majorana fermion code to efficiently correct a single excitation~\cite{Bravyi10,Hastings17,Vijay2017}.

\begin{figure*}[t!]
\begin{center}
	\begin{minipage}{\columnwidth}
		\centering
		\includegraphics[width=\columnwidth]{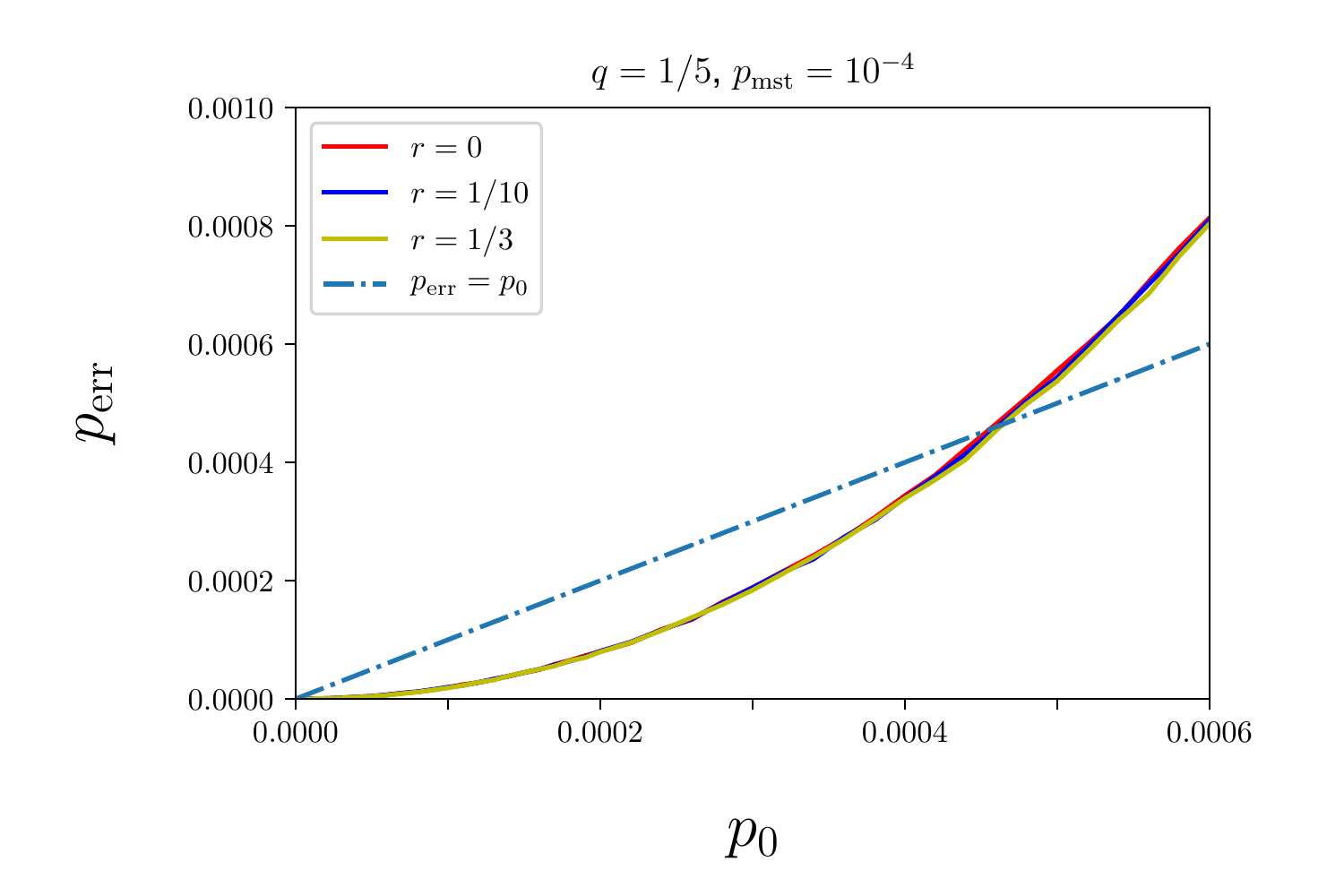}
	\end{minipage}
	\begin{minipage}{\columnwidth}
		\centering
		\includegraphics[width=\columnwidth]{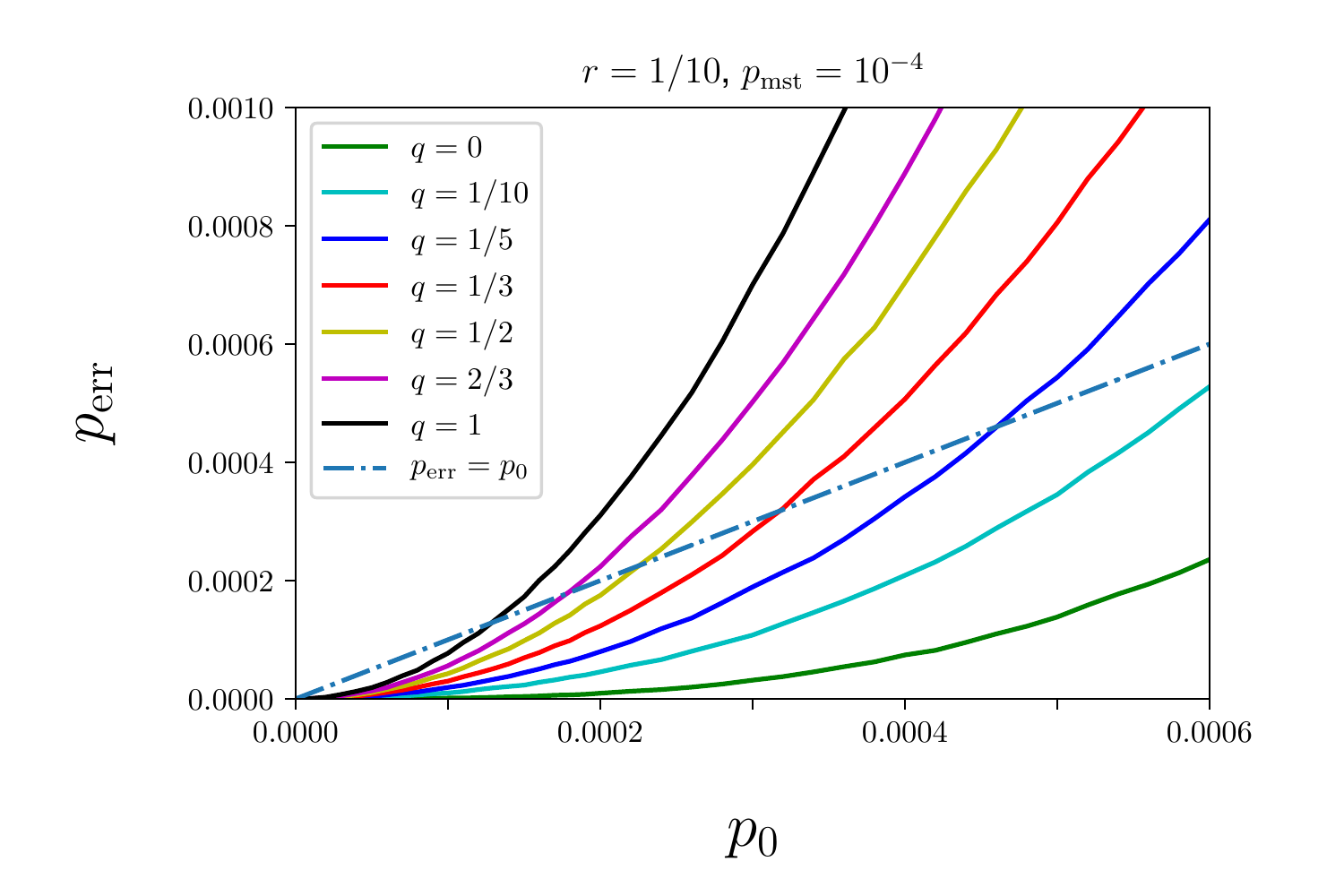}
	\end{minipage}
	\begin{minipage}{\columnwidth}
		\centering
		\includegraphics[width=\columnwidth]{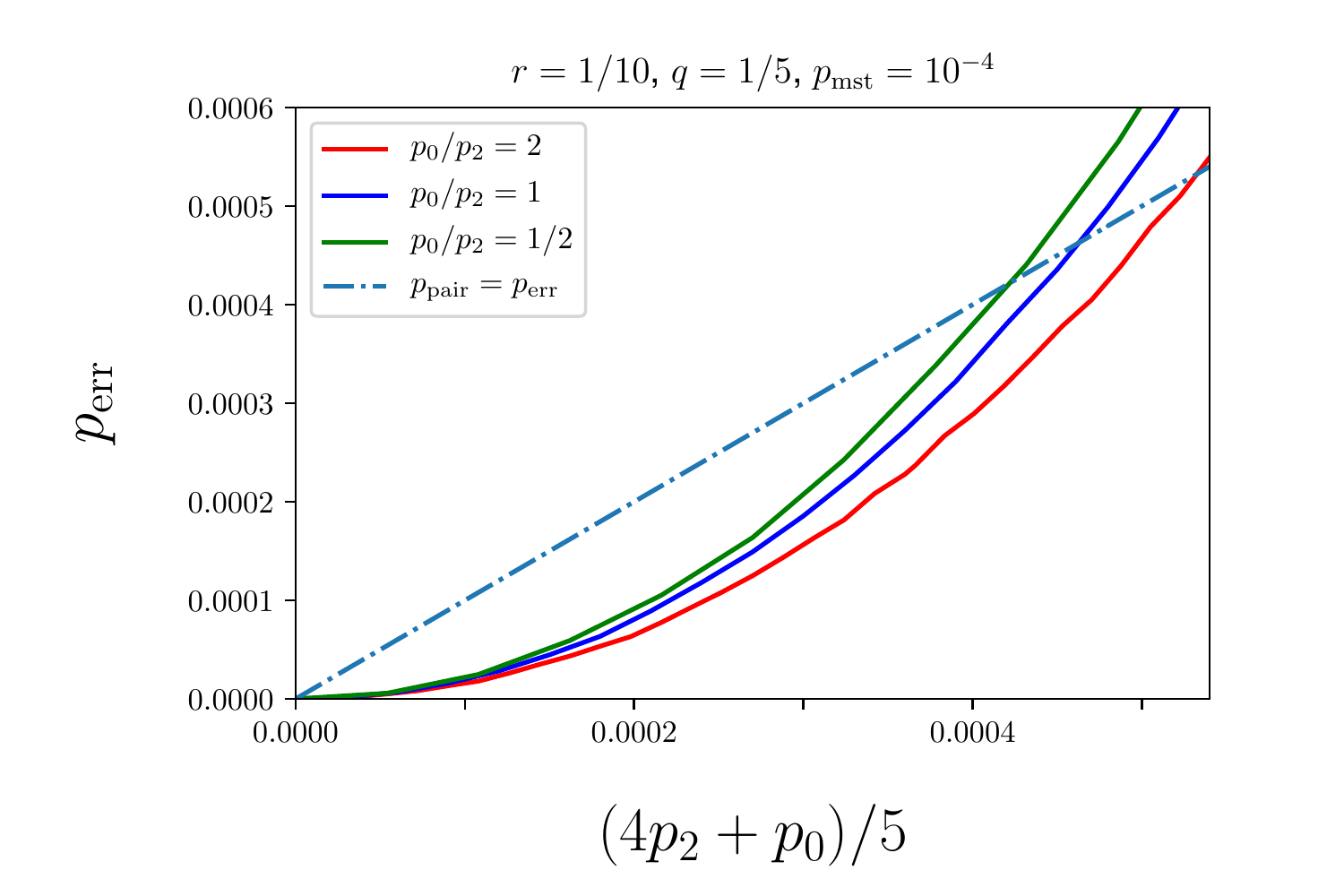}
	\end{minipage}
	\begin{minipage}{\columnwidth}
		\centering
		\includegraphics[width=\columnwidth]{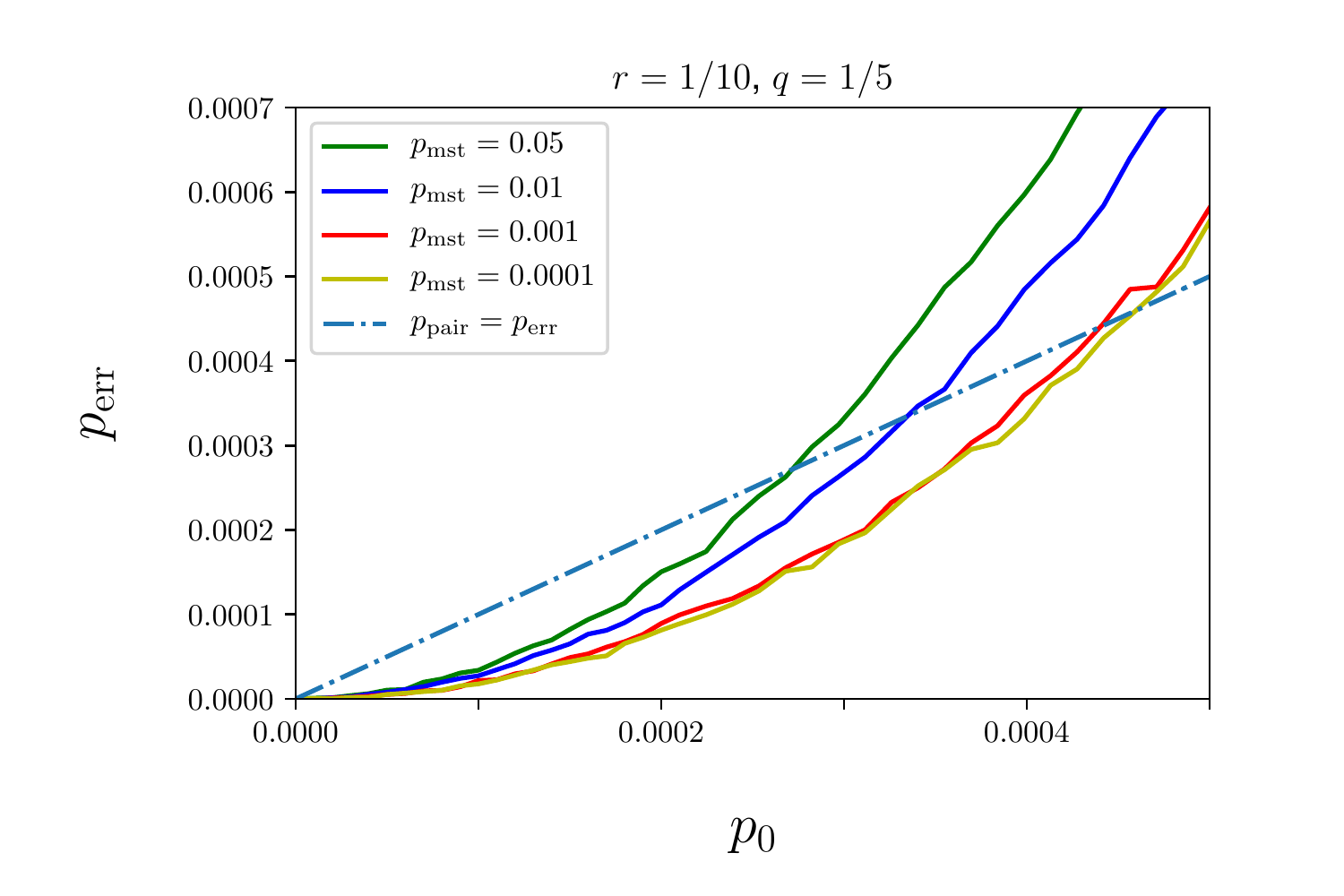}
	\end{minipage}
	\caption{Pseudo-threshold simulation results for the MC noise model varying $r$ ({\it top left}), $q$ ({\it top right}), $p_2$ ({\it bottom left}), and $p_\text{mst}$ ({\it bottom right}) (see Table~\ref{fig:prob-summary}).  Solid curves are the logical error rate as a function of the noise event rate per qubit (averaged over the 20 measured qubits and 5 idle qubits in the system), dot-dashed lines indicate no error correction, and their intersections denote the pseudo-threshold for those parameter choices. In the top right panel, the pseudo-threshold for $q=1/10$ is $\approx 6.5\times 10^{-4}$, and for $q=0$ is $\approx 9.8\times 10^{-4}$.  See text for discussion of the results.  The first three plots were generated for $10^8$ effective Monte Carlo trials per data point and the bottom right plot was generated for $10^7$ effective Monte Carlo trials per data point, see Appendix~\ref{app:code} for simulation details.
	}\label{fig:MC}
\end{center}
\end{figure*} 

The bottom panel of Fig.~\ref{fig:QpBf} shows the pseudo-threshold dependence on $p_\text{mst}$ for $r=1/10$.  The logical error rate, and therefore the pseudo-threshold, have much weaker dependence on $p_\text{mst}$ than on $p_0$.  Intuitively, this insensitivity to measurement bit-flip can be understood as a direct consequence of the repetition of stabilizer measurements: a measurement bit-flip must be repeated in order to show up in the accepted syndrome, whereas a quasiparticle or pair-wise dephasing event occurring in the first time step will affect all subsequent syndromes.  We expect $p_\text{mst}$ to contribute significantly to $p_\text{err}$ when ${p_\text{mst}\gtrsim \mathcal{O}\left(\sqrt{p_0}\right)}$\footnote{One example scenario in which measurement bit-flips contribute to a failure of the error correction protocol is if two single qubit errors occur in the first time step, {\it e.g.}, $x$ errors on qubits in different columns, and a measurement bit-flip occurs for a qubit in a different column in {\it both} the second and third time step (or both the third and fourth time steps).  The same logical error could instead occur if three $x$ errors affecting qubits in different columns occur in the first, second, or third time steps.  The two scenarios are similarly probable when $p_\text{mst}^2\sim \mathcal{O}\left(p_0\right)$. }.  This estimate is consistent with the approximately constant value ${p_\text{th}^\text{QpBf}\approx 8\times 10^{-3}}$ for $p_\text{mst}=10^{-4}-10^{-2}\ll \sqrt{p_\text{th}^\text{QpBf}}$, and the sharp pseudo-threshold decrease when ${p_\text{mst}=0.08\approx \sqrt{p_\text{th}^\text{QpBf}}}$.  The insensitivity to $p_\text{mst}$ is independent of whether or not there are MZMs in the system and should apply generally to any error correction scheme that relies on repetition to correct for measurement bit-flips.  A similar result was found in the context of the surface code built from qubits subjected to Pauli noise with bit-flip measurement in Ref.~\onlinecite{Harrington04}: for small Pauli error probability ($p_0<0.01$), the probability of a stabilizer measurement bit-flip could approach 50\% without strongly affecting the logical error rate.  In comparison, the probability of a stabilizer measurement bit-flip here is $1-\left(1-p_\text{mst}\right)^{10}$, which equals 50\% for $p_\text{mst}\approx 0.07$.  

Finally, we note that in the regime of weak $p_\text{mst}$ dependence, the pseudo-threshold estimate for QpBf is approximately consistent with the simulation of Qp if the logical error rate is plotted as a function of $p_0/4$.  If we ignore measurement bit-flips, the simulation for QpBf mainly differs from Qp in that there are four time steps before applying the decoder, thus the QpBf simulation roughly quadruples the probability of a noise event compared to the Qp simulation\footnote{This argument is not exact as only errors occuring during the first three time steps will be actively corrected by the error correction protocol.}.

\subsubsection{Majorana circuit noise (MC)}

The error probabilities for MC are given in Table~\ref{fig:prob-summary} (here we write $q_2$ as $q$).  Because we are restricting our analysis to the operations necessary for error correction of quantum memory, during any given time step every island is either idle or involved in a four-MZM measurement with another island.  Measurements are still defined according to Eq.~(\ref{eq:Pauli-M-mapping}).  It is now important to keep track of which islands are involved in a given measurement, therefore we can no longer assume that all stabilizers are applied simultaneously.  Rather, each set of stabilizer measurements  (see Fig.~\ref{fig:measurement-steps}) is preceded by a noisy time step, so that a single error correction round is comprised of four time steps and four measurement steps.  As for QpBf, faults in the measurement process require repetition and we use the same protocol as before: we accept the last repeated syndrome, or if no syndrome repeats, we assume the syndrome for the last round.  We show in Appendix~\ref{app:EC} that this repetition scheme satisfies the conditions for fault tolerance.  

The simulation results are plotted in Fig.~\ref{fig:MC}.  The top left panel shows the dependence on $r$ for $q=1/5$ (single-qubit errors four times more likely than two-qubit errors), $p_0=p_2$, and $p_\text{mst}=10^{-4}$.  The pseudo-threshold dependence on $r$ is even weaker than was the case for QpBf, and is within noise of the simulation for realistic choices of $r$ (including unphysical values of $r=1/2$ and $1$ does reveal a weak $r$-dependence).   This indicates that the code is insensitive to a redistribution of errors involving an even or odd number of Majorana operators.  We stress that the insensitivity of the pseudo-threshold to changes in $r$ is highly dependent on the assumption $e^{-1/r_k}< p_k$ (see Section~\ref{sec:prob-summary}) that went into the construction of the circuit noise models. If this were not the case, a single excitation in one island could spread throughout the system through correlated events occurring with probability $q\sim \mathcal{O}(1)$.  We comment further on architectures where this inequality is not satisfied in Section~\ref{sec:extensions} and Appendix~\ref{app:long-lived}. 

The pseudo-threshold's dependence on $q$ is shown in the top right panel of Fig.~\ref{fig:MC} for $r=1/10$, $p_0=p_2$, and $p_\text{mst}=10^{-4}$.  As is expected from similar studies using qubit noise models, we see that a higher weight of correlated events significantly lowers the pseudo-threshold~\cite{Tomita14}.   When all initial excitations in an island lead to correlated events, {\it i.e.}, $q=1$, the pseudo-threshold $p_\text{th}^\text{MC}\approx 1.2\times 10^{-4}$.  When half of the errors on a measured island are from correlated event processes, {\it i.e.}, $q=1/2$, the pseudo-threshold increases to the ${p_\text{th}^\text{MC}\approx 4.5\times 10^{-4}}$.  Finally, in the limit of no correlated events, {\it i.e.}, $q=0$, the pseudo-threshold is $p_\text{th}^\text{MC}\approx 9.8\times 10^{-4}$.  This last limit  agrees with the simulation results for Qp and QpBf by scaling the $x$-axes of Figs.~\ref{fig:Qp} and \ref{fig:QpBf} by factors of $1/16$ and $1/4$, respectively, to account for the factor of four increase in the number of time steps that can contribute to a logical error compared to the QpBf simulation.

The bottom left panel of Fig.~\ref{fig:MC} plots the pseudo-threshold for $p_2=p_0/2,\,p_0,\,$and $2p_0$ with $r=1/10$, $q=1/5$, and $p_\text{mst}=10^{-4}$.  In order to keep the total probability of an error fixed, the $x$-axis is now given by $\left(p_0+4p_2\right)/5$, the average rate of creating an excitation in the $d=5$ system.  The pseudo-threshold is lower for $p_2>p_0$, which is consistent with our expectation that a higher percentage of correlated events should decrease the pseudo-threshold.  This dependence is not especially strong, which can be understood as resulting from the small ratio (1/5) of islands that are idle in any given time step. 

Finally, we show the pseudo-threshold dependence on $p_\text{mst}$ in the bottom right panel of Fig.~\ref{fig:MC} for $r=1/10$, $q=1/5$, and $p_0=p_2$.  The curves for $p_\text{mst}=10^{-4}$ and ${p_\text{mst}=10^{-3}}$ are essentially within noise of each other, and only for ${p_\text{mst}\gtrsim 10^{-2}\sim\mathcal{O}\left( \sqrt{p_0}\right)}$ does the pseudo-threshold depend noticeably on $p_\text{mst}$.  The intuition is the same as for model QpBf: measurement bit-flips must be repeated over successive stabilizer measurements in order to affect the accepted syndrome and influence the correction procedure.

\begin{figure}[t!]
	\includegraphics[width=\columnwidth]{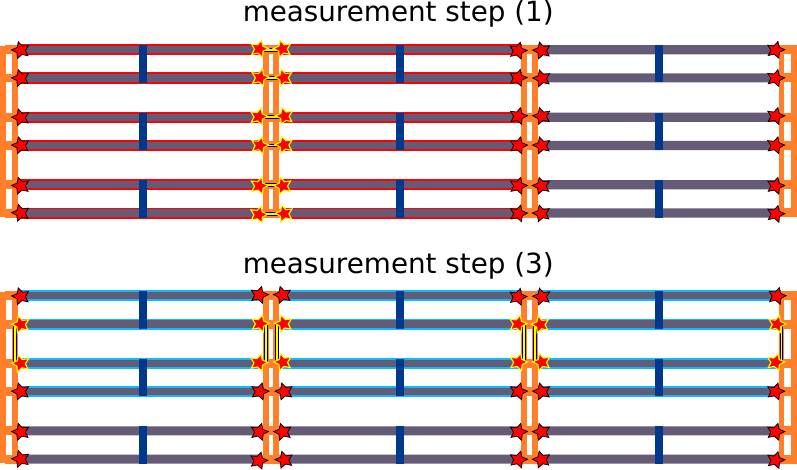}
	\caption{Measurement steps (1) ({\it top panel}) and (3) ({\it bottom panel}), see Fig.~\ref{fig:measurement-steps}, for the $d=3$ Bacon-Shor code implemented for a system of tetrons.  Measured MZMs are highlighted in yellow.  Tetrons involved in $XX$ measurements between horizontal nearest neighbors are highlighted in red, tetrons involved in $ZZ$ measurements between vertical nearest neighbors are highlighted in blue.  The PMC simulation takes into account the geometrical arrangement of MZMs on a tetron, so that different MZMs on the left and right islands are measured in an $XX$ measurement, and different MZMs on the top and bottom islands are measured in a $ZZ$ measurement, see Eqs.~\eqref{eq:PMC-XX} and \eqref{eq:PMC-ZZ}.
	}
	\label{fig:tetronBS}
\end{figure}

\subsubsection{Physical Majorana circuit noise (PMC)}

\begin{figure*}[t!]
\begin{center}
	\begin{minipage}{\columnwidth}
		\centering
		\includegraphics[width=\columnwidth]{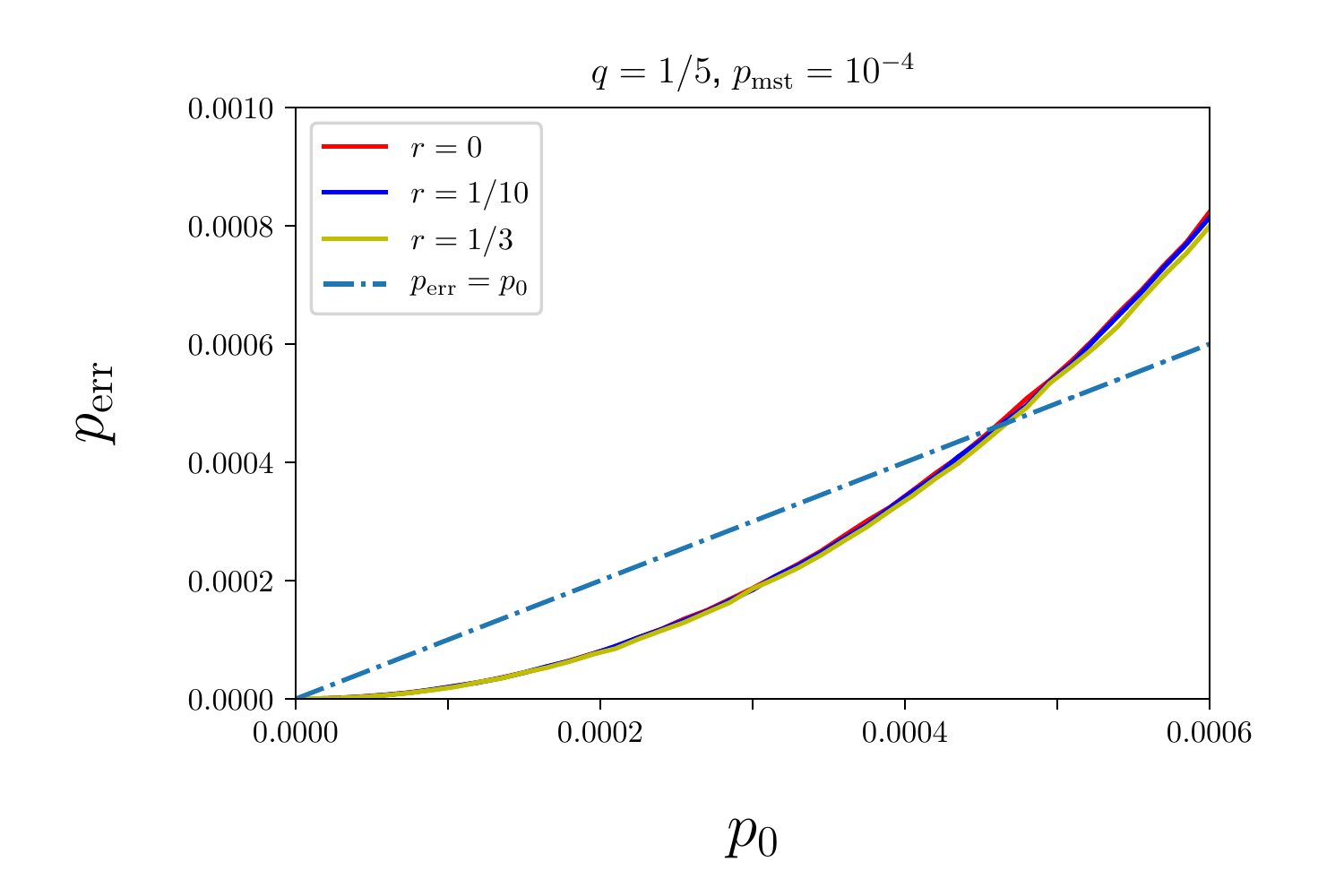}
	\end{minipage}
	\begin{minipage}{\columnwidth}
		\centering
		\includegraphics[width=\columnwidth]{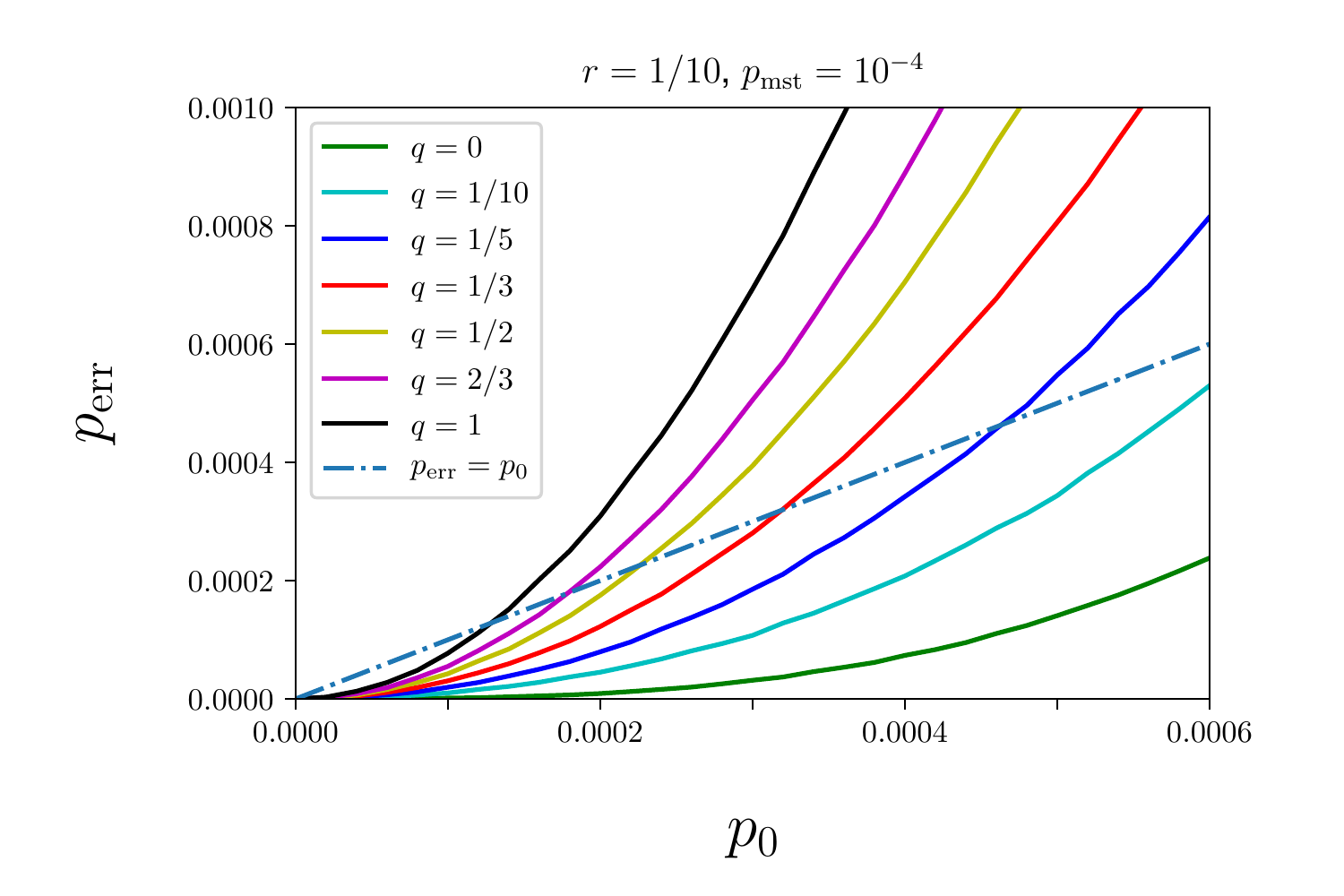}
	\end{minipage}
	\begin{minipage}{\columnwidth}
		\centering
		\includegraphics[width=\columnwidth]{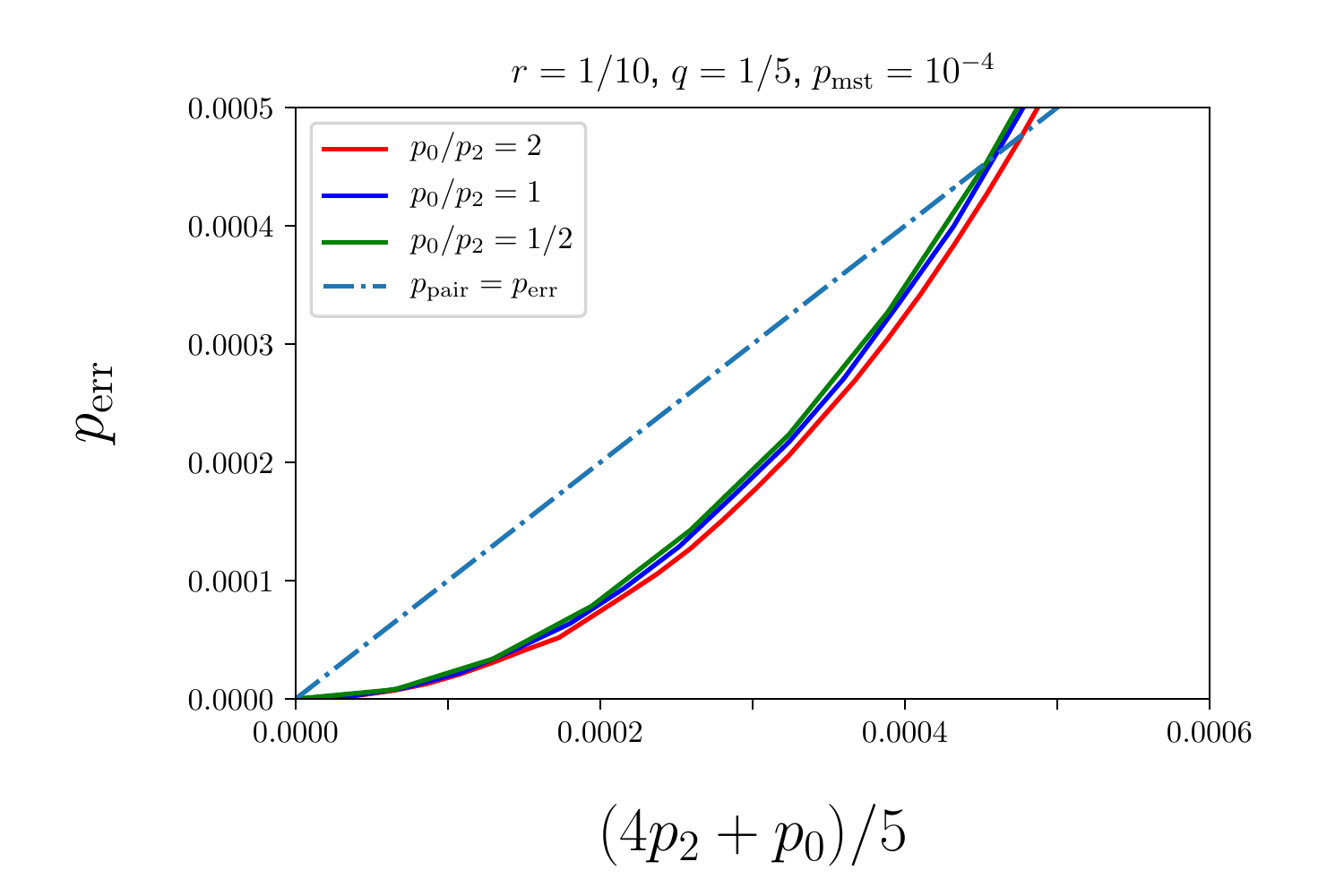}
	\end{minipage}
	\begin{minipage}{\columnwidth}
		\centering
		\includegraphics[width=\columnwidth]{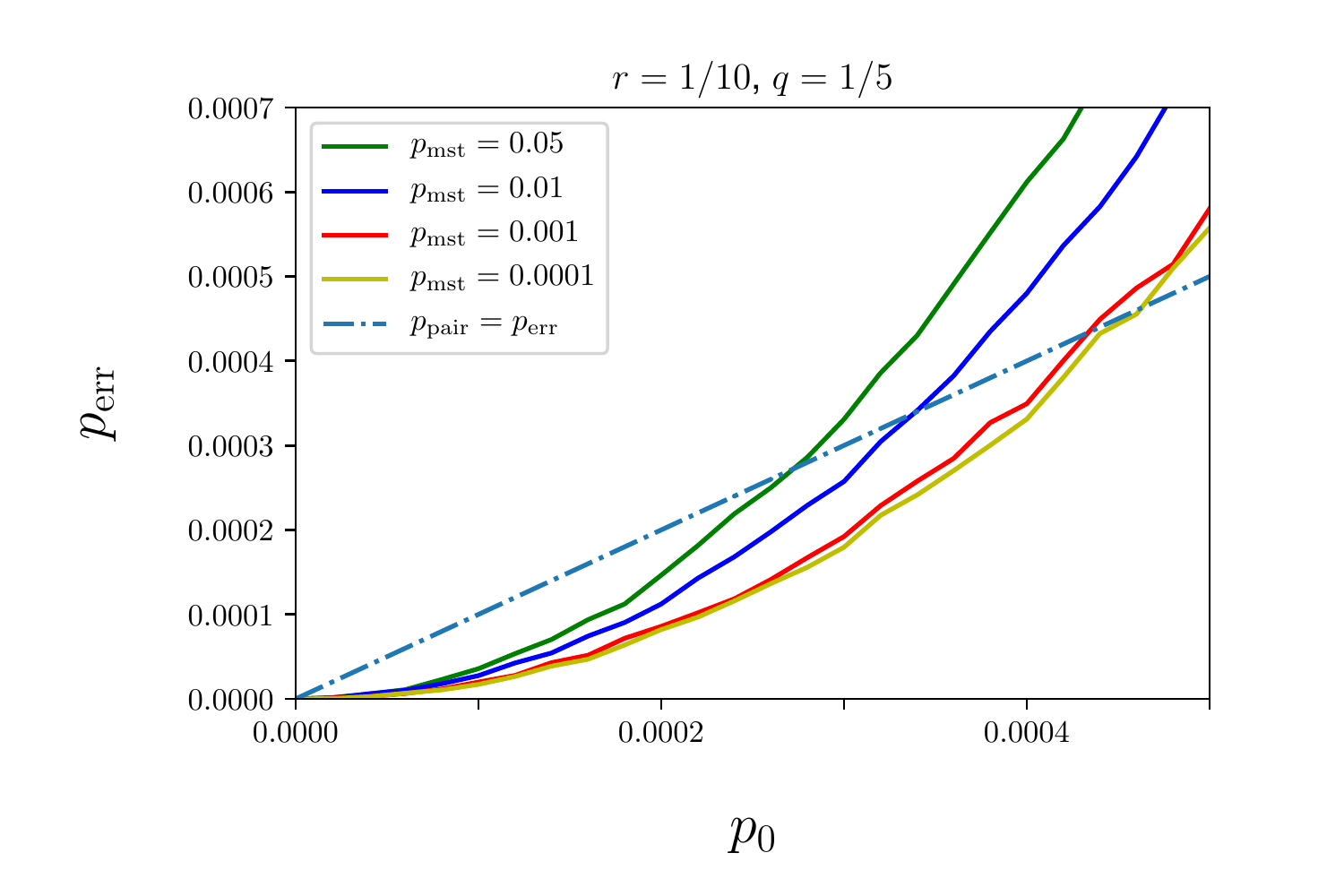}
	\end{minipage}
	\caption{Pseudo-threshold simulation results for the PMC noise model varying $r$ ({\it top left}), $q$ ({\it top right}), $p_2$ ({\it bottom left}), and $p_\text{mst}$ ({\it bottom right}) see Table~\ref{fig:prob-summary}).  Solid curves are the logical error rate as a function of the noise event rate per qubit (averaged over the 20 measured qubits and 5 idle qubits in the system), dot-dashed lines indicate no error correction, and their intersections denote the pseudo-threshold for those parameter choices. See text for discussion of the results.  The first three plots were generated for $10^8$ effective Monte Carlo trials per data point and the bottom right plot was generated for $10^7$ effective Monte Carlo trials per data point, see Appendix~\ref{app:code} for simulation details. }\label{fig:PMC}
\end{center}
\end{figure*}

For PMC, we follow the same protocol as for MC: every time step is followed by a stabilizer measurement and we repeat the error correction round four times (for a total of 16 time steps, 16 stabilizer measurements, and four sets of syndromes).  We assume the last syndrome which repeats twice is correct; if no two syndromes are repeated, we assume the last syndrome. 

As noted at the end of Section~\ref{sec:prob-summary}, models PMC and MC applied to a system of tetrons with $p_0=p_2$ only differ in a slight redistribution of the odd correlated events.   However, since PMC is defined with a particular physical implementation in mind and keeps track of which MZMs are involved in a measurement, we now take into account the geometric arrangement of MZMs on an island to define the measurements.  We replace Eq.~(\ref{eq:Pauli-M-mapping}) with the following definitions of an $XX$ measurement of horizontal nearest neighbor tetrons and a $ZZ$ measurement of vertical nearest neighbor tetrons (see Fig.~\ref{fig:tetronBS}):
\begin{align}
X_jX_{j+1} &= \gamma_{j,2}\gamma_{j,3}\gamma_{j+1,1}\gamma_{j+1,4} \label{eq:PMC-XX}
\\ Z_j Z_{j+5} &= \gamma_{j,3}\gamma_{j,4}\gamma_{j+5,1}\gamma_{j+5,2}. \label{eq:PMC-ZZ}
\end{align}
Note that now, for every tetron not located at a corner, each MZM on that tetron is involved in at least one measurement. This is a departure from the case where one MZM (e.g., $\gamma_{j,4}$) is not touched by any measurement and prevents a one-to-one mapping of quasiparticle events and Pauli errors.  In particular, a single-MZM error can lead to measurement syndromes that are not described by a local Pauli error.   
Equations~\eqref{eq:PMC-XX} and \eqref{eq:PMC-ZZ} are more convenient measurements to implement physically; however, there is no fundamental limitation to implementing measurements described by Eq.~\eqref{eq:Pauli-M-mapping} if the tetron architecture includes additional ancilla islands to facilitate longer-range measurements~\cite{Karzig17}.

The PMC simulation results are shown in Fig.~\ref{fig:PMC}.  The $r$-dependence,  $q$-dependence, and $p_\text{mst}$-dependence of the logical error rate are shown in the top-right, top-left, and bottom-right panels, respectively.  These plots are not noticeably different from those in Fig.~\ref{fig:MC} for the MC noise model, indicating that the redistribution of odd correlated errors (due to the difference between models MC and PMC) and the measurement redefinition have little effect on the pseudo-thresholds.  

The bottom-left panel of Fig.~\ref{fig:PMC} plots the logical error for $p_0/p_2=2,1,1/2$ with $r=1/10$, $q=1/5$, and $p_\text{mst}=10^{-4}$.  The $x$-axis, $(4 p_2+p_0)/5$, is chosen for easy comparison to model MC but is no longer the average rate of creating an excitation (the probability some of the noise events in PMC on a measured island are independent of $p_2$).   As expected, increasing $p_2$ worsens the logical error rate because it increases the relative distribution of correlated events, but it does so to a smaller degree than for MC since fewer of the noise events depend on $p_2$.

\subsection{Experimental Implications} \label{sec:experimental-implications}

The pseudo-threshold calculations in the previous section have important implications for experimental realizations of a tetron-based quantum computing architecture.  In the limit of a hard superconducting gap and large charging energy, {\it i.e.}, $T\ll\text{min}(\Delta, E_C)$, quasiparticle events are rare and the noise model PMC is a reasonable description of the most-likely error sources in the system.  The weak dependence of the pseudo-threshold on $r$ in this regime indicates that once quasiparticle events are sufficiently rare and the relaxation time is less than one tenth of the four-MZM measurement time, it is not essential to further suppress the quasiparticle poisoning rate.   

A second important experimental implication regards the tradeoff between correlated and measurement errors.  Statistical errors in the measurement contribute to the probability $p_\text{mst}$ in our noise models.  For the same length of measurement time, larger measurement visibility, which is controlled by the tunneling amplitude between the quantum dot and MZMs (see Section~\ref{sec:physical}), results in smaller statistical error and hence smaller $p_\text{mst}$.  Conversely, the parameter $q$ controlling the relative distribution of single-qubit and two-qubit errors increases with tunneling amplitude [see Eqs.~\eqref{eq:ex,g-g,ex1}, \eqref{eq:ex,g-g,ex2} and \eqref{eq:q}].  Thus, we see a tradeoff in which reducing the tunneling amplitude can suppress correlated events at the expense of increasing measurement errors.  However, the pseudo-threshold results indicate that, at least for Shor error correction of quantum memory, correlated events have a much stronger effect on the pseudo-threshold than do measurement bit-flips.  Therefore, it is desirable to work with a smaller tunneling amplitude from the point of view of the pseudo-threshold.  Reference~\onlinecite{Tomita14} reached a similar conclusion about the relative importance of two-qubit errors compared to measurement errors for small-distance surface codes. The tradeoff of measurement and correlated errors is particularly interesting when also considering different error correcting codes that might provide certain advantages but introduce stronger correlated errors due to higher-weight measurements. 

As noted earlier, the pseudo-threshold estimates in this paper prioritize simplicity of simulation over optimal magnitude, and should therefore not be interpreted as quantitative experimental targets.  However, if the correlation parameter $q$ can be reasonably estimated for a given system, then previous studies using the qubit-based model Pauli circuit noise are applicable.  For instance, if $q\sim 1/15$, then Ref.~\onlinecite{Cross09} suggests a pseudo-threshold for the $d=5$ Bacon-Shor code $\mathcal{O}\left(10^{-3}\right)$ (with Steane error correction).

\section{Extensions}\label{sec:extensions} 

We now discuss how the noise models can be modified to describe other Majorana-based systems. The physical error sources we considered (thermal excitation, extrinsic quasiparticle poisoning, fluctuations in the MZM hybridization energies, error in the measurements) are ubiquitous to all Majorana-based quantum computing architectures. Therefore, quasiparticle events, pair-wise dephasing events, and measurement bit-flips, as well as the corresponding probabilities $p_{\text{qp}}^{(k)},p_{\text{pair}}^{(k)},$ and $p_{\text{mst}}^{(k)}$, respectively, should be present in any realistic  stochastic Majorana noise model.  The maximal number of MZMs involved in a parity measurement will vary depending on the proposal, therefore the allowed integer values of $k$ will be system-dependent.  Furthermore, the degree to which a particular proposal is susceptible to these errors will depend on experimental parameters of the system, including the size and gap of the superconducting islands, whether or not the island has a charging energy, and what potential quasiparticle reservoirs ({\it e.g.}, metal or superconducting leads, quantum dots) are present.  

The main assumption of all four Majorana noise models introduced in Section~\ref{sec:noise-models} regards the energy separation between even and odd parity states of the underlying quantum computing architecture.  More specifically, we assume the system is divided into superconducting islands and that the environment relaxes islands with odd Majorana parity  to the even parity subspace within a time step with high probability.  When this is not the case, {\it e.g.}, for grounded superconducting islands~\cite{Hassler2011,vanHeck2011,Hyart2013}, step 0 of the Majorana noise models needs to be modified.  In such systems, a single quasiparticle excitation can be long-lived, and multi-island stabilizer measurements can propagate the excitation throughout the system, see Appendix~\ref{app:long-lived}.  

Qubit-based error correcting codes are less effective in a MZM system with long-lived quasiparticle excitations.  If the error correcting measurements introduce connected paths of islands spanning the system, a single quasiparticle excitation can cause a logical error within a time step.  Therefore, in this scenario, the measurements must be carefully designed to minimize the spread of high-weight errors through quasiparticle excitations.   Provided there is still a finite relaxation rate, the probability for an excitation not to relax over a full time step is $\text{exp}\left\{-1/r\right\}$.  Thus, if measurements are designed so that excitations can only travel to neighboring islands in a time step\footnote{Note that the Bacon-Shor code already utilizes this clever design by only using two-qubit measurements.}, the probability for a single excitation to travel through $d$ (the code distance) islands (possibly resulting in a logical error) is $\text{exp}\left\{ -(d-1)/2r\right\}$.  For $d$-large, this probability becomes vanishingly small and the qubit-based code retains some error correcting ability.  

Alternatively, one could use a Majorana fermion code~\cite{Bravyi10,Hastings17,Vijay2017} to distinguish odd and even parity states, thereby correcting quasiparticle excitations before they can spread throughout the system. Such codes require the ability to either (1) measure the total parity of the MZMs on an island, or (2) dynamically adjust the number of MZMs on an island, both of which are experimentally challenging for the following reasons.  (1) Proposals to date are not able to distinguish the total fermion parity of an island from the total parity associated with the MZMs.  The measurements proposed in Refs.~\onlinecite{Hassler2011,vanHeck2011,Hyart2013,Aasen2016} measure the total charge on the island, and are thus unable to identify the presence of a thermally excited quasiparticle. Charge measurements of the islands are therefore helpful only if the charge excitations are long-lived while thermal excitations relax sufficiently fast. (2) Dynamically adjusting the number of MZMs on an island requires sufficient tunability of certain experimental parameters ({\it e.g.}, Josephson energy) to transition from the fully disconnected regime to the fully connected regime: residual coupling in the former can lead to increased probability of correlated events; non-fully connected regions within an island in the latter can result in mutual capacitances leading to higher weight errors, see Appendix~\ref{app:higher-order}.  Furthermore, the tuning procedure must be done sufficiently slowly and smoothly to avoid introducing diabatic errors into the system~\cite{Cheng11,Karzig15,Scheurer13, Hell16, Knapp16}.

Model PMC was built on the further assumption of single or two-island quantum dot-based measurements involving at most two MZMs per island~\cite{Plugge2017,Karzig17}.  Additional correlated events can arise for alternate measurement schemes~\cite{Hassler2011, vanHeck2011, Hyart2013,Litinski2017,Vijay2017} or if more than two islands are connected during a measurement.  For instance, surface code implementations that rely on eight-MZM stabilizer measurements connecting four islands \cite{Landau2016} could result in a correlated event involving any subset of the four islands.  

Finally, the pseudo-threshold calculations in Section~\ref{sec:threshold} are specific to Shor error correction of quantum memory for a system with four MZMs per island.  The pseudo-threshold estimates would likely change for a Steane error correction analysis, if logical operations on the system are included, or if islands have more than four MZMs ({\it e.g.}, a hexon architecture~\cite{Karzig17}).  The benefit of the Bacon-Shor code is lost for Steane error correction, since the stabilizers could no longer be implemented from two-island measurements. Generally, for a stabilizer code, Steane error correction will estimate higher pseudo-thresholds, however we do not anticipate the change from Shor to Steane error correction to affect the relative importance of quasiparticle and correlated events ($r$ and $q$ dependence). 
Including logical operations would probably result in a greater dependence on the relative magnitudes of $p_0,\,p_1,\,$ and $p_2$, as the majority of islands would no longer be involved in two-island measurements for each time step. 
With hexons ($m=3$), the percentage of correlated events that do not commute with all stabilizer measurements is greater than for the case of tetrons; therefore we expect the pseudo-threshold to have a stronger $q$ dependence.  Furthermore, since physical considerations can constrain the number of allowed correlated events in PMC relative to MC, more substantial differences in the pseudo-threshold estimates are possible than were found for the tetron case considered here.

\section{Conclusions and Outlook}\label{sec:conclusions}

The primary goal of this paper is to connect the underlying physical processes causing errors in a Majorana-based quantum computing architecture to the noise models used for fault tolerance analysis of the system.  We developed stochastic Majorana noise models in close analogy to the standard qubit-based noise models generally used in threshold calculations.  These Majorana noise models allow for errors involving an odd number of Majorana operators to occur, and have different probability distributions depending on whether a MZM island begins a time step in an even or odd parity state.  The result of our analysis is that for quasiparticle-poisoning-protected qubits, the pseudo-threshold estimate for the $d=5$ Bacon-Shor code under each of the Majorana noise models is well-approximated by the estimate using the analogous qubit-based model.  Essentially, when the relaxation parameter $r\lesssim 1/5$, a quasiparticle event in one time step relaxes with high probability in the subsequent time step;  the cumulative effect is to apply a Pauli error broken-up over two time steps, which does not dramatically alter the error correcting code's performance.  This result does not depend on any particular feature of the Bacon-Shor code, thus we expect it to apply more generally to larger error correcting codes, as well as to error correction of logical gates. It is a positive result that MZM systems with short-lived quasiparticle excitations ({\it e.g.}, charging-energy-protected qubits) can be analyzed with the simpler qubit-based noise models, as this simplifies numerical simulations and allows for studies of larger error-correction schemes.  

Conversely, a MZM system with long-lived quasiparticles cannot be accurately described by a qubit-based noise model.  In such a system, an excitation could travel between islands connected by measurements over several time steps with $\mathcal{O}(1)$ probability, thereby spreading throughout the system.  
In order to prevent excitations traveling long distances, it is therefore beneficial when performing measurements in parallel to avoid creating connected paths of islands spanning the system.  
The Majorana noise models developed here could be extended to systems with long-lived excitations (see Appendix~\ref{app:long-lived}). The latter make qubit-based codes less efficient and might require a Majorana fermion code for error correction~\cite{Bravyi10,Hastings17,Vijay2017} (see discussion in Section~\ref{sec:extensions}).

A useful observation is that in a Shor error correction scheme, in which stabilizer measurements are repeated to protect against a single measurement bit-flip, the error correcting code can sustain high probabilities of measurement errors without affecting the pseudo-threshold.  This result applies equally well to conventional qubit systems.  While generally Steane error correction results in higher pseudo-thresholds, if measurement error is a limiting obstacle, Shor error correction could be an attractive alternative. 

Our pseudo-threshold analysis further demonstrates that there is a strong dependence on the relative distribution of single-island and correlated events.  It is therefore essential that any realistic noise analysis of a given system carefully estimate the rate of transferring excitations between islands so as to choose an appropriate distribution of single-island and correlated events.  Furthermore, proposals involving higher-weight measurements ({\it e.g.}, four-island stabilizer measurements in the surface code) will be subject to higher-weight correlated events that could significantly affect the threshold estimate. The relative importance of correlated errors compared to measurement errors also indicates that it is worth optimizing the measurement processes so that correlated errors are suppressed even if this reduces the measurement fidelity. 

This work is a first step towards comprehensive analysis of the fault tolerance of Majorana-based quantum computing architectures. Future directions of study include understanding to what extent our results hold for both larger error correcting codes and logical gate error correction.  Additionally, an important open question is to determine the optimal error correction procedure for Majorana-based quantum computation.  This analysis should weigh the experimental feasibility of the quantum error correcting codes under consideration ({\it e.g.}, ability to perform stabilizer measurements and whether the underlying system needs to be charging-energy-protected) and the physical noise sources affecting the underlying architecture in addition to the usual fault tolerance criteria ({\it e.g.}, threshold values and ratio of physical to logical qubits).  We believe the connections elucidated here between physical error processes in MZM systems and noise models would aid in such an analysis.

\section*{Acknowledgments}

We are grateful to Bela Bauer, Chris Chamberland, Nicolas Delfosse, Daniel Litinski, Tom O'Brien, and Krysta Svore for helpful discussions and comments on a draft of this paper.  C.K. acknowledges support from the NSF GRFP under Grant No. DGE $114085$.


\appendix

\section{Higher order errors}\label{app:higher-order}

The probabilities defining PMC correspond to the lowest order error processes occuring in a tetron architecture.  These lowest order processes include measurement bit-flips and all noise events that involve only a single excitation (one factor of $p_k$ in the language of Fig.~\ref{fig:prob-tree} and Table~\ref{fig:prob-summary}): quasiparticle, pair-wise dephasing, and correlated events.  All other errors in the model occur at higher order, that is, with a probability that is the product of probabilities explicitly defined in the model ({\it i.e.}, a probaiblity $\mathcal{O}\left(p_k^2\right)$).  We now justify why other error processes (transitions into higher excited states, other correlated events, or mutual capacitance terms) that are physically present in the system can be absorbed into these higher order terms.   

Throughout our discussion, we have only considered the excited states $\ket{e_\Delta}$ and $\ket{e_{C,\pm}}$.  In particular, we have neglected an extrinsic quasiparticle tunneling into an above-gap state, states with multiple above-gap quasiparticles, or states with charge $2e$ or higher.  Let ${\Gamma_{g\to e}=\max\left(\Gamma_{g\to e_{\Delta}},\Gamma_{g \to e_{C,\pm}} \right).}$  Transitions into all higher energy states are additionally suppressed compared to $\Gamma_{g\to e}$ by a factor $\exp\{-\delta \varepsilon/T\}$, where $\delta \varepsilon$ is the energy difference between the higher excited state and $\max\left( \Delta,E_C \right)$.  As ${\delta \varepsilon\geq \text{min}\left(\Delta,E_C\right)}$, this additional exponential factor makes such a transition a higher order process.  For the same reason, a correlated event involving the process ${\ket{e_x}\otimes \ket{g}\to \ket{e_x}\otimes \ket{e_x}}$ ($\ket{e_x}\in \{\ket{e_\Delta},\ket{e_{C,\pm}}\}$), is also a higher order process, as it requires two excitations. 

For a noise model that gives conservative threshold estimates, it is sometimes beneficial, for the sake of keeping the model simple, to overestimate the probability of lowest order errors, so that multiple lowest order errors capture the effects of additional error processes. As an example, we note that the presence of interactions in the MZM system can lead to an additional error source that we have not explicitly included in our model.  So far we have neglected mutual charging energies within a superconducting island, {\it e.g.}, between the two MZM nanowires comprising a tetron.  In the case of tetrons, mutual capacitances are exponentially suppressed in the number of channels in the superconducting backbone (blue line in the left panel of Fig.~\ref{fig:tetron}) connecting the two MZM nanowires. The remaining small but finite mutual capacitances, however, would add four-(or more)-MZM terms to Eq.~(\ref{eq:tetron-Ham}) that involve pairs of MZMs belonging to different tetrons. This capacitance can, depending on the screening properties of the system, fall off as a powerlaw with distance, and could therefore give rise to high-weight errors (errors involving products of $2n$ MZMs with $n\geq 2$) in the MZM system.  To justify ignoring these high-weight errors at lowest order in the noise model, we could overestimate the probability $p_{\text{pair}}^{(k)}$ of pair-wise dephasing events so that the probability of an error involving $2n$ MZMs is given by $\left( p_{\text{pair}}^{(k)}\right)^n$. 


\section{Long-lived excitations}
\label{app:long-lived}

\begin{figure}
\begin{center}
	\includegraphics[width=\columnwidth]{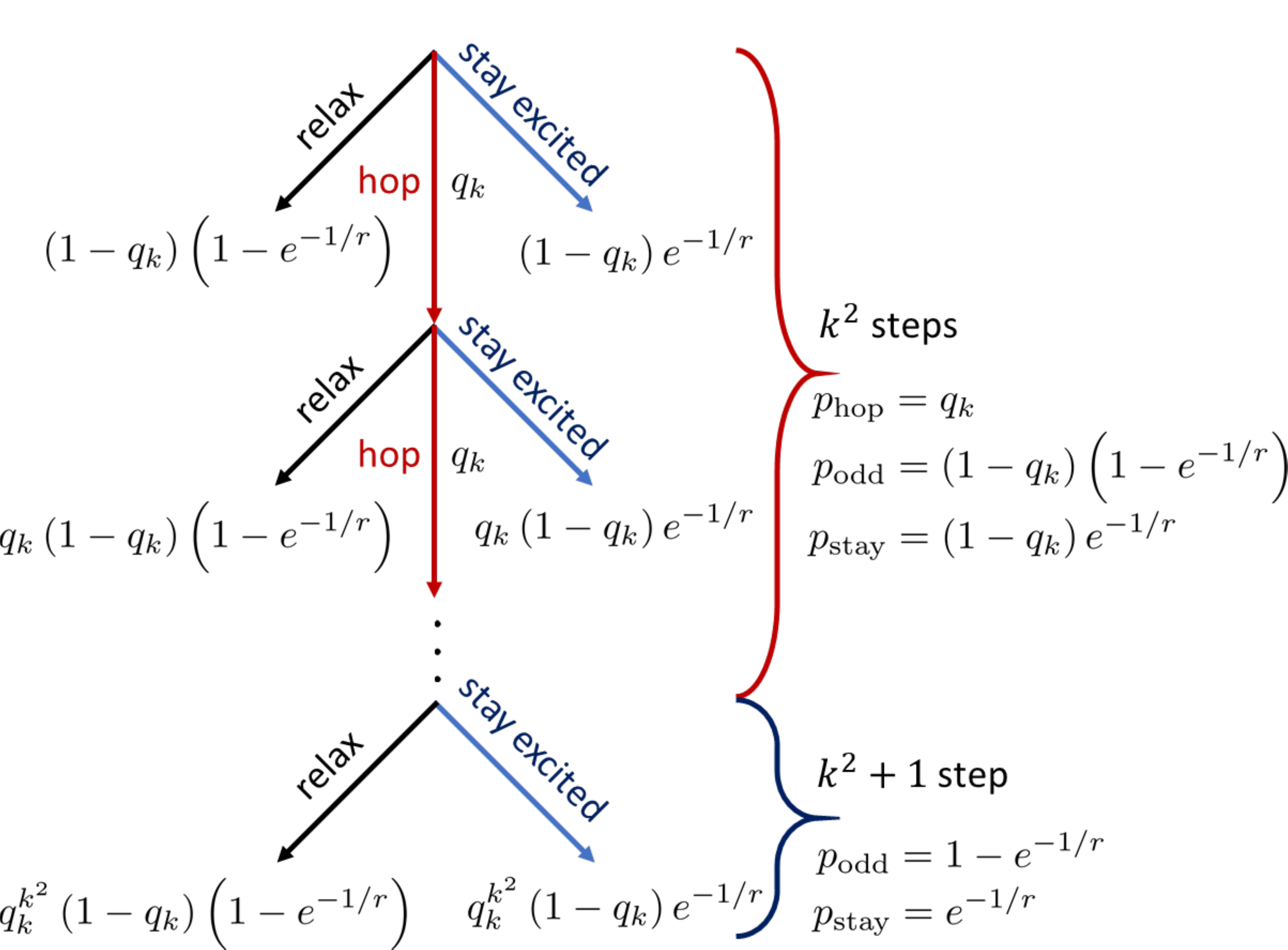}
	\caption{ Decision tree for the generalized step~\ref{step:0-mod} for models MC and PMC in the case of long-lived excitations.  If an island involved in a $k$-island measurement begins in the odd parity state, it can either relax (left black line) with probability $p_\text{odd}$, hop to another island involved in the measurement (center red line) with probability $p_\text{hop}$, or remain excited throughout the time step (right blue line) with probability $p_\text{stay}$.  In the first or the last case, the noise model proceeds as described in the main text to allow for single island noise events, correlated events, or measurement bit-flips.  In the latter case, if at least one of the islands involved in the measurement is still in the odd parity state, it again proceeds to the next step of the decision tree (relax, hop, or stay excited). Even for $q_k\sim 1$, the tree can be truncated after $k^2$ steps where all $k$ island have been visited by the excitation. In the final step the excitation can then be assumed to either relax or stays excited.  The probabilities for $p_\text{odd}$ and $p_\text{stay}$ are modified for the final step of the decision tree as shown.
	}
	\label{fig:tree}
\end{center}
\end{figure}

The Majorana noise models presented in Section~\ref{sec:noise-models} need to be modified when the system has long-lived excitations ({\it e.g.}, in the case of islands with small charging energies) to include an additional type of correlated event:
\begin{itemize}
\item {\it Hopping event}: application of two Majorana operators involving two islands connected by a measurement.  A hopping event for islands $i$ and $j$ is of the form $\gamma_{i,a}\gamma_{j,b}$.
\end{itemize}
A hopping event can occcur when an island involved in a $k$-island measurement begins a time step in the odd parity state and transfers its excitation to another island involved in the same measurement.  Such a process is more likely to occur when an excitation is long-lived.
 Models MC and PMC can be generalized to account for long-lived excitations by modifying step 0 as follows:
\begin{enumerate}
\setcounter{enumi}{-1}
\item For the set of islands involved in the same $k$-island measurement, if at least one of the islands begins the time step with odd parity, do one of the following: \label{step:0-mod}
\begin{enumerate}
\item Apply a hopping event involving an odd parity island with probability $p_\text{hop}$, and return to the beginning of step 0. \label{step:hop}
\item Apply a quasiparticle event to an island with odd parity with probability $p_\text{odd}$, and proceed to step 1. \label{step:relax}
\item Do not apply any noise events with probablity $p_\text{stay}=1-p_\text{hop}-p_\text{odd}$, and proceed to step 1. \label{step:remain}
\end{enumerate}
\end{enumerate}
The generalized step~\ref{step:0-mod} results in the decision tree depicted in Fig.~\ref{fig:tree}. The chance for a large number $n_\text{hop}$ of hopping events is exponentially suppressed by $q_k^{n_\text{hop}}$. Moreover, even for large $q_k\sim 1$, the tree can be truncated, with the final step modified to not include step~\ref{step:hop}.  For instance, if every island is only directly connected to at most two other islands (as in a tetron measurement), then a hopping event is a random walk through the connected path of islands.  After $k^2$ steps, the average path includes the path involving all islands in the measurement; therefore all possible noise events occuring after $k^2+1$ steps can be captured by a more-probable process in a smaller number of steps.

In the language of Section~\ref{sec:probabilities}, the probabilities for the modified step~\ref{step:0-mod} are given by 
\begin{align}
p_\text{hop} &= q_k,
\\ p_\text{odd} &= \left(1-q_k\right) \left( 1-e^{-1/r}\right),
\\ p_\text{stay} &= \left(1-q_k\right) e^{-1/r}.
\end{align}
(If truncating the modified step~\ref{step:0-mod}, the probabilities for the final step do not involve $q_k$, see Fig.~\ref{fig:tree}.)
Recall that $q_k$ is the probability of an energy transfer between islands connected by a measurement, while $r$ is the probability that an excitation occuring during a time step does not relax before the end of the time step.  The probability that an island beginning a time step in the odd parity state does not relax is $e^{-1/r}$.  

In the main text, we set $p_\text{hop}=0$.  The justification for this approximation is that we assume $r$ is small enough (quasiparticle excitations are short-lived enough) that noise events resulting from a sequence of hopping events are already captured by a sequence of correlated and single qubit noise events.  More concretely, one might worry that not including hopping events neglects the possibility that a single excitation can result in a high-weight error spread over several time steps from a quasiparticle traveling between different islands connected by subsequent measurements.  Consider, for instance, the following two scenarios leading to a weight-three error: (1) An idle island is initially excited in the first time step, the excitation is transferred to a second island in the second time step and does not relax, and the excitation is transferred to a third island in the third time step and then relaxes. This process applies a pair of Majorana operators on all three islands involved. In an alternative scenario (2) an idle island experiences no noise events in the first time step, the first and second islands are involved in an even correlated error in the second time step, and the third island undergoes a pair-wise dephasing event in the third time step.  Process (1) only requires a single excitations to cause a weight-three error, while process (2) is the most probable implementation of a weight-three error over three time steps that does not involve hopping events. Process (1) occurs with probability 
\begin{equation}\label{eq:prob1}
p_\text{qp}\left(p_\text{hop}p_\text{stay}\right)\left( p_\text{hop}p_\text{odd}\right)
\end{equation}
 and process (2) occurs with probability 
\begin{equation}\label{eq:prob2}
 \left(1-p_\text{qp} -\frac{3}{4}p_\text{pair}\right)p_\text{cor,even}^{(2)} p_\text{pair}^{(2)}.
 \end{equation} 
 (We assume the first island is idle in the first time step for simplicity, the argument naturally extends if the island is instead involved in a measurement.)  Processes (1) and (2) result in the same computational effect for the MC simulation discussed in Section~\ref{sec:results}, because the net application of Majorana operators at the end of the third time step are equivalent for (1) and (2), and intermediate steps are interpreted as measurement errors.
We are therefore justified in neglecting process (1) when the probability~\eqref{eq:prob1} is much less than \eqref{eq:prob2}.  This is the case provided
\begin{align}\label{eq:condition}
p_0&\gg \frac{q}{1-q} \frac{r}{(1-r)^2} e^{-1/r} \left(1-e^{-1/r}\right)\approx q r e^{-1/r}.
\end{align}
The simplifying approximation holds for small $q$ and $r$. For $q=1/5$ and $p_0=10^{-3}$ condition~\eqref{eq:condition} is valid provided $r<1/5$, and for $q=1/5$ and $p_0=10^{-4}$ is valid provided $r=1/7$.  Note that the $r$ values in Fig.~\ref{fig:MC} were chosen to exagerate the $r$ dependence of the noise model and do not strictly fall into this regime.  Including additional time steps does not result in a stronger condition on the relative size of $p_0$ and $r$. 

When $r$ is too large to satisfy Eq.~\eqref{eq:condition}, we need to use the more complicated step~\ref{step:0-mod}.  In this regime, a single excitation can result in a high-weight error by continually spreading to new islands with each multi-island measurement. 

In scenarios with long-lived excitations it becomes crucial to devise a careful measurement protocol to avoid introducing paths of connected islands spanning the system. If errors can only hop to $\mathcal{O}(k)$ nearby islands in a single time step, an excitation will eventually relax before traversing the system if the system size becomes sufficiently large. This implies that, while being less efficient, large qubit-based codes will still be able to correct for long-lived excitations. At some point it might, however, be more practical to use Majorana fermion codes to be able to detect whether an island is in the odd parity state before the excitation can propagate.

Finally, if the quantum computing architecture is built from grounded superconducting islands, an island can spend an equal amount of time in the even and odd parity states.  In this case, the probabilities of different noise events are the same regardless of in which parity state the system begins.  In this regime, excitations are dominated by extrinsic quasiparticle poisoning events and Eq.~\eqref{eq:p} is modified to 
\begin{align}
p_k=2m  \left( \text{exp} \left\{ \Gamma_{g\to e_C}^{(k)} \tau \right\}-1\right).
\end{align}
In particular, $p_k$ is no longer exponentially suppressed and all noise events become much more likely. 

\section{Pseudo-threshold code}\label{app:code}

In this appendix, we give further details of the code used to simulate the pseudo-thresholds reported in the main text Section~\ref{sec:threshold}. 
\\

\noindent {\bf Perfect measurement Monte Carlo algorithm}: In both the qubit-based model Pauli noise and the Majorana model Qp, the measurements are perfect and the algorithm to model error correction is as in Algorithm~\ref{alg:perfmeas}. 

\begin{figure}
	\begin{algorithm}[H]
		\footnotesize
		\caption{Perfect Measurement Monte Carlo}
		\begin{algorithmic}[1]
			\State n\text{Fail} = 0
			\State \textbf{for} i = 1 \textbf{to} i =  N \textbf{do}
			\State ~~~ MList = ZeroArray[n]
			\State ~~~ err = Noise[MList]		
			\State ~~~ MList = MList + err
			\State ~~~ gaugeOutcome = Dot[gaugeMatrix,MList]	
			\State ~~~ stabOutcome = Dot[stabGaugeMatrix,gaugeOutcome]		
			\State ~~~ corr = LookupBaconShor[stabOutcome]	
			\State ~~~ MList = MList + corr
			\State ~~~ logicalErr = Dot[logicalMatrix,MList]
			\State ~~~ \textbf{if} Total[logicalErr] $>$ 0 \textbf{then}
			\State ~~~ ~~~ n\text{Fail} = n\text{Fail} +1
			\State ~~~ \textbf{end if}
			\State \textbf{end for}
			\State \textbf{return} n\text{Fail}/N
		\end{algorithmic}
		\label{alg:perfmeas}
	\end{algorithm}
\end{figure}

The algorithm estimates $p_\text{err}$
by sampling simulated trials of faulty quantum error correction and counting the fraction of trials which result in a logical error. 
In each trial, errors occur according to the noise model (for a given set of noise parameters), and are recorded in a binary vector `err' of length $100 = 4 \times 25$, {\it i.e.}, one for each MZM. 
A bit value of $1$ indicates that the corresponding Majorana operator was applied to the system, otherwise the bit value is $0$. 
This error is added bitwise to the length 100 binary vector `MList' (initially trivial), which tracks the net set of Majorana operators which have been applied to the system. 
After the noise has been applied, the gauge measurement outcomes are inferred, in practice by storing each gauge operator as a row of a binary array `gaugeMatrix' (with a bit value of $1$ for each Majorana operator appearing in the gauge operator), and multiplying it by `MList'.
The results are stored as a binary vector `gaugeOutcome'.
The stabilizers `stabOutcome' are found by multiplying the appropriate subset of gauge measurement outcomes and a correction `corr' is inferred from a lookup table for the Bacon-Shor code, which would apply the appropriate Pauli correction for two or fewer single qubit Pauli errors (again encoded in a bit string) as described in Section~\ref{sec:BS}. 
The final result of the error correction attempt is obtained by adding `corr' to `MList'. If the error correction was successful `MList' is a binary representation of a trivial operator. This is checked by taking a binary dot product of the list of all possible non-trivial logical operators `logicalMatrix' with `MList' which yields $0$ ($1$) when `MList' shares an even (odd) number of Majorana operators with a certain logical operator and thus (anti)commutes. Anticommutation with any local operator is then counted as a failure. In our simulations, the number of samples used to generate each data point was $N=10^5$.
\\ 

\begin{figure}
\begin{algorithm}[H]
\footnotesize
	\caption{ Perfect Measurement Importance \\
Sampling Monte Carlo }
	\begin{algorithmic}[1]
		\State pErr = 0
		\State \textbf{for} v = 2 \textbf{to} v = vTrunc \textbf{do}
		\State ~~~  n\text{Sample} = 0
		\State ~~~  n\text{Fail} = 0
		\State ~~~  \textbf{while} n\text{Fail} $<$ N \textbf{do}
		\State ~~~  ~~~ w = RandomSubset[W,v]
		\State ~~~  ~~~ n\text{Sample} = n\text{Sample} + 1
		\State ~~~  ~~~ MList = ZeroArray[n]
		\State ~~~  ~~~ err = PostSelectedNoise[MList,w]		
		\State ~~~  ~~~ MList = MList + err
		\State ~~~  ~~~ gaugeOutcome = Dot[gaugeMatrix,MList]	
		\State ~~~  ~~~ stabOutcome = Dot[stabGaugeMatrix,gaugeOutcome]		
		\State ~~~  ~~~ corr = LookupBaconShor[stabOutcome]	
		\State ~~~  ~~~ MList = MList + corr
		\State ~~~  ~~~ logicalErr = Dot[logicalMatrix,MList]
		\State ~~~  ~~~ \textbf{if} Total[logicalErr] $>$ 0 \textbf{then}
		\State ~~~  ~~~ ~~~ n\text{Fail} = n\text{Fail} +1
		\State ~~~  ~~~ \textbf{end if}
		\State ~~~  \textbf{end while}
		\State ~~~  pErr = pErr + $\text{Pr}(v)$ n\text{Fail}/n\text{Sample}		
		\State \textbf{end for}		
		\State \textbf{return} pErr
	\end{algorithmic}
	\label{alg:perfmeasIS}
\end{algorithm}
\end{figure}

\noindent {\bf  Importance sampling with perfect measurement}:
The above Monte Carlo procedure would become slow for small error rates, since the number of repetitions becomes very large, requiring huge computational resources.
This is largely because, for very low error rates, almost every run has only zero or one faults.
In practice, the error rates required for our purposes were sufficiently large for Pauli noise and the Majorana model Qp to allow us to use Monte Carlo, but it is useful to describe this case for illustrative purposes. 
Since the protocol is fault tolerant (as proven in Appendix~\ref{app:EC}), zero or one faults will never cause a logical failure.  Therefore, we can save time by only generating samples in which at least two faults occur.  
In order to do this, we must be able to both calculate the probability that at least one fault occurs, and efficiently sample from the distribution formed by post selecting on there being at least one fault. 
This procedure is known as {\it importance sampling}.  Here, we can apply importance sampling relatively simply by considering each tetron independently, and by separately considering errors on all subsets of $v$ tetrons, for $v=2,3,4\dots$.  We can truncate $v$ at $v_\text{trunc}$ when the 
probability of $v_\text{trunc}$ tetrons having an error becomes negligible.  

Let $W$ be the set of all tetrons, and let $\text{Pr}(v)$ be the probability of precisely $v$ of the tetrons having an error.  Then, the importance sampling Monte Carlo algorithm is implemented as in Algorithm~\ref{alg:perfmeasIS}. 
In each run, a subset $w$ containing $v$ tetrons is randomly selected; noise is drawn from the post-selected distribution where errors occur on each tetron in the subset $w$ and nowhere else.
\\

\noindent {\bf Imperfect measurement Monte Carlo algorithm}: We now consider the noise models for which the measurements can be imperfect ({\it i.e.}, QpBf, MC, PMC, and their analogous qubit-based models). 
For these models, our quantum error correction protocol repeats measurements four times to achieve fault-tolerance, essentially by increasing the reliability of the inferred stabilizer outcomes.
Algorithm~\ref{alg:imperfmeas} therefore includes four error correction rounds $t$, and the correction protocol is implemented as normal, but by using the last repeated set of stabilizer outcomes (or the last stabilizer outcome if none are repeated). 
We also apply an additional round of perfect error correction to ensure that the system is returned to the code space.

\begin{figure}
\begin{algorithm}[H]
\scriptsize
	\caption{Imperfect Measurement Monte Carlo} 
	\begin{algorithmic}[1]
		\State n\text{Sample} = 0
		\State n\text{Fail} = 0
		\State \textbf{while} n\text{Fail} $<$ N \textbf{do}
		\State ~~~ n\text{Sample} = n\text{Sample} + 1
		\State ~~~ MList = ZeroArray[n]
		\State ~~~ \textbf{for} t = 1 \textbf{to} 4 \textbf{do}
		\State ~~~ ~~~ err = Noise[MList]		
		\State ~~~ ~~~ MList = MList + err
		\State ~~~ ~~~ gaugeOutcome = Dot[gaugeMatrix,MList]	
		\State ~~~ ~~~ stabOutcome$^*$[t] = Dot[stabGaugeMatrix,gaugeOutcome]
		\State ~~~ \textbf{end for}
		\State ~~~ t=4;	
		\State ~~~ \textbf{while} t$>$1 \textbf{and} stabOutcome$^*$[t] $\neq$ stabOutcome$^*$[t-1] \textbf{do}
		\State ~~~ ~~~ t = t-1	
		\State ~~~ ~~~ stabOutcome = stabOutcome$^*$[t]	
		\State ~~~ \textbf{end while}
		\State ~~~ \textbf{if} t $==$ 1 \textbf{then}
		\State ~~~ ~~~ stabOutcome = stabOutcome$^*$[4]
		\State ~~~ \textbf{end if}			
		\State ~~~ corr = LookupBaconShor[stabOutcome]	
		\State ~~~ MList = MList + corr
		\State ~~~ GaugeOutcome = Dot[gaugeMatrix,MList]	
		\State ~~~ resStabOutcome = Dot[stabGaugeMatrix,GaugeOutcome]		
		\State ~~~ resCorr = LookupBaconShor[resStabOutcome]	
		\State ~~~ resMList = MList + resCorr
		\State ~~~ logicalErr = Dot[logicalMatrix,resMList]
		\State ~~~ \textbf{if} Total[logicalErr] $>$ 0 \textbf{then}
		\State ~~~ ~~~ n\text{Fail} = n\text{Fail} +1
		\State ~~~ \textbf{end if}
		\State \textbf{end while}
		\State \textbf{return} n\text{Fail}/n\text{Sample}
	\end{algorithmic}
	\label{alg:imperfmeas}
\end{algorithm}
\end{figure}

We have used $S$ to represent the number of stabilizer generators.
The function `Noise[MList]' here is very general -- in the MC noise model for example, it actually entails a sequence of four time steps, during each of which a different set of gauge generators is (noisily) measured.
\\

\noindent {\bf  Importance sampling with imperfect measurement}:
A key feature of the importance sampling in the perfect measurement case described above is that each possible fault is independent. 
However, now that we have multiple time steps, and since in some of our noise models, the noise can depend on the current state of the system, we can no longer treat the faults as being independent. 
Moreover, importance sampling really is necessary for these cases since the error rates in the region of interest are very small rendering Monte Carlo extremely time consuming.
The number of samples that would be needed for the regime of interest for MC and PMC would require extremely large sample numbers without any kind of importance sampling. 
Fortunately, we can implement a partial importance sampling: we run importance sampling for the first fault that occurs, and then after that we run the system forward in time with usual Monte Carlo.  The algorithm is a straightforward combination of Algorithms~\ref{alg:perfmeasIS} and \ref{alg:imperfmeas}.
The effective number of total samples used per data point for imperfect measurements was $10^7$ for the plots in which $p_\text{mst}$ was varied, and and $10^8$ samples for the rest of the simulations (although many of these were not actually run due to the importance sampling step).
\\

\begin{table*}
\begin{center}
\scriptsize
	\begin{tabular}{|c|c|c|} \hline 
		~Bit Strings~&~Even Probabilities~& ~ Odd Probabilities ~ \\ \hline 
		$(0,0,0,0)$ & $1- p_\text{qp} - \frac{3}{4} p_\text{pair}$ & $\left(1-p_\text{odd}\right) \left(1- p_\text{qp} - \frac{3}{4} p_\text{pair}\right) +\frac{1}{4} p_\text{odd} p_\text{qp}$  \\ \hline 
		$(1,0,0,0),\, (0,1,0,0),\,(0,0,1,0),\, (0,0,0,1)$ & $\frac{1}{4}p_\text{qp}$ & $\frac{1}{4}\left(1-p_\text{odd}\right)p_\text{qp} + \frac{1}{4}p_\text{odd}\left( 1-p_\text{qp} - \frac{1}{4}p_\text{pair}\right) + \frac{3}{16} p_\text{odd}p_\text{pair}$ \\ \hline
		$(1,1,0,0), \, (1,0,1,0),\, (1,0,0,1) $ & $ \frac{1}{4}p_\text{pair}$ &  $\frac{1}{4}\left( 1- p_\text{odd} \right) p_\text{pair}+ \frac{1}{4}p_\text{odd} p_\text{qp} $ \\  \hline
	\end{tabular}
	\caption{The eight computationally distinct single-island bit strings (left column) and their corresponding probabilities when the island begins the time step with even parity (middle column) and odd parity (right column).  The probabilities are written for models Qp and QpBf, but are easily generalized to MC by including the superscript $(k)$ when the island in question is involved in a $k$-island measurement.  The first row corresponds to no error, the second row to errors involving an odd number of MZMs, and the last row to errors involving two MZMs. }
	\label{table:Qp}
\end{center}
\end{table*}

\noindent {\bf Single island error probabilities}:
Lastly, in Table~\ref{table:Qp} we explicitly write the eight computationally distinct bit strings for a single island and the corresponding probability of applying that bit string to `MList' in a time step.   This table is used to generate single-island error strings in models Qp, QpBf, and MC.
For the $j$th tetron, the bit string $(1,0,1,0)$ corresponds to $\gamma_{j,1}\gamma_{j,3}$, or equivalently to $\gamma_{j,2}\gamma_{j,4}$ (because the total parity of an island is fixed).  For simplicity, we write the probabilities for models Qp and QpBf.  Model MC is straightforwardly generalized by adding the appropriate superscript to each of the probabilities: $(0)$ for an idle island or $(k)$ for a $k$-island measurement.  Model PMC is slightly more complicated for a measured island as we need to distinguish which MZMs are being measured.  Note that since the parity of a tetron is fixed, bit strings that differ by $(1,1,1,1)$ are computationally equivalent.    Models MC and PMC will additionally have two-island errors, whose probabilities depend on $p_\text{cor,even}$ and $p_\text{cor,odd}$.

\section{Fault tolerant conditions}\label{app:EC}

In this appendix, we detail how fault tolerance conditions (EC A)-(EC B) are satisfied for noise models Qp, QpBf, MC, and PMC.
\\

\noindent {\bf Qp}:  We simulate model Qp with a noisy time step followed by perfect application of all stabilizer measurements followed.  Measurements are defined according to Eq.~(\ref{eq:Pauli-M-mapping}).  We use the minimum weight perfect matching decoder for the resulting syndrome to choose which correction operator to apply.  

Model Qp assumes perfect measurements, thus (EC~A) and (EC~A') are automatically satisfied.  To check that (EC~B) holds, we know that the $d=5$ Bacon-Shor code can correct all single- and two-qubit Pauli errors, and that in a tetron system all pair-wise dephasing events can be expressed as Pauli errors.  Therefore, we just need to check that quasiparticle errrors are similarly correctable.  A single fault corresponds to either a quasiparticle excitation in one time step followed by relaxation in the next time step, or a pair-wise dephasing event.  The former process has the same net effect as the latter, and is thus correctable.  To see this more explicitly, consider the sequence of the quasiparticle event $\gamma_{j,a}$ occuring in one time step and the island relaxing back to the even parity subspace in the subsequent time step with application of $\gamma_{j,b}$.  The first error has the same syndrome as the Pauli error $\gamma_{j,a}\gamma_{j,4}$ (because $\gamma_{j,4}$ is not involved in any of the stabilizer measurements), thus the error correction will inadvertently apply the operator $\gamma_{j,4}$.  The net operator applied to the island at the end of the next time step is $\gamma_{j,b}\gamma_{j,4}$, which is simply a Pauli error.  More generally, once the island has returned to the even parity subspace, the net operator applied to the island is a Pauli error, which can be corrected in the standard way.  Intermediate corrections (while the island remains in the odd parity subspace) do not worsen the error.
\\

\noindent {\bf QpBf}:  Model QpBf is simulated with a noisy time step followed by application of all stabilizer measurements, repeated four times ({\it i.e.}, one time step per error correction round).  Measurements are defined according to Eq.~(\ref{eq:Pauli-M-mapping}).  We apply the minimum weight perfect matching decoder to the last syndrome to be repeated.  If no syndrome is repeated, we select the last syndrome.

To check that conditions {(EC~A)} and {(EC~A')} hold, first consider a single measurement bit-flip occuring, which flips one of the four syndromes.  Our protocol ensures that we select the correct syndrome, thus we correctly ignore the measurement bit-flip.  Now, suppose a pair-wise dephasing event occurs in the first, second, or third time step.  The third and fourth syndromes will then agree, and correctly reflect the error applied to the system, thus the error correction step at the end of the protocol will return the system to the error-free code state.  If the pair-wise dephasing event occurs in the fourth time step, we will select the syndrome indicating no error has occurred: this will neither correct nor worsen the error, so that in the subsequent error correction round the error can be corrected. Finally, a quasiparticle event followed by a relaxation has the net effect of applying a Pauli operator and is thus correctable (either in the same error correction round or the subsequent round depending during which time step the initial excitation occurs); the intermediate syndrome will be interpreted as a measurement error and thus does not worsen the error.

For condition (EC~B), we need to show that any two faults occuring within an error correction round can be corrected either in the same error correction round or in the next.  
 If neither of the faults are measurement bit-flips, then they are correctable for the same reasons given for model Qp (although possibly not until the next error correction round if they occur in the fourth time step). 
If only one fault is a measurement bit-flip, it will not appear in the selected syndrome and will not affect the correction protocol, and the remaining fault is correctable for the same reasons given in Qp.
Finally, if both faults are measurement bit-flips, they will only affect the correction protocol if they result in the same syndrome in the second and third, or the third and fourth measurement repetitions.  The correction step will then at most apply two Pauli operators to the system, which can be corrected in the subsequent error correction round if no additional faults occur.  
\\

\noindent {\bf MC}:  Model MC is simulated with the error correction round shown in Fig.~\ref{fig:measurement-steps}, with a noisy time step preceding each stabilizer measurement.  The entire error correction round is then repeated four times (total of 16 time steps).  Measurements are defined according to Eq.~(\ref{eq:Pauli-M-mapping}).   We apply the minimum weight perfect matching decoder to the last syndrome to be repeated.  If no syndrome is repeated, we select the last syndrome.   We use the $d=5$ Bacon-Shor code as a $d=3$ code when including correlated events.

Conditions (EC~A)-(EC~B) are satisfied for the same reasons given for QpBf.  Even correlated events map to two-qubit Pauli errors and can thus be corrected for the same reasons as in the qubit system.  Odd correlated events $\gamma_{i,a}\gamma_{i,b}\gamma_{j,c}$ result in the same syndrome as the even correlated event $\gamma_{i,a}\gamma_{i,b}\gamma_{j,c}\gamma_{j,4}$, with $i\neq j$. After relaxation back to the even parity subspace, the net Majorana operator string applied to islands $i$ and $j$ is equivalent to a two-qubit Pauli operator applied in a single time step.  Because there are more time steps for each error correction round, it is even less likely for measurement bit-flips to be repeated in a way that affects the accepted error syndrome.
\\

\noindent {\bf PMC}:  Model PMC is simulated according to the same protocol as MC, with the only difference being that measurements are now defined according to Eqs.~(\ref{eq:PMC-XX}) and (\ref{eq:PMC-ZZ}).

The fault tolerance conditions (EC~A)-(EC~B) are satisfied for the same reasons given in the other noise models, with only a slight modification.  Here, for all non-corner islands, every MZM on the island is involved in some stabilizer measurement, and the syndrome corresponding to a quasiparticle event on that island will match the syndrome corresponding to multiple pair-wise dephasing events.   However, once the island relaxes back to the even parity ground state, the net operator applied to the island will be equivalent to a Pauli operator.  Thus, the syndrome will change, and the intermediate step at which the island was excited will be interpreted as a measurement bit-flip.  

%

\end{document}